\documentclass[10pt]{report}

\usepackage{times}
\usepackage{epsf}
\usepackage[numbers,sort&compress]{natbib}
\usepackage{amsmath}
\usepackage{amsfonts}
\usepackage{amssymb}
\usepackage{bm}
\usepackage{bbm}
\usepackage{graphicx}
\usepackage{epsfig}
\usepackage{latexsym}
\usepackage{nicefrac}
\usepackage{xr}

\usepackage{fancyhdr}
\renewcommand{\chaptermark}[1]%
{\markboth{\MakeUppercase{\chaptername~\thechapter: #1\\[2ex]}}{}}

\def\bbox#1{\mbox{\boldmath$#1$}}

\def\corresponds{{\lower.2ex\hbox{=}}{\rm\kern-.75em^\triangle}}
\def\succsim{\succ\kern-.9em_\sim\kern.3em}
\def\precsim{\prec\kern-1em_\sim\kern.3em}
\def\slantfrac#1#2{\kern1em^{#1}\kern-.3em/\kern-.1em_{#2}}
\def\lfrac#1#2{{}^{#1\!}\kern-.0em/_{#2}}

\def\buildrel#1\under#2{\mathrel{\mathop{\kern0pt #2}\limits_{#1}}}
\def\Res#1{{\buildrel {\scriptstyle #1} \under {\textstyle \rm Res}}\,}
\def\tr{{\rm Tr}\,}
\def\d{{\rm d}}
\def\e{{\rm e}}
\def\ud{{\textstyle{1\over 2}}}

\begin{document}

\pagestyle{empty}

{\bf 

\Large

\vspace*{1.5in}

\flushright{\Large \bf \sf Multi--Instantons and Exact Results I:}
\flushright{\Large \bf \sf Conjectures, WKB Expansions, and 
Instanton Interactions}

\vspace{.5in}

\large
\flushright{Jean Zinn--Justin\\
{\em DAPNIA/CEA Saclay}\\
{\em F-91191 Gif-sur-Yvette, France and}\\
{\em Institut de Math\'{e}matiques}\\
{\em de Jussieu--Chevaleret,}\\
{\em Universit\'{e} de Paris VII, France}\\[2ex]
Ulrich D.~Jentschura \\
{\em University of Freiburg, Germany and} \\
{\em NIST Division 842: Atomic Physics,} \\
{\em Physics Laboratory, USA} }

}

\vfill

\pagestyle{empty}

\newpage

\vspace*{0.0cm}
\begin{center}
\begin{tabular}{c}
\hline
\rule[-3mm]{0mm}{12mm}
{\large \sf Multi-Instantons and Exact Results I:}\\
\rule[-5mm]{0mm}{12mm}
{\large \sf Conjectures, WKB Expansions, and Instanton Interactions}\\
\hline
\end{tabular}
\end{center}
\vspace{0.0cm}
\begin{center}
Jean Zinn--Justin\\
\vspace{0.2cm}
\scriptsize
{\em DAPNIA/DSM$^{*,**}$\\
Commissariat \`{a} l'\'{E}nergie Atomique, Centre de Saclay,
F-91191 Gif-sur-Yvette, France and\\
Institut de Math\'{e}matiques
de Jussieu--Chevaleret, Universit\'{e} de Paris VII, France}
\end{center}

\normalsize
\begin{center}
Ulrich D. Jentschura\\
\vspace{0.2cm}
\scriptsize
{\em 
Physikalisches Institut der Universit\"{a}t Freiburg,\\
Hermann--Herder--Stra\ss{}e 3, 79104 Freiburg im Breisgau, Germany and\\
National Institute of Standards and Technology,
Gaithersburg, MD20899-8401, Maryland}
\end{center}
\vspace{0.3cm}
\begin{center}
\begin{minipage}{14.0cm}
{\underline{Abstract}}
We consider specific quantum mechanical
model problems for which perturbation theory fails to explain physical
properties like the eigenvalue spectrum even qualitatively, 
even if the asymptotic perturbation series is augmented by
resummation prescriptions to ``cure'' the divergence in large orders of
perturbation theory. Generalizations of perturbation theory
are necessary which include instanton configurations, characterized by
nonanalytic factors $\exp(-a/g)$ where $a$ is a constant 
and $g$ is the coupling. 
In the case of one-dimensional quantum mechanical
potentials with two or more degenerate minima, the
energy levels may be represented as an infinite sum of terms 
each of which involves a certain power of a 
nonanalytic factor and represents itself an infinite divergent series. 
We attempt to provide a unified representation of 
related derivations previously found scattered
in the literature. 
For the considered quantum mechanical problems, we discuss the derivation
of the instanton contributions from a semi-classical calculation of the
corresponding partition function in the path integral formalism.  We also
explain the relation with the corresponding WKB expansion of the solutions
of the Schr\"odinger equation, or alternatively of the Fredholm
determinant $\det(H-E)$ (and some explicit calculations that verify this
correspondence). We finally recall how these conjectures naturally emerge
from a leading-order summation of multi-instanton contributions to the
path integral representation of the partition function. The same strategy
could result in new conjectures for problems where our present
understanding is more limited.
\end{minipage}
\end{center}

\vspace{0.6cm}

Asymptotic problems and properties\\
\vfill
\begin{center}
\begin{minipage}{15cm}
\begin{center}
\hrule
{\bf \scriptsize
\noindent electronic mail: zinn@spht.saclay.cea.fr.\\[2ex]
\noindent ${}^{*}$D\'{e}partment d'astrophysique, de physique des particules,
de physique nucl\'{e}aire et de l'instrumentation associ\'{e}e\\
\noindent ${}^{**}$ Laboratoire de la Direction des
Sciences de la Mati\`ere du Commissariat \`{a} l'Energie Atomique}
\end{center}
\end{minipage}
\end{center}

\newpage

\tableofcontents

\newpage

\clearpage\fancyhead[R]{\normalsize \rightmark}
\pagestyle{fancy}

%
%
\chapter{Introduction}
\label{intro}

The divergence of perturbation theory in higher orders and the asymptotic
character of the perturbation series have led to fairly 
extensive investigations on large-order perturbation 
theory~\cite{LGZJ1990long}. 
Here, we are
concerned with time-dependent,
classically allowed field configurations (or classical
particle trajectories that solve the Euclidean equations of motion) for
which the contribution to the field energy (the classical Euclidean
action) is finite. These configurations give dominant contributions the
generating functionals (partition functions) in quantum field theory
(quantum mechanics). The instantons manifest themselves in a particularly
clear manner in the path integral representation from which they were
originally derived. However, the analysis suffers, even in comparatively
simple cases, from a number of mathematical subtleties, and confusion may
easily arise with regard to a number of necessary analytic continuations
that assign a well--defined meaning to the resulting infinite series.
Here, we are concerned with a number of quantum mechanical model problems
in which instantons play a crucial role: in the cases considered,
perturbation theory fails to account for the eigenvalue spectrum even
qualitatively; two or more distinct energy levels are characterized by one
and the same perturbation series. The degeneracy which persists after the
application of perturbation theory is lifted only when instanton
configurations are additionally taken into account. The example to be
considered is the quantum mechanical double-well oscillator in which the
particle trajectory may describe an arbitrary number $n$ of oscillations
between the two minima (and we also consider generalizations of this
problem with more than two minima). The Euclidean action of two
oscillations between the minima is found to differ slightly from the sum
of two single oscillations, the difference giving rise to a modification
of the contribution to the path integral and the partition function: this
is the origin of the instanton interaction and the multi-instantons. The
result of the investigations is that the energy eigenvalue is given as an
infinite sum over series each of which characterizes a particular
configuration in which the particle oscillates $n$ times between the
minima, including the modifications to the resulting Euclidean action (the
$n$-instanton configuration).

Let us consider a double-well oscillator with a coupling $g$ whose inverse
is proportional to the spatial separation of the two minima. An expansion
in powers of $g$ then corresponds to an expansion about the case of
infinite separation of the two minima (i.e., about the case where the
other minimum at $x = \infty$ can be ignored and a single minimum
results). This cannot be assumed to be satisfactory because a situation
with two degenerate minima is qualitatively different from a potential
with a single minimum. Instanton configurations provide the necessary
supplement to perturbation theory, which is a pure expansion in powers of
$g$. The expansion of the eigenvalue in the instanton formalism is
systematic and leads to a well-defined representation as a triple series
in (i) the nonperturbative expression $\exp(-1/g)$, raised to the power
$n$ of the number of instanton oscillations, (ii) logarithms of the form
$\ln(-g)$ which are caused by proportionality factors multiplying the
instanton interaction, and (iii) powers of the coupling $g$.
The logarithms $\ln(-g)$, for positive coupling $g > 0$, give rise to imaginary
contributions which are compensated by imaginary parts of generalized 
Borel sums of lower-order instanton configurations, resulting in a real
energy eigenvalue, as it should be. The necessary analytic continuations have
given rise to some discussion and confusion in the past; today, we can 
verify the validity of this procedure by comparing the ``predictions'' of 
the multi-instanton expansion, including the inherent analytic continuations,
with numerically determined energy eigenvalues of up to 180 decimal figures.

Let us briefly expand on the connection between the generalized 
expansions discussed here and the theory of differential equations
as developed by Riemann. It is well known that the hypergeometric equation
is a second-order differential equation with three singularities at
$z=0,1,\infty$. The solutions of this equation, the hypergeometric functions,
can be expressed as power series in $z$. If one assumes further 
singularities, the solutions are no longer expressible in terms of
simple power series. Instead, they may be expressed as series of the 
form $c \, z^a \, (\ln z)^p$ ($c$ constant, $a \in C$, $p$ integer,
$p \geq 0$ confined). The Riemann--Hilbert theorem shows that 
they are essentially characterized by the way in which they reproduce 
themselves under analytic continuations along closed paths in the 
complex plane. An equation with essential singularities admits solutions
which are expressible as a sum of terms of the form 
$c \, \exp[q(z)\, z^{-d}] \, z^a \, (\ln z)^p$ 
($c$ constant, $q$ a polynomial, $d$ positive and integer,
$a \in C$, $p$ integer, $p \geq 0$ confined). Inspired by the 
ideas expressed in the volume \cite{Bo1994long}, we call expansions of
this type ``resurgent expansion.''

Here, we discuss in detail a set of conjectures about the complete form of the
perturbative expansion of the spectrum of a quantum Hamiltonian $H$, in
situations where the potential has degenerate minima
\cite{ZJ1981jmplong,ZJ1981npb,ZJ1983npb,ZJ1984jmp,Bo1994long}. 
It will be shown that the coupling parameter $g$
takes the formal role of the natural unit of action $\hbar$, and 
the systematic expansion about the classical problem (expansion in 
power of $\hbar$ in the form of a WKB expansion) will be used in order
to derive the resurgent expansions in a systematic way.
Perturbative expansions are obtained by
first approximating the potential by a harmonic potential near its
minimum. They are expansions for $\hbar \to 0$ valid for energy
eigenvalues of order $\hbar$, in contrast with the WKB expansion where
energies are non-vanishing in this limit (WKB approximation applies to
large quantum numbers). When the potential has degenerate minima,
perturbation series can be shown to be non-Borel summable. Moreover,
quantum tunneling generates additional contributions to eigenvalues of
order $\exp(-{\rm const.} \ /\hbar)$, which have to be added to the
perturbative expansion (for a review and more detail about barrier
penetration in the semi-classical limit see for 
example~\cite{ZJ1996ch43}).

Therefore, the determination of eigenvalues starting from their
expansion for $\hbar$ small is a non-trivial problem. The conjectures we
describe here, give a systematic procedure to calculate eigenvalues, for
$\hbar$ finite, from expansions which are shown to contain powers of
$\hbar$, $\ln\hbar$ and $\exp(-{\rm const.}\ /\hbar)$, i.e.~resurgent
expansions. Moreover,
generalized Bohr--Sommerfeld formulae allow us to derive the many series
which appear in such formal expansions from only two of them, which can be
extracted by suitable transformations from the corresponding WKB
expansions. Note that the relation to the WKB expansion is not
completely trivial. Indeed, the perturbative expansion corresponds (from
the point of view of a semi-classical approximation) to a situation with
confluent singularities and thus, for example, the WKB expressions for
barrier penetration are not uniform when the energy goes to zero. 

These concepts will be applied to the following 
classes of potentials,
\begin{itemize}
\item the double-well potential,
\item more general symmetric potentials with degenerate 
minima,
\item a potential with two equal minima but asymmetric wells,
\item a periodic-cosine potential,
\item resonances of the ${\mathcal O}(\nu)$-symmetric anharmonic oscillator,
for negative coupling,
\item eigenvalues of the ${\mathcal O}(\nu)$-symmetric anharmonic oscillator,
for negative coupling but with the Hamiltonian 
endowed with nonstandard boundary conditions,
\item a special potential which has the property 
that the perturbative expansion of the ground-state energy 
vanishes to all orders of the coupling constant (Fokker--Planck 
Hamiltonians).
\end{itemize}

Based on the extensive mathematical work carried out on the 
subject \cite{Bo1994long}, it is perhaps rather natural to remark 
that several of these conjectures have meanwhile 
found a natural explanation in
the framework of Ecalle's theory of resurgent 
functions~\cite{Ec1981} and have now been
proven by Pham and his collaborators 
\cite{Ph1988,DeDiPh1990,DeDi1991,De1992}.
In what follows, we always assume that the potential is an analytic entire
function. This is also the framework in which the available
mathematical proofs exist.
We first explain the conjecture in the case of the so-called {\it
double-well}\/ potential.
Note that, in what follows, the symbol $g$ plays the role of $\hbar$ and
the energy eigenvalues are measured in units of $\hbar$, a normalization
adapted to perturbative expansions.
Although several conjectures have now been
proven, we believe that heuristic arguments based on instanton
considerations are still useful because, suitably generalized, they could
lead to new conjectures in more complicated situations.

Instantons have played a rather significant role in explaining
unexpected behaviour in several aspects of quantum field theory,
for example the nontrivial structure of the quantum chromodynamic
vacuum or the absence of the ``superfluous'' $\eta'$ meson. Instantons
effectively eliminate the the $U(1)$ symmetry that would generates the 
$\eta'$ particle, eliminating the problem. In the case of the one-dimensional
problems studied in the current review, the instanton contributions can 
be subjected to a systematic study, including an investigation of the 
interaction among instantons which leads to a systematic expansion 
of the energy eigenvalues in $\hbar$, $\ln\hbar$ and 
$\exp(-{\rm const.}\ /\hbar)$.

This review is organized as follows. 
In chapter \ref{BSqf}, we describe the conjectures. In chapter
\ref{BSWKB}, we discuss the connection with the WKB expansion as
derived by different methods from Schr\"odinger equation. In chapter
\ref{ssninstdw}, we explain how these conjectures were suggested by a
summation of the leading order {\em multi-instanton}\/ contributions to
the quantum partition function $\tr \exp(-\beta H)$ in the path integral
representation. In chapter \ref{ssninstcos}, we discuss in the same way
the periodic cosine potential, and in chapter \ref{ssninstg} more
general potentials with degenerate minima. In chapter \ref{ssOnuanhar},
related problems in the example of the ${\mathcal O}(\nu)$ quartic anharmonic
oscillator are considered. Finally, the
appendix contains additional technical details and explicit calculations
verifying some implications of the conjectures at second WKB order.

%
%
\chapter{Generalized Bohr--Sommerfeld Quantization Formulae}
\label{BSqf}

%
%
\section{Double--Well Potential}
\label{sDouble}

%
%
\subsection{Resurgent Expansion (Double--Well)}
\label{ssResurgent}

\begin{figure}[htb!]
\begin{center}
\begin{minipage}{12.0cm}
\begin{center}
\epsfxsize=121.mm
\epsfysize=85.mm
\centerline{\epsfbox{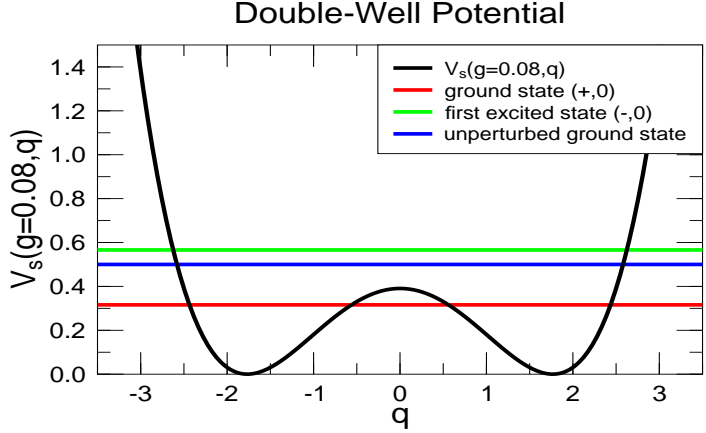}}
\caption{\label{figpot}
Double-well potential for the case $g = 0.08$ with 
the two lowest eigenvalues and the unperturbed 
eigenstate. The energy of the unperturbed ground state is 
the energy of the ($N=0$)-state of the harmonic oscillator 
ground state with $E=1/2$. The quantum
numbers of the ground state in the potential
$V_s(g=0.08,q)$ are $(+,0)$ (positive parity and 
unperturbed oscillator occupation quantum number $N=0$).
The potential $V_s$ is defined in (\ref{origpot}).
The energy eigenvalue corresponding to the 
state $(+,0)$ is $E_{+,0} = 0.317\,851\,364\,6\dots$. 
The first excited state $(-,0)$
has an energy of $E_{-,0} = 0.566\,114\,759\,4\dots$ 
The arithmetic mean of the 
energies of the two states $(\pm,0)$ is $0.441\,983\,062\,0\dots$
and thus different from the value $1/2$. This is a consequence of the 
two-instanton (more general: even-instanton) shift of the energies.}
\end{center}
\end{minipage}
\end{center}
\end{figure}

Let us consider a quantal particle bound
in a double-well potential which we write
in the symmetric form 
\begin{equation}
V_s(g,q) =
\frac{g}{2}\,
\left(\frac{1}{2 \sqrt{g}} + q\right)^2 \,
\left(\frac{1}{2 \sqrt{g}} - q\right)^2.
\label{origpot}
\end{equation}
The minima are at $q = \pm 1/\sqrt{g}$. The potential 
is represented graphically in figure~\ref{figpot}, with an indication of the 
ground state and  the first excited state ($g = 0.08$).
Further calculations profit from a translation of the coordinate 
\begin{equation}
q \to q + \frac{1}{2\sqrt{g}}.
\end{equation}
This translation preserves the spectrum of the
Hamiltonian and 
leads to the quantum mechanical eigenvalue problem 
\begin{equation}
H \psi(x) = E \, \psi(x)
\end{equation}
in $L^2(R)$. The Hamiltonian after the translation is
\begin{equation}
H = -{1 \over 2}\, \left({ \d  \over \d  q} \right)^2  + V(g,q)\,,
\qquad V(g,q) = {1 \over 2}\, q^2 \, (1 - \sqrt{g} \, q)^2. 
\label{ehamgq}
\end{equation}
For $g \to 0$, we recover the unperturbed harmonic oscillator.
Consequently, the unperturbed ground state in figure
\ref{figpot} has an energy of $E = 1/2$.
When Rayleigh--Schr\"{o}dinger
perturbation theory is applied 
to the $N$th harmonic oscillator state,
the energy shift due to the terms of order
$g^3$ and $g^4$ in $V(g, q)$
[see equation (\ref{ehamgq})] gives rise to 
a formal power expansion in $g$ (not $\sqrt{g}$),
\begin{equation}
E_N \sim \sum_{l=0}^\infty E^{(0)}_{N,l} \, g^l\,.
\label{esim}
\end{equation}
For the ground state ($N=0$), the first terms read
\begin{equation}
E_0 \sim {1 \over 2} - g - {9 \over 2} \, g^2 - {89 \over 2} \, g^3
-{5013 \over 8} g^4 - \dots
\label{esimground}
\end{equation}
The coefficients in this series grow factorially, and the series is 
not Borel summable (all coefficients except the leading term have negative
sign (a complete list of the leading 300
perturbative coefficients for the ground
state can be found at~\cite{JeHome}).
In higher orders of perturbation theory, the numerators and 
denominators of the coefficients exhibit rather prime factors. 
For instance, we have 
\begin{equation}
\label{eE017}
E_{0,17} = - {3^2 \over 2^{17}}\,
130\,116\,860\,668\,372\,133\,614\,952\,623.
\end{equation}
Recently, connections between number theory and physical perturbation
theory have attracted considerable attention. We will not pursue
these questions any further in the current work.

We continue the discussion by observing
that for negative $g$, the series (\ref{esim}) and in particular
(\ref{esimground}) are alternating and 
indeed Borel summable. 
One might now expect that a suitable analytic continuation
of the Borel sum from negative to positive $g$ gives the desired 
result for the energy levels of the double-well problem.
That is not the case. Perturbation theory fails to account for the 
spectrum even qualitatively. For every one eigenstate of the unperturbed
problem ($g=0$), two eigenstates emerge in the case of nonvanishing
coupling $g > 0$. For example, the unperturbed 
ground state with an energy $E= 1/2$ splits into two states with 
different energy for $g > 0$, as shown in figure \ref{figpot}.
The new quantum number that differentiates the 
two states is the parity $\varepsilon = \pm 1$ 
[see (\ref{ehamdw}) below].
Indeed, as is well known, the ground state with $N = 0$ ``splits up'' into 
two states each of which has opposite parity. The two levels
that emerge from the ground state as the coupling $g$ is ``switched on''
are separated by an energy interval of 
\begin{equation}
\approx {2 \over \sqrt{\pi\,g}} \, \exp\left( -{1 \over 6\,g} \right)\,.
\label{separation}
\end{equation}
This term is nonperturbative in $g$, i.e. its formal power series
in $g$ vanishes to all orders.

%
%

For all further discussions, we scale the coordinate
$q$ in (\ref{ehamgq}) as
\begin{equation}
q \to {1 \over \sqrt{g}} \, q.
\end{equation}
This scale transformation again conserves all eigenvalues.
The Hamiltonian corresponding to the double-well potential can 
therefore be written alternatively as
\begin{subequations}
\label{ehamdw}
\begin{eqnarray} 
\label{econvention}
H &=& -{g\over2} \left({ \d  \over \d  q} \right)^2 + 
{1\over g} \, V(q),\\
\label{epotdw}
V(q) &=& {1\over2} \, q^2 (1-q )^2\,. 
\end{eqnarray}
\end{subequations}
This representation shows that $g$ takes over the formal
role of $\hbar$.
The Hamiltonian is symmetric in the exchange $q \leftrightarrow  1-q$
and thus commutes with the corresponding parity operator $P$,
whose action on wave functions is
\begin{equation}
P\,\psi(q)=\psi (1-q)\ \Rightarrow [H,P]=0\,. 
\label{edwparity}
\end{equation}
The eigenfunctions of $H$ satisfy
\begin{equation}
H\,\psi_{\varepsilon,N}(q) =
E_{\varepsilon ,N}(g)\psi_{\varepsilon,N}(q)\,,
\quad
P\,\psi_{\varepsilon,N}(q) =
\varepsilon \, \psi_{\varepsilon,N}(q), 
\end{equation}
where $\varepsilon =\pm1$ is the parity 
and the quantum number $N$ can be uniquely assigned
to a given state by the 
requirement $E_{\varepsilon ,N}(g)=N+1/2+{\mathcal O}(g)$,
i.e.~by identifying the specific unperturbed harmonic oscillator eigenstate
to whose energy eigenvalue the state energy converges in the 
limit $g \to 0$. We have conjectured~\cite{ZJ1981npb,ZJ1983npb} 
that the eigenvalues 
$E_{\varepsilon ,N}(g)$ of the Hamiltonian (\ref{ehamdw}) have a 
complete semi-classical resurgent expansion of the form
\begin{equation} 
\label{ecomexp}
E_{\varepsilon,N}(g)= 
\sum^{\infty}_{l = 0} E^{(0)}_{N,l} \, g^{l} + 
\sum^{\infty}_{n=1} \left(2\over g\right)^{Nn}\,
\left( - \varepsilon {\e^{-1/6g}\over\sqrt{\pi g}} \right)^{n} \,
\sum^{n-1}_{k=0} \left\{ \ln\left(-\frac{2}{g}\right) \right\}^k \,
\sum^{\infty}_{l=0} e_{N,nkl} \, g^{l}. 
\end{equation} 
The expansion (\ref{ecomexp}) is the sum of the perturbative
expansion which is independent of the parity,
\begin{equation}
\label{ePert}
E^{(0)}_N (g )= \sum^\infty_{l=0} E^{(0)}_{N,l} \, g^{l}
\end{equation}
and of the $n$-instanton contributions, each of which is
given by
\begin{equation}
E^{(n)}_{\epsilon,N}(g) = 
\left(2\over g\right)^{Nn}\,
\left( - \varepsilon {\e^{-1/6g}\over\sqrt{\pi g}} \right)^{n} \,
\sum^{n-1}_{k=0} \left\{ \ln\left(-\frac{2}{g}\right) \right\}^k \,
\sum^{\infty}_{l=0} e_{N,nkl} \, g^{l}\,.
\label{eInstn}
\end{equation}
Here, $E^{(0)}_N(g)$ is the perturbation series
whose coefficients $E^{(0)}_{N,l}$ 
only depend on the quantum number $N$ (and 
the order of perturbation theory $l$), but not on the 
parity $\varepsilon$ [see also (\ref{esim})]. The index $n$ gives the order
of the ``instanton expansion''. One-instanton effects are 
characterized by a single power of the nonperturbative 
factor $\exp(-1/6g)$, two-instanton effects
are multiplied by  $\exp(-1/3g)$, etc. 
Actually, as shown below in chapters~\ref{ssninst} and~\ref{sSpecific}, 
the one-instanton effect $E^{(1)}_{\epsilon,N}(g)$
involves a summation over $n$-instanton
configurations, however neglecting the instanton interaction.
The two-instanton shift $E^{(2)}_{\epsilon,N}(g)$ and higher-order
corrections involve the instanton interaction.

The logarithms
of the form $\ln(-2/g)$ also enter into the formalism, their
maximum power (for given $n$) is $n-1$. Thus, the 
two-instanton effect carries one and only one logarithm etc.
Each of the infinite series
\begin{equation}
\sum^{\infty}_{l=0} e_{N,nkl}g^{l}
\end{equation}
is characterized by the state quantum number $N$, the 
instanton order $n$ and the power of the logarithm $k$.
The series $\sum e_{N,nkl}g^{l}$ in powers of $g$ are not Borel 
summable for $g>0$ and have to be summed for $g$ negative. 
The values for $g$ positive are then obtained by analytic 
continuation, consistently with the determination of $\ln(-g)$. 
In the analytic continuation from $g$ negative to $g$ positive, 
the Borel sums become complex with imaginary parts exponentially smaller 
by about a factor $\exp(-1/3g)$ than the real parts. 
These imaginary contributions are canceled by the imaginary parts 
coming from the function $\ln(-2/g)$. For instance, the imaginary part
incurred by analytic continuation of the perturbation series 
[equation (\ref{ePert}) below] is
canceled by the ``explicit'' imaginary part originating from the 
two-instanton effect [the term with $n=2$ in (\ref{ecomexp})].

The instanton terms $E^{(n)}_{\epsilon,N}(g)$ are parity-dependent for odd $n$.
In \cite{JeZJ2001long}, a different notation was used
for the coefficients that enter into the expansion
(\ref{ecomexp}). When comparing formulas,
observe that in~\cite{JeZJ2001long},
$\epsilon$ denotes the coefficient, whereas here,
the slightly different symbol
$\varepsilon = \pm$ denotes the parity, and we have
\begin{equation}
\epsilon^{(N,\varepsilon)}_{nkl} \equiv (-\varepsilon)^n \, e_{N,nkl}\,.
\end{equation}
The $e$-coefficients allow for a more compact notation.
Their introduction is based on the observation that the odd-instanton
coefficients have different sign for states with different parity,
whereas the even-instanton contributions have the same sign for
states with opposite parity. Thus, odd instantons contribute
to the separation of energy levels with the same $N$, but opposite
parity, whereas the energy shift due to even instantons
is the same for both levels with the same $N$, irrespective of the 
parity. To give an example,
we quote here
from~\cite{JeZJ2001long}, without proof,
several known coefficients for the 
one- and two-instanton contributions, 
from which the parity/sign pattern is apparent:
\begin{subequations}
\label{epsiloncoeff}
\begin{eqnarray}
\epsilon^{(0,+)}_{100} &=& - \epsilon^{(0,-)}_{100} = -1 \,, \quad
\epsilon^{(0,+)}_{101} = - \epsilon^{(0,-)}_{101} = \frac{71}{12} \,,
\\
\epsilon^{(0,+)}_{101} &=& - \epsilon^{(0,-)}_{101} = \frac{6299}{288}
\,, \quad
\epsilon^{(0,+)}_{210} = \epsilon^{(0,-)}_{210} = 1 \,,
\\
\epsilon^{(0,+)}_{211} &=& \epsilon^{(0,-)}_{211} = -\frac{53}{6} 
\,, \quad
\epsilon^{(0,+)}_{212} = \epsilon^{(0,-)}_{212} = -\frac{1277}{72} \,,
\\
\epsilon^{(0,+)}_{200} &=& \epsilon^{(0,-)}_{200} = \gamma
\,, \quad
\epsilon^{(0,+)}_{201} =
\epsilon^{(0,-)}_{201} = -\frac{23}{2} - \frac{53}{6} \, \gamma \,,
\\
\epsilon^{(0,+)}_{202} &=&
\epsilon^{(0,-)}_{202} = \frac{13}{12} - \frac{1277}{72} \, \gamma \,.
\end{eqnarray}
\end{subequations}
Here, $\gamma = 0.57221\dots$ is Euler's constant.
Utilizing the $e$-coefficients which enter into
equation (\ref{ecomexp}),
these results can be summarized in a much more compact form as
\begin{subequations}
\label{ecoeff}
\begin{eqnarray}
e_{0,100} &=& 1 \,, \quad
e_{0,101} = -\frac{71}{12} \,, \quad
e_{0,101} = -\frac{6299}{288}\,, 
\\
e_{0,210} &=& 1 \,, \quad
e_{0,211} = -\frac{53}{6} \,, \quad
e_{0,212} = -\frac{1277}{72} \,,
\\
e_{0,200} &=& \gamma\,,\quad
e_{0,201} = -\frac{23}{2} - \frac{53}{6} \, \gamma \,,\quad
e_{0,202} = \frac{13}{12} - \frac{1277}{72} \, \gamma \,.
\end{eqnarray}
\end{subequations}
The coefficient $e_{0,100}$ gives the one-instanton separation
of the levels with quantum numbers $(\pm,0)$ which 
can be approximated as
\begin{equation}
E_{-,0} - E_{+,0} \approx -{2 \over \sqrt{\pi\,g}} \,
\exp\left( -{1 \over 6\,g} \right)\,e_{0,100},
\end{equation}
in agreement with (\ref{separation}). 
The well-known, classic treatment of the double-well
problem is based on the ``matched addition'' or
``matched subtraction'' of the two solutions that characterize
the two solutions in each single well. Only the leading term
in the quantity $\exp(-1/6g)$ can be recovered in this
way \cite{LaLi1958}. The advantage of the expansion
(\ref{ecomexp}) is that it generalizes the classic treatment
to all orders in $\exp(-1/6g)$, thereby giving a clear interpretation
of the way in which the many-instanton contributions manifest themselves
in a generalized resurgent expansion in higher orders in $\exp(-1/6g)$.
Also, the expansion (\ref{ecomexp}) gives a systematic
understanding of the error incurred upon 
terminating the instanton expansion at a given order.

%
%
\subsection{Quantization Condition (Double--Well)}

We have conjectured~\cite{BeWu1973long}
that all series that enter into the 
generalized resurgent 
expansion (\ref{ecomexp}) can be obtained by a
small-$g$ expansion of 
a generalized Bohr--Sommerfeld quantization
formula.
In the case of the double-well potential,
this quantization condition reads
\begin{equation}
{1 \over \sqrt{2\pi}} \, \Gamma\left( {1 \over 2} - B_{\rm dw}(E,g) \right) \,
\left(- {2 \over g} \right)^{B_{\rm dw}(E,g)} \, 
\exp\left[-{A_{\rm dw}(E,g) \over 2}\right] = 
\varepsilon {\rm i}\,,
\label{equantization}
\end{equation}
or alternatively
\begin{equation}
\left[\Gamma\left(\frac12 - B_{\rm dw}(E,g)\right)\right]^{-1} +
{\varepsilon {\rm i} \over \sqrt{2\pi}} \left(-{2 \over g}
\right)^{B_{\rm dw}(E,g)}\,\exp\left[-{A_{\rm dw}(E,g) \over 2}\right]=0\,, 
\label{egenisum}
\end{equation}   
($\varepsilon=\pm 1$) where the $A$ and the $B$-functions
enjoy symmetry properties and can be expressed as formal
series in their arguments,
\begin{subequations}
\label{defABdw} 
\begin{eqnarray}
B_{\rm dw}(E,g) &=& -B_{\rm dw}(-E,-g) =
E + \sum^{\infty}_{k=1} g^{k} \, b_{k+1}(E), 
\label{defB} 
\\
A_{\rm dw}(E,g) &=& -A_{\rm dw}(-E,-g) = 
{1 \over 3g}+ \sum^{\infty}_{k=1} g^{k} \, a_{k+1}(E) \,. 
\label{defA}
\end{eqnarray} 
\end{subequations}
The coefficients $b_k(E)$ and $a_k(E) $ are odd or even 
polynomials in $E$ of degree $k$,
as necessitated by the symmetry properties. 
The three first orders, for example, are 
\begin{subequations}
\begin{eqnarray}
B_{\rm dw}(E,g) &=& E+
g \, \left(3 \, E^2 + {\displaystyle{1\over4}}\right)+
g^2 \, \left(35 \, E^3+{\displaystyle{25 \over4}}E\right) + 
{\mathcal O}\left(g^3 \right),
\label{eBdble} \\ 
A_{\rm dw}(E,g) &=& {\displaystyle{1 \over 3}}\,g^{-1}+ 
g\, \left(17\,E^2+{\displaystyle{19 \over 12}} \right)
\nonumber\\
& & + g^2\, \left(227\,E^3+{\displaystyle{187\over4}}E\right)+
{\mathcal O}\left(g^3 \right).
\label{eAdble} 
\end{eqnarray}  
\end{subequations}
The function $B_{\rm dw}(E,g)$ has been inferred from 
the perturbative expansion, and indeed, the 
perturbative quantization condition reads
\begin{equation}
\label{equantpertdw}
B_{\rm dw}(E,g) = N + \ud\,.
\end{equation}
Here, $N$ is a nonnegative integer.
The perturbative expansion for an energy level is obtained 
by solving this equation for $E$.
The function $A_{\rm dw}(E,g)$ had initially been determined at this order 
by a combination of analytic and numerical calculations.

The formula (\ref{ecomexp}) is an exact expansion for the 
energy eigenvalues of the double-well anharmonic oscillator
at finite $g$, which can be inferred by a systematic 
expansion of the quantization condition (\ref{egenisum}) for small $g$. 
The generalized resurgent expansion (\ref{ecomexp})
can be derived either by a systematic
expansion of the path integral representing the partition function
about nontrivial saddle points of the Euclidean action, or by a
systematic investigation of the WKB expansion. Both of 
these methods will be discussed in the sequel.

We illustrate the statements made 
by considering some example calculations.
Specifically, we start by investigating
approximations to the energies of the
states $(+,0)$ and $(-,0)$ which emerge from 
the unperturbed oscillator ground state.
The resurgent expansion is explicitly 
given in (\ref{ecomexp}),
with the first perturbative terms provided in (\ref{esimground})
and the leading instanton coefficients (\ref{ecoeff}).
Neglecting higher-order perturbative 
(in $g$) and higher-order instanton
effects [in $\exp(-1/6g)$], the energies of the two lowest-lying 
states are given by
\begin{equation}
E_{\varepsilon,0} \approx {1 \over 2} - {\varepsilon \over \sqrt{\pi g}} \,
\exp\left[ -{1 \over 6\,g} \right].
\end{equation}
Here, $\varepsilon = \pm$ is again the parity.
We define the abbreviations
\begin{equation}
\xi(g) = \frac{1}{\sqrt{\pi g}} \, \exp\left[ -{1 \over 6\,g} \right]
\label{defxi}
\end{equation}
and 
\begin{equation}
\label{defchi}
\chi(g) = \ln\left( - \frac{2}{g} \right)\,.
\end{equation}
An expansion of the quantization conditions
(\ref{equantization}) and
(\ref{egenisum}) for small $\xi(g)$ 
[note that a small positive $g$ implies a small $\xi(g)$] then
involves an evaluation of the expression
\begin{equation}
\Gamma\left( {1 \over 2} - B_{\rm dw}(E_{\varepsilon,0},g) \right) \approx
\Gamma\left( \varepsilon \, \xi(g) \right) \approx 
\varepsilon \, {1 \over \xi(g)} =
\varepsilon \, {\sqrt{\pi g} \over \exp\left[ -{1 \over 6\,g} \right]}.
\end{equation}
The quantization condition (\ref{equantization}) then 
takes the form
\begin{equation}
{1 \over \sqrt{2\pi}} \,
\left( \varepsilon \,  {\sqrt{\pi g}} \over 
\exp\left[ -{1 \over 6\,g} \right]\right) \,
\left( - {2 \over g} \right)^{1/2} \,
\exp\left[ -  {1 \over 2} \, {1 \over 3\,g} \right] \approx
\varepsilon {\rm i}.
\end{equation}
This equation is approximately fulfilled provided that 
the analytic continuation of the square root is carried out such that
for $g > 0$
\begin{equation}
\left( - {2 \over g} \right)^{1/2} \to 
{\rm i}\, \left({2 \over g} \right)^{1/2} \,.
\end{equation}
A systematic expansion of the quantization conditions 
(\ref{equantization}) or alternatively (\ref{egenisum}) 
in powers of $\chi(g)$, 
$\xi(g)$ and $g$ then leads in a natural way
to the generalized resurgent expansion
(\ref{ecomexp}). We recall that some of the known coefficients for the 
ground state have been given in the form of the 
$\epsilon^{(N,\pm)}_{nkl}$ (\ref{epsiloncoeff}) or 
in the form of the $e_{N,nkl}$-coefficients 
(\ref{ecoeff}). We now turn to a discussion of states with general $N$.
The treatment is simplified 
if we restrict the discussion
to the dominant effects in each order of the 
instanton expansion, i.e. if we restrict the discussion
to the terms with $l=0$ in the resurgent expansion (\ref{ecomexp}).
We employ the approximation 
\begin{equation}
B_{\rm dw}(E,g) \to E
\end{equation}
in (\ref{defB}) and obtain the following approximative 
quantization condition
\begin{equation}
{\e^{-1/6g}\over \sqrt{2\pi}}\left(-{2\over g}\right)^{E} =
-{ \varepsilon {\rm i} \over \Gamma({1 \over 2} - E)} \ 
\Leftrightarrow \ 
{\cos \pi E \over \pi} = \varepsilon {\rm i}\,
{\e^{-1/6g} \over \sqrt{2\pi}} \,
\left(-{2\over g}\right)^{E}{1 \over \Gamma({1 \over 2} + E)}\, . 
\label{ssninstlead}
\end{equation}
We show in chapter~\ref{ssninstdw} that, from the point of 
view of the path integral representation, 
the successive terms correspond to {\em multi-instanton}\/ contributions,
i.e.~to contributions which are generated by the modification
of the instanton action due to the overlap region which results
when ``gluing together'' two (or more) instanton solutions to 
a multi-instanton configuration. 

For example, the term $n=1$ in (\ref{ecomexp}), which is also the 
one-instanton contribution,  is
\begin{equation} 
E^{(1)}_{N} (g)=-{\varepsilon \over N!} \left({2 \over g}
\right)^{N+1/2}{\e^{-1/6g} \over \sqrt{ 2\pi}} \bigl(1+{\mathcal O}(g)\bigr). 
\end{equation} 
This result generalizes the coefficient $e_{0,100} = 1$ 
(\ref{ecoeff}) to the case of arbitrary $N$,
\begin{equation}
\label{eN100}
e_{N,100} = {1 \over N!}.
\end{equation}
The term $n=2$, which is the two-instanton contribution,  is 
\begin{equation} 
E^{(2)}_{N}(g)={1 \over \left(N! \right)^2}
\left({2 \over g} \right)^{2N+1}{\e^{-1/3g} \over 2\pi}
\left[\ln(-2/g)-\psi (N+1 )+
{\mathcal O}\left(g \ln g \right) \right] , 
\end{equation}
where $ \psi $ is the logarithmic derivative of the $\Gamma $-function. 
This result generalizes the coefficient $e_{0,210} = 1$ 
(\ref{ecoeff}) to the case of arbitrary $N$,
\begin{equation}
\label{eN210}
e_{N,210} = \left({1 \over N!}\right)^2.
\end{equation}
and the result $e_{0,200} = \gamma$ to
\begin{equation}
\label{eN200}
e_{N,200} = -\left({1 \over N!}\right)^2 \psi (N+1 ).
\end{equation}

More generally, it can be easily be verified 
upon inspection of (\ref{ecomexp}) 
that the {$ n $-instanton} 
contribution has at leading order the form 
\begin{equation} 
E^{(n)}_{N}(g)=
\left( 2\over g\right)^{n(N+1/2)}
\left( -\varepsilon \frac{\e^{-1/6g}}{\sqrt{ 2\pi}} \right)^{n} 
\left[ P^N_{n} \bigl(\ln(-g/2)\bigr)+
{\mathcal O}\left(g \left( \ln g \right)^{n-1} \right) \right] , 
\label{einstEnn}
\end{equation}
in which $ P^N_{n}(\sigma) $ is a polynomial of degree
$ n-1 $ (not the Legendre polynomial). For example, for $ N=0 $  
one finds 
\begin{equation}
P^0_1(\sigma)= 1 \,, \qquad
P^0_2(\sigma)= \sigma + \gamma\, , \qquad 
P^0_3(\sigma)= {3 \over 2} \, \left(\sigma+\gamma \right)^2 
+{\pi^2 \over 12}\,,  
\label{ePthree} 
\end{equation} 
in which $ \gamma $ is Euler's constant: 
$\gamma =-\psi(1) =0.577215...$.

%
%
\subsection{Large--Order Behaviour (Double--Well)}
\label{ssLargeOrder}

The perturbation series fails to describe the 
energy levels at the one-instanton level.
However, the mean of the energies
of the levels $(+,N)$ and $(-,N)$ may be described
to greater accuracy by the perturbation 
series than each of the 
two opposite-parity 
energy levels, because the one-instanton effect has opposite
sign for opposite-parity state.
The perturbation series, which is 
nonalternating, may be resummed using 
standard techniques; the only difficulty 
which persists is the singularity along the 
integration axis in the Laplace--Borel integral.
However, one may easily define a generalized Borel
transform. The calculation of such a 
generalized transform is rather trivial
and discussed for example in~\cite{Je2000prd}.
In particular, the Borel sum naturally develops an imaginary part
for $g > 0$ which represents a fundamental 
ambiguity. In the case of the quantum electrodynamic
effective action, the imaginary part which describes
pair production emerges from an appropriate analytic 
continuation. In the case of the double-well 
potential, the energy values are real, and 
there are compensating imaginary contributions from
higher-order instanton effects which compensate the 
imaginary part incurred by the Borel sum of the 
nonalternating perturbation series.

The mean of the energies
of the levels $(+,N)$ and $(-,N)$ is described by the 
perturbation series up to a common shift received 
by both levels $(\pm,N)$ via the
{\em two}-instanton effect. Thus, it is up to the 
{\em two}-instanton level that the perturbation series
has a physical meaning: indeed, perturbation
theory describes approximately the mean of the 
energies of the levels with the same $N$, but 
opposite parity. The mean of the energy of these levels,
to better accuracy, is then given as the sum
of the real part of the (complex) Borel sum of the perturbation 
series, and of the two-instanton effect. Thus, 
at the level of the two-instanton effect, we expect 
the asymptotic perturbation series to go 
through its minimal term which, for a factorially
divergent series, is of the same order-of-magnitude 
as the imaginary part of its Borel sum and also 
of the  same order-of-magnitude as the difference
of the real part of the generalized Borel sum 
of the perturbation series and the result obtained
by optimal truncation of that same series.

Thus, after an analytic continuation of the 
series $\sum^{\infty}_{l=0} e_{N,nkl} \, g^{l}$
in (\ref{ecomexp}) from $g$ negative to $g$ positive, 
two things happen: the Borel sums become
complex with an imaginary part exponentially smaller by about 
a factor $\exp(-1/3g)$ than the real part. Simultaneously, the 
function $\ln(-2/g) $ that enters into the 
two-instanton contribution in equation
(\ref{ecomexp}) also becomes complex and gets an imaginary 
part $ \pm {\rm i}\,\pi $. Since the sum of all contributions is real, the 
imaginary parts must cancel. This property relates, for example, the 
non-perturbative imaginary part of the Borel sum of the perturbation 
series to the perturbative imaginary part of the two-instanton 
contribution.
We have identified (in the sense of a formal power series) the 
quantity (\ref{ePert})
\begin{equation}
E^{(0)}_N(g) = \sum^\infty_{l=0} E^{(0)}_{N,l} \, g^{l}
\end{equation}
as the perturbative expansion.
We now specialize to the ($N=0$)-states with 
positive and negative parity.
The relation to the two-instanton contribution
as given by the polynomial $P^N_2$ implicitly defined
in (\ref{einstEnn}) is
\begin{equation}  
{\rm Im}\, E^{(0)}_0 (g ) \mathop{\sim}_{g\to0} 
{ 1 \over \pi g} \, \e^{-1/3g} \,
{\rm Im}\,\left[ P^0_2 \bigl( \ln ( -g/2)\bigr) \right] = 
-{1 \over g} \e^{-1/3g}\,.
\label{einstImii}
\end{equation} 
An introductory discussion 
of dispersion relations and imaginary parts generated 
by the summation of nonalternating factorially divergent 
series is given in appendix~\ref{sDisp},
especially in chapter~\ref{ssDispersion}.
The coefficients $E^{(0)}_{N,k}$
are related to the imaginary part by a Cauchy integral
[see equation (\ref{edispersion})]:
\begin{equation}
E^{(0)}_{0,k} ={ 1 \over \pi} \int^{\infty}_{0} 
{\rm Im}\,\left[ E^{(0)}_0 (g)\right] {\d g \over g^{k+1}}.
\end{equation}
For $ k \to \infty $, the integral is dominated by 
small $g$ values. From the asymptotic estimate 
(\ref{einstImii}) of ${\rm Im}\, E^{(0)}$ for $g\to0$, 
one then derives the large order behaviour of the 
perturbative expansion~\cite{LGZJ1990long}:
\begin{equation} 
E^{(0)}_{0,k}
\mathop{\sim}_{k\to\infty } 
-{1 \over \pi} 3^{k+1} k!\,.  
\label{eEzerok}
\end{equation}  
Let us us now consider the perturbative expansion
about the one-instanton shift in the sense of 
equation (\ref{eInstn}), with the 
convention $E^{(1)}_{k,N} \equiv e_{N,10k}$:
\begin{equation} 
E^{(1)}_{\varepsilon,N} (g )=
-{\varepsilon \over \sqrt{ \pi g}} \e^{-1/6g}
\left( 1 + \sum^{\infty}_{k=0} E^{(1)}_{k,N} \, g^{k} \right) \,.
\end{equation}
One may derive, from the imaginary part of $ P^N_3 $,
the large-order behaviour of this expansion,
by expressing that the imaginary part of 
$E^{(1)}_0(g)$ and $E^{(3)}_0(g)$ must cancel at
leading order:  
\begin{equation} 
{\rm Im}\, E^{(1)}_0 (g) \sim 
- \varepsilon \,
\left({ \e^{-1/6g} \over \sqrt{ \pi g}} \right)^3 
\, {\rm Im}\,\left[ P_3 \bigl( \ln ( -g /2 )\bigr)  \right]\,. 
\label{eImEung} 
\end{equation}
The coefficients $ E^{(1)}_{k,0} $ again are given by a dispersion
integral: 
\begin{equation} 
\varepsilon \, E^{(1)}_{k,0} = 
{1 \over \pi} \, \int^{\infty}_{0} 
\left\{ {\rm Im}\,\left[ E^{(1)}_0 (g )\right] 
\sqrt{ \pi g} \, \e^{1/6g} \right\} \,
{\d g \over g^{k+1}}. 
\end{equation}
Using then equations (\ref{ePthree}) and (\ref{eImEung}), one finds
\begin{equation} 
E^{(1)}_{k,0}\sim -{1 \over \pi} \int^{\infty}_{0} \,
3 \left( \ln {2 \over g}+\gamma \right) \,
\e^{-1/3g}{ \d  g \over g^{k+2}}. 
\end{equation} 
At leading order for $ k $ large,  $ g $  can be replaced  by its saddle
point value $ 1 /3k $ in $ \ln g $ and, finally, one obtains 
\begin{equation} 
E^{(1)}_{k,0} = e_{0,10k} = -{3^{k+2} \over \pi} k! 
\left[ \ln (6\,k)+\gamma +
{\mathcal O}\left({ \ln  k \over k} \right) \right]\,. 
\label{eEunk}
\end{equation}
Both results (\ref{eEzerok}) and (\ref{eEunk}) have been checked against the
numerical behaviour of the corresponding series for which  
300 terms can easily be calculated~\cite{JeZJ2001long,JeHome}.

%
%
\subsection{The Function $\Delta$ (Double--Well)}

A specific function $\Delta$ of the energy levels
of the double-well potential, expressed
as a function of the coupling $g$, has been the subject of 
rather intensive investigations in the 
past~\cite{ZJ1983npb}. Here, we discuss the motivation for 
the definition of $\Delta$, as well as its evaluation
and the comparison to the instanton expansion.

We define the quantity ${\mathcal B}\left\{E^{(0)}_N (g)\right\}$ as the 
real part of the Borel sum of the the perturbation series (\ref{ePert}),
\begin{equation}
\label{defcalBE0}
{\mathcal B}\left\{E^{(0)}_N (g)\right\} = 
{\rm Re}\,\left[ {\rm Borel\ sum\ of}\
E^{(0)}_N(g) \equiv \sum_{l=0}^\infty E^{(0)}_{N,l} \, g^l \right]\,.
\end{equation}
In~\cite{Je2000prd}, some explicit examples are
given for the calculation of Borel sums in cases
where the perturbation series is nonalternating
(as in our case). In the same sense, we denote by 
${\mathcal B}\left\{E^{(1)}_N (g)\right\}$ the
real part of the Borel sum of the the perturbation series 
about one instanton in the notation (\ref{eInstn}),
evaluated for the state with negative parity 
(so that the result, for $g > 0$, is positive):
\begin{equation}
\label{defcalBE1}
{\mathcal B}\left\{E^{(1)}_N (g)\right\} = 
\frac{\e^{-1/6 g}}{\sqrt{\pi\,g}} \, \left\{ 
{\rm Re}\,\left[ {\rm Borel\ sum\ of}\
\sum_{l=0}^\infty e_{N,10l} \, g^l \right] \right\}\,.
\end{equation}
An illustrative example is provided by equation 
(\ref{eE1N0}) below, where the case $N=0$ is considered
and the first terms ($l = 0,\dots,8$) are given explicitly.
The quantity ${\mathcal B}\left\{E^{(1)}_N (g)\right\}$
is of importance for the entire discussion in
chapter~\ref{sThreeInstanton} below. 

We now consider the ratio
\begin{equation} 
\label{defDelta}
\Delta (g) = 4 { \left\{ { 1 \over 2} \,
\left(E_{+,0}+E_{-,0} \right) - 
{\mathcal B}\left\{E^{(0)}_0 (g)\right\} \right\} \over 
\left(E_{+,0}-E_{-,0} \right)^2
\left[ \ln(2g^{-1}) + \gamma \right]}. 
\end{equation}
In this expression $ E_{+,0} $ and $ E_{-,0} $ are the two 
lowest eigenvalues of $H$. In the sum 
$(E_{+,0}+E_{-,0})$ 
the contributions corresponding to an odd number of
instantons cancel. Therefore, the numerator is dominated for 
$ g $ small by the real part of the two-instanton contribution 
proportional to ${\rm Re}\, P_2$. Note that, if the perturbation series
were able to describe the mean of the energies 
$ E_{+,0} $ and $ E_{-,0} $ exactly, then the numerator would vanish.
The difference 
$ (E_{+}-E_{-} ) $, as we know, is dominated by the one-instanton 
contribution which in turn is characterized by 
a nonperturbative factor $\exp[-1/6 g]$. Squaring the
energy difference, we obtain an expression of the 
order of $\exp[-1/3 g]$ which is characteristic of the 
two-instanton effect. Forming the ratio of the 
two-instanton effect, divided by the squared one-instanton,
we obtain an expression of order one (the additional 
logarithm appears because of the nonvanishing
$e_{0,210}$ coefficient).
Using the expansion (\ref{ecomexp}) together
with the explicit results for the coefficients (\ref{ecoeff}), 
one verifies easily
that $ \Delta(g ) $ indeed tends to unity as $ g \to 0$. 

Moreover, performing an expansion in powers of $g$ and inverse powers
of $\ln(2/g)$ and keeping only the first few terms in 
$\{1/\ln(2/g)\}$ in each term in the $g$-expansion, one finds 
\cite{JeZJ2001long}
\begin{scriptsize}
\begin{eqnarray}
\label{asympDelta}
\lefteqn{\Delta(g) \sim 1 + 3 \, g - \frac{23}{2} \frac{g}{\ln(2/g)} \,
\left[1 - \frac{\gamma}{\ln(2/g)} + \frac{\gamma^2}{\ln^2(2/g)} +
\mathcal{{\mathcal O}}\left(\frac{1}{\ln^3(2/g)}\right) \right]}
\nonumber\\
& & + \frac{53}{2} g^2 - 135 \, \frac{g^2}{\ln(2/g)} \,
\left[1 - \frac{\gamma}{\ln(2/g)} + \frac{\gamma^2}{\ln^2(2/g)}  +
\mathcal{{\mathcal O}}\left(\frac{1}{\ln^3(2/g)}\right) \right]
+ \mathcal{{\mathcal O}}\left(g^3\right) \,.
\end{eqnarray}
\end{scriptsize}
The higher-order corrections, which are only
logarithmically suppressed with respect to
the leading terms $1 + 3\,g$, change the numerical values 
quite significantly, even for small $g$.  

%
%
\begin{table}[tbh]
\begin{center}
\begin{minipage}{15cm}
\begin{center}
\caption{\label{tabcirmi} The ratio $ \Delta (g) $ as a function
of $ g $.}
\vspace*{0.3cm}
\begin{scriptsize}
\begin{tabular}{cr@{.}lr@{.}lr@{.}lr@{.}lr@{.}lr@{.}l%
r@{.}lr@{.}lr@{.}lr@{.}lr@{.}l}
\hline
\hline
\rule[-3mm]{0mm}{8mm} coupling $g$ &
 $0$ & $005$ &
 $0$ & $006$ &
 $0$ & $007$ &
 $0$ & $008$ &
 $0$ & $009$ &
 $0$ & $010$ \\
\hline
$\Delta(g)$ num. &
\rule[-3mm]{0mm}{8mm}
 $1$ & $0063(5)$ &
 $1$ & $0075(5)$ &
 $1$ & $00832(5)$ &
 $1$ & $00919(5)$ &
 $1$ & $00998(5)$ &
 $1$ & $01078(5)$ \\
$\Delta(g)$ asymp. &
\rule[-3mm]{0mm}{8mm}
 $1$ & $00640$ &
 $1$ & $00739$ &
 $1$ & $00832$ &
 $1$ & $00919$ &
 $1$ & $01001$ &
 $1$ & $01078$ \\
\hline
\hline
\end{tabular}
\end{scriptsize}
\end{center}
\end{minipage}
\end{center}
\end{table}

%
%
\begin{figure}[htb!]
\begin{center}
\begin{minipage}{12.0cm}
\begin{center}
\epsfxsize=121.mm
\epsfysize=85.mm
\centerline{\epsfbox{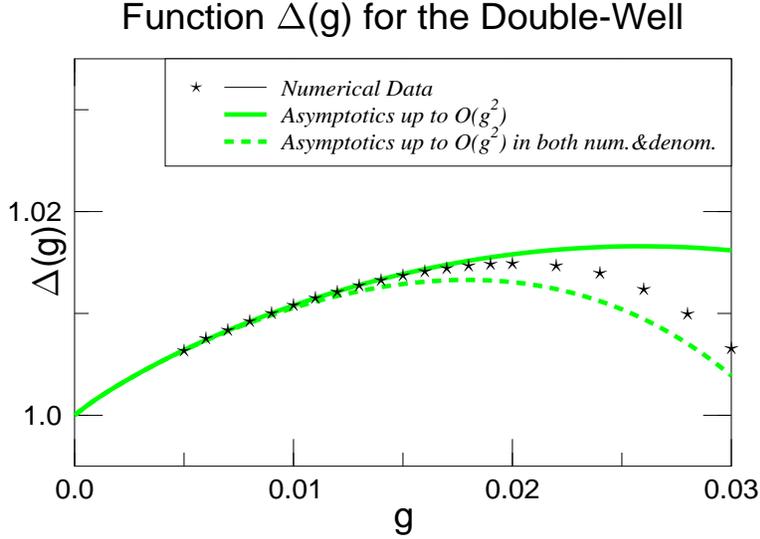}}
\caption{\label{figdelta}
Double-well potential: Comparison of numerical 
data obtained for the function $\Delta(g)$ with the sum of 
the terms up to the order of $g^2$
of its asymptotic expansion  
for $g$ small [see equation (\ref{asympDelta})],
solid line. If we express both the numerator 
and the denominator of (\ref{defDelta}) as a power series 
in $g$ [with the help of the instanton coefficients
(\ref{epsiloncoeff})]
and keep all terms up to the order of $g^2$,
the dashed curve results. This latter approach
is also used for the curves in figures~\ref{figdelta2},~\ref{figdelta3},
and~\ref{figd} below.}
\end{center}
\end{minipage}
\end{center}
\end{figure}

Table~\ref{tabcirmi} displays numerical results, 
obtained by solving the Schr\"odinger equation  
in a range of small values of $ g $ for which
the evaluation is still reasonably precise; there are
in agreement with the first few asymptotic terms up to 
numerical precision~\cite{JeZJ2001long}.

Of course, 
for larger values, significant deviations
from the leading asymptotics must be expected due to higher-order
effects; these are indeed
observed as seen in figure \ref{figdelta}. 
For example, at $g=0.1$ the numerically
determined value reads $\Delta(0.1) = 0.87684(1)$ whereas the first
asymptotic terms sum up to a numerical value of $0.86029$.  
Further calculations are described in chapter~\ref{Explicit},
where also reference values for 
$\Delta(g)$ are given. 

%
%
\section{Other Potentials}

%
%
\subsection{Symmetric Wells}
\label{sssym}

We consider a slight generalization of (\ref{equantization}),  
i.e.~a potential with degenerate minima at $q = 0$ and $q = q_0$, and
\begin{equation}
\label{eomega}
V(q) = \ud \, \omega^2 \, q^2 + {\mathcal O}(q^3)\,,\quad 
V(q) = \ud \, \omega^2 \, (q-q_0)^2 + {\mathcal O}\left((q-q_0)^3\right)\,.
\end{equation}
The potential $V$ is supposed to fulfill
the symmetry condition $V(q) = V(q_0 - q)$.
Without loss of generality, one may set
\begin{equation}
\label{eobdasym}
\omega = 1 
\end{equation}
in (\ref{eomega}), by a simple scaling
argument. Indeed, a Hamiltonian [see equation~(\ref{econvention})]
\begin{equation}
H(g, \omega) 
= -{g\over2} \left({ \d  \over \d  q} \right)^2 + {1\over g} 
\left(
{1\over2} \, \omega^2 \, q^2 + 
{1\over2} \, \omega^2 \, (q - q_0)^2 
\right) 
\end{equation}
satisfies
\begin{equation}
H(g, \omega) = 
\omega \, H\left(\frac{g}{\omega}, 1\right)\,,
\end{equation}
which maps the spectrum of the general case with arbitrary
$\omega$ onto a seemingly more special case, with 
$\omega = 1$ and a redefined coupling $g$.
The constant $C$ is given as [see also equation (\ref{edefCii}) below],
\begin{equation}
\label{edefC}
C = q_0^2 \,
\exp\left[\int_0^{q_0} \,
\left({1\over \sqrt{2 V(q)}} - {1\over q} - {1\over q_0-q}\right)\right]\,.
\end{equation}
As in the case of the double-well potential, the function
$A_{\rm sym}(E,g)$, which described the instanton-related effects,
has no term of order $g^0$ [cf.~equation~(\ref{defA})]
\begin{equation} 
A_{\rm sym}(E,g) = a/g + {\mathcal O}(g) \,.
\end{equation}
The constant $a$ is given by [see equation~(\ref{edefAii}) below]
\begin{equation}
\label{edefA}
a = 2 \, \int_0^{q_0}\d q\,\sqrt{2V(q)}\,.
\end{equation}  

There is, just as in the case of the double-well problem, 
degeneracy on the level of perturbation theory which is lifted by
instantons. The quantization condition is conjectured to be
\begin{equation}
{1 \over \sqrt{2\pi}} \, 
\Gamma\left( {1 \over 2} - B_{\rm sym}(E,g) \right) \,
\left(- {2 C \over g} \right)^{B_{\rm sym}(E,g)} \, 
\exp\left[-{A_{\rm sym}(E,g) \over 2}\right] = 
\varepsilon {\rm i}\,,
\label{equantsym}
\end{equation}
with $\varepsilon$ denoting the parity. The function 
$B_{\rm sym}(E, g) = E + {\mathcal O}(g)$ 
is related to the perturbative expansion
about each of the two symmetric wells, as given in (\ref{equantCii}) 
below and explained in detail in chapter~\ref{ssPerturbative}.
The perturbative quantization condition reads
\begin{equation}
\label{equantpertsym}
B_{\rm sym}(E, g) = N + \ud
\end{equation}
with integer positive $N$.
Explicit calculations of the first few terms in the expansion of 
$B_{\rm sym}(E, g)$ in $g$ are carried out in appendix~\ref{appPerturbative},
and relevant formulas can be found in equation (\ref{einstBgengii}).
For $A_{\rm sym}(E, g)$, see appendix~\ref{sssymAEg}.

It is important to remark that
the $A$ and $B$-functions (\ref{eBEgj}) and (\ref{eAEgj})
are determined by the formal properties of the
potential alone. Specifically, these functions may be inferred,
for a given potential, via an evaluation of the 
contour integrals of the WKB expansion [left-hand side 
of equation (\ref{econj})]. We distinguish here 
several different $A$ and $B$ functions by assigning
indices to these symbols, depending on the 
potential in question. However, it is necessary to remember 
that the potential alone determine the instanton-related $A$-function
and the perturbative $B$-function are determined by the 
potential alone [which we always use in the convention 
(\ref{econvention})]. 

%
%
\subsection{Asymmetric Wells}
\label{ssasym}

In the case of a potential with two asymmetric wells, 
we have to modify the {\em quantization condition}
(\ref{equantization}). Specifically, we will 
discuss the case of a potential with two asymmetric wells
about which one may expand 
\begin{equation}
\label{eomega1omega2}
V(q) = \ud \, \omega_1^2 \, q^2 + {\mathcal O}(q^3)\,,\quad 
V(q) = \ud \, \omega_2^2 \, (q-q_0)^2 + {\mathcal O}\left((q-q_0)^3\right)\,.
\end{equation}
Of course, the  asymmetry manifests itself in 
the relation $V(q) \neq V(q_0 - q)$.
The generalized
Bohr--Sommerfeld quantization formula takes the form
[see also equations~(\ref{econj}),
(\ref{edegenminima}), 
(\ref{edefCgenii}), 
(\ref{eCunCde}),
(\ref{egenpole}), 
(\ref{econstbC}), and~(\ref{econjii}) below]
\begin{eqnarray}
& & \frac{1}{\Gamma(\ud-B_1(E, g)) \, 
\Gamma(\ud-B_2(E, g))} \nonumber\\
& & \qquad + {1 \over 2\pi}\,
\left(-{2 C_1 \over g} \right)^{B_1(E,g)}\,
\left(-{2 C_2 \over g} \right)^{B_2(E,g)} \,
\e^{-A(g,E)}=0\,, 
\label{egeneral}
\end{eqnarray}   
where $B_1(E,g)$ and $B_2(E,g)$
are determined by the perturbative expansions 
(see chapter~\ref{ssPerturbative}) around 
each of the two minima of the potential 
\begin{equation}
B_1(E,g) = E/\omega _1 + {\mathcal O}(g),\quad 
B_2(E,g) = E/\omega _2 + {\mathcal O}(g)\,,
\end{equation}
The perturbative quantization conditions are
\begin{equation}
\label{equantpertasym}
B_1(E,g) = N_1 + \ud\,,\quad
B_2(E,g) = N_2 + \ud\,,
\end{equation}
with nonnegative integer $N_1$ and $N_2$.
In equation (\ref{egeneral}),
the quantities
$C_1$ and $C_2$ are numerical constants, adjusted in such a way that
$A(E,g)$ has no term of order $g^0$:
\begin{equation} 
A(E,g) = a/g + {\mathcal O}(g) \,.
\end{equation}
Again, $a$ is given by
\begin{equation}
\label{edefAi}
a = 2 \, \int_0^{q_0}\d q\,\sqrt{2V(q)}\,.
\end{equation}

Equation (\ref{egeneral}) has two series of energy eigenvalues, 
\begin{equation} 
E_{N} = \left(N+\ud \right) \, \omega_1 + {\mathcal O}(g)\,,  \quad
E_{N} = \left(N+\ud \right) \, \omega_2 + {\mathcal O}(g)   
\end{equation}
close for $g\to0$ to the poles of the two $ \Gamma $-functions. 
The same expression contains the instanton contributions to 
the two different sets of eigenvalues. 

The quantization condition (\ref{egeneral}) is qualitatively different
from (\ref{equantization}) and (\ref{equantsym}) because parity is 
explicity broken in the asymmetric case. 
One verifies that multi-instanton contributions are
singular for $\omega =1$. But, if one takes the symmetric limit, one obtains 
a form consistent with the product of the two equations (\ref{equantsym}).

Without loss of generality, one may set 
\begin{equation}
\label{eobda}
\omega_1 = 1 \,, \qquad \omega_2 = \omega
\end{equation}
in (\ref{eomega1omega2}), by a simple scaling 
argument. Indeed, a Hamiltonian [see equation~(\ref{econvention})]
\begin{equation}
H(g, \omega_1, \omega_2) 
= -{g\over2} \left({ \d  \over \d  q} \right)^2 + {1\over g} 
\left(
{1\over2} \, \omega^2_1 \, q^2 + 
{1\over2} \, \omega^2_2 \, (q - q_0)^2 
\right) 
\end{equation}
satisfies
\begin{equation}
H(g, \omega_1, \omega_2) = 
\omega_1 \, H\left(\frac{g}{\omega_1}, 1, \frac{\omega_2}{\omega_1}\right)
= H(\tilde{g}, 1, \tilde{\omega})\,,
\end{equation}
a relation that maps the spectrum of the general case with arbitrary
$g$, $\omega_1$, and $\omega_2$ onto the more special case 
$\tilde{g}$, $\omega_1 = 1$, $\omega_2 = \tilde{\omega}$. 
One may than redefine $\tilde{g} \to g$ and $\tilde{\omega} \to \omega$.
The constant $C$, defined as [see equation (\ref{edefCgenii}) below],
\begin{equation} 
\label{edefCgen}
C_\omega = q^2_{0} \, \omega^{2/(1+\omega)} \,
\exp \left\{{ 2\omega \over 1+\omega} \,
\left[ \int^{q_{0}}_{0} \d q \,
\left({ 1 \over \sqrt{ 2V (q )}}-{1 \over q}-
{1 \over\omega (q_{0}-q )} \right)\right] \right\}\,,
\end{equation}
tends to (\ref{edefC}) in the limit $\omega \to 1$.
The constants $C_1$ and $C_2$ are then 
given by [see equation (\ref{eCunCde}) below]:
\begin{equation} 
\label{econsequence}
C_1 = C_\omega\,, \qquad C_2 = C_\omega/\omega\,. 
\end{equation}
These enter into the quantization condition (\ref{egeneral}).
We note also that the quantization condition (\ref{egeneral}) may be
derived, to leading order in $g$, by an exact evaluation of the 
multi-instanton contribution to the path integral [see (\ref{egenpole})].

We now consider the 
imaginary part of the leading-instanton contribution and 
large-order estimates.
The expression (\ref{egeneral}) can be used to determine the 
large order behaviour of perturbation theory by calculating the imaginary part 
of the leading instanton contribution and writing a dispersion integral. 
Setting $\omega _1=1$, $\omega _2=\omega $, for the
energy $E_N(g)=N+\ud + {\mathcal O}(g)$  one finds 
the imaginary part 
\begin{equation}
{\rm Im}\, E_N(g)\mathop{\sim}_{g\to 0} 
K_N \, g^{-(N+1/2)(1+1/\omega)} \, \e^{-a/g} 
\end{equation}
with [equation (\ref{edefCgen})]
\begin{eqnarray}
K_N &=& \frac{(-1)^{N+1}}{2\pi \, N!}\,
\omega^{-(N+1/2)/\omega}\,
\left(2 \, C_\omega \right)^{(N+1/2)(1+1/\omega)}
\\
& & \quad \times \sin\left[\pi(N+\ud)(1+{1/\omega})\right]\,
\Gamma\left[\ud -(N+\ud)/\omega\right].
\end{eqnarray}
From ${\rm Im}\, E_N (g) $, one infers that the coefficients  
$ E^{(0)}_{N,k}$  of the perturbative expansion of $E_N(g)$ behave, 
for  order $k\to\infty $, like
\begin{equation} 
E^{(0)}_{N,k}
\mathop{=}_{k\to \infty} K_N\,
{\Gamma[k+(N+1/2)(1+1/\omega)] \over 
a^{k+(N+1/2)(1+1/\omega)}} \,
\bigl[ 1+ {\mathcal O}\left(k^{-1} \right) \bigr]\,. 
\end{equation} 
Note that this expression, in contrast to the instanton contribution
to the real part, is uniform in the limit $\omega=1$ in which the result
(\ref{eEzerok}) is recovered.
%

%
%
\subsection{Periodic Cosine Potential}
\label{ssPeriodicCosine}
 
We first consider the {\em periodic cosine potential.}
This example differs from the preceding ones because the potential 
is still an entire function but no longer a polynomial. 
On the other hand the periodicity
of the potential simplifies the analysis, because it allows
classifying eigenfunctions according to their behaviour 
under a translation of one period:
\begin{equation} 
\psi_{\varphi}(q+T)=\e^{i\varphi} \psi_{\varphi}(q),
\end{equation}
where $T$ is the period. For the cosine potential 
${1 \over16}(1- \cos 4q)$ (and thus $T=\pi/2$),
the conjecture then takes the form 
\begin{equation}
\label{eQuantCos}
\left({2 \over g}\right)^{-B_{\rm pc}(E,g)}{\e^{A_{\rm pc}(E,g)/2}\over 
\Gamma[\ud - B_{\rm pc}(E,g)]} +
\left({-2 \over g}\right)^{B_{\rm pc}(E,g)}\,
{\e^{-A_{\rm pc}(E,g)/2}\over \Gamma[\ud + B_{\rm pc}(E,g)]} =
{2 \cos\varphi \over \sqrt{2\pi}}.
\end{equation} 
The first few terms of the
perturbative expansions of the functions 
$ A_{\rm pc} $ and $ B_{\rm pc} $ 
for the periodic potential are
\begin{subequations}
\label{ecos} 
\begin{eqnarray} 
B_{\rm pc}(E,g) &=& E + 
g \, \left(E^2+ \frac{1}{4} \right)
+g^2 \, \left(3\,E^3 +
\frac{5}{4}\,E\right) + 
{\mathcal O}\left(g^3 \right),
\label{ecosb} 
\\
A_{\rm pc}(E,g) &=& g^{-1} + 
g \left(3\,E^2+ \frac{3}{4} \right)+
g^2\left(11\,E^3+ \frac{23}{4}\,E\right) 
+ {\mathcal O}\left(g^3 \right)\,,
\label{ecosa} 
\end{eqnarray} 
\end{subequations}
where again $A$ has been initially obtained by a 
combination of analytic and numerical techniques.
The perturbative quantization condition is 
\begin{equation}
\label{equantpertpc}
B_{\rm pc}(E,g) = N + \ud\,.
\end{equation}
The conjecture has been verified numerically up to four-instanton order 
with decreasing precision. The theory of resurgence has 
allowed to also investigate this potential, 
as well as general trigonometric potentials~\cite{De1992}. 

%
%
\subsection{${\mathcal O}(\nu)$--Anharmonic Oscillator 
and Fokker--Planck Equation}
\label{ssAOFP}

We consider now the {\em ${\mathcal O}(\nu)$-symmetric anharmonic oscillator}
with the Hamiltonian
\begin{equation}
H = -\ud \, \bm{\nabla}^2+\ud \, {\bf q}^2 +g \left({\bf q}^2\right)^2, 
\label{ehamOnu}
\end{equation}
where $\bf q$ belongs to $\mathbb{R}^{\nu}$.  The analytic continuations of the
eigenvalues to $g<0$ are complex, the imaginary part determining 
the decay rate by barrier penetration of states concentrated initially 
in the well at ${\bf q}=0$.  Moreover, the behaviour of the imaginary part 
for $g\to 0_-$ is directly related to the large-order 
behaviour of the perturbative expansion.

The eigenfunctions of $H$ can be classified according to the 
irreducible representations of the orthogonal ${\mathcal O}(\nu)$ group. 
We work in $\mathbbm{R}^\nu$; at 
fixed angular momentum $l$, the Hamiltonian (\ref{ehamOnu}), 
expressed in terms of the radial variable $r=|{\bf q}|$, takes the form 
\begin{equation}
\label{eradialschr}
H_l \equiv H_l(g) = - {1\over 2}\left(\d\over\d r\right)^2
- {1\over 2}\,{\nu-1\over r}{\d\over\d r} 
+ {1\over 2}\,{l\,(l+\nu-2)\over r^2}
+ {1\over 2}\,r^2 +g\,r^4.
\end{equation}
After the change of variables
\begin{equation}
H_l \mapsto \tilde H_l = r^{(\nu-1)/2}\, H_l \, r^{(1-\nu)/2}\,,
\end{equation}
one finds
\begin{equation}
\label{esymj}
\tilde H_l =
-{1\over 2} \, \left(\d\over\d r\right)^2
+ {1 \over 2} \, {j^2 - 1/4 \over r^2}
+ {1 \over 2} \, r^2 
+ g \, r^4
\end{equation}
with
\begin{equation}
\label{edefj}
j=l+\nu/2-1\,.
\end{equation}
The Hamiltonian $H_l$ is one-dimensional with a non-polynomial potential.
Eigenvalues depend on the parameter $j \equiv j(l)$ (but with the 
constraint $j(l) \ge \nu/2-1$). Note also the formal symmetry 
$j \mapsto -j$.

Writing the perturbative quantization condition as 
\begin{equation}
\label{equantpertOnu}
B_{\nu}(E,g,j)=1+j +2N\,, \quad N\ge 0\,,
\end{equation}
one finds [see also equations (\ref{eBdble}), 
(\ref{enudw}), and (\ref{eBEgjg4})]
\begin{subequations}
\label{eABnu}
\begin{equation}
\label{eBEgj}
B_{\nu}(E,g,j) = E
- \frac12 \, \left[3 \, E^2 - j^2 + 1 \right] \, g
+ \frac{5}{4} \, \left[7 \, E^3 + (5 - 3\,j^2) \, E\right] \, g^2
+ {\mathcal O}\left(g^3 \right)\,.
\end{equation}
The first few terms of the expansion  
of the function $A_\nu$ are
[see also (\ref{eAEgjg4})]:
\begin{eqnarray}  
\label{eAEgj}
A_{\nu}(E,g,j) &=& 
-{1 \over 3}\,g^{-1} + 
\left({3 \over 4}\,j^2-{19 \over 12}-{17 \over 4} \, E^2 \right) \, g
\nonumber\\
& & + \frac{1}{8} \, 
\left( 227 \, E^2 - 77\,j^2 + 187\right) E \, g^2 +
{\mathcal O}\left(g^3 \right). 
\end{eqnarray}
\end{subequations}
The $A$ and $B$ functions are determined exclusively by the 
semi-classical expansion of the solutions of 
the spectral equation of the Hamiltonian (\ref{eradialschr}).

In order to write down a quantization condition, it is necessary
to specify boundary conditions for the Hamiltonian.
In the current chapter, we will take advantage of a
well-known connection between the 
${\mathcal O}(\nu)$-anharmonic oscillator at negative (!)
coupling on the one hand to the double-well potential 
at positive coupling with a symmetry-breaking
term on the other hand~\cite{SeZJ1979,BuGr1993}.
(This result, initially conjectured on the basis of
numerical evidence, was first proven later by path integral
manipulations~\cite{An1982}
or recursion formulae~\cite{DaPrMa1984}, and later 
generalized to arbitrary $j$~\cite{BuGr1993}.)

We briefly recall here the central statement
regarding the connection between
``broken-double-well and the ${\mathcal O}(\nu)$--symmetric quartic potential.
Indeed, as shown below in chapter~\ref{ssOnudw}, the 
spectral equation implied by the Hamiltonian
\begin{subequations}
\label{ediff}
\begin{equation}
\label{ediffstart}
H_l(-g) = - {1\over 2}\left(\d\over\d r\right)^2
- {1\over 2}\,{\nu-1\over r}{\d\over\d r}  
+ {1\over 2}\,{l\,(l+\nu-2)\over r^2}
+ {1\over 2}\,r^2 - g\,r^4\,.
\end{equation}
may be reformulated under appropriate substitutions into 
\begin{equation}
\label{ediffeq}
-{ g\over2} \, \varphi''(p) +
{1\over 2g} \, \left(p^2 - \frac{1}{4}\right)^2 \, \varphi(p) - 
j \, p \, \varphi(p) =
{E\over 2} \, \varphi(p) \,. 
\end{equation}
\end{subequations}
The spectrum of this equation is bound from below and 
leads naturally to a self-adjoint extension of the ${\mathcal O}(\nu)$
anharmonic oscillator at negative coupling. This spectrum is not the same 
as the resonances implied for negative coupling if the
Hamiltonian is endowed with boundary conditions that imply 
a vanishing of the wave function in complex directions of the 
parameters, which would lead to resonances discussed below in 
chapter~\ref{ssAOresonances}.
As is evident from (\ref{ediffstart}) and (\ref{ediffeq}),
the Hamiltonian $H_l$ for $g <0$, with proper boundary conditions, thus 
has a self-adjoint extension, which in turn is (almost) equivalent
to the double-well potential, but with a linear symmetry-breaking term.

In the case $j=0$ the linear symmetry-breaking term 
in (\ref{ediffeq}) vanishes.
Setting $q = p + 1/2$, equation (\ref{ediffeq}) becomes 
identical to the spectral 
equation of the double-well potential. This consideration 
leads immediately to the following correspondence
\begin{subequations}
\label{ecorrespondence}
\begin{eqnarray}
B_{\nu}(E,-g,j=0) &=& 2 \, B_{\rm dw}(E/2,g) \,,  
\\
A_{\nu}(E,-g,j=0) &=& A_{\rm dw}(E/2,g)\,.
\end{eqnarray}
\end{subequations}
The correspondence between the $j=0$-component of the 
${\mathcal O}(\nu)$-anharmonic oscillator
and the (symmetric) double-well potential~\cite{SeZJ1979,BuGr1993},
thus finds a natural extension to instanton effects described 
by the function 
$A_{\nu}(E,-g,j=0) = A_{\rm dw}(E/2,g)$.
The perturbation series implied by the condition 
$B_{\rm dw}(E/2,g) = N+\ud$ leads to a non-Borel summable 
series for $E$ as a function of $g$; the imaginary parts are compensated by 
instantons.

In view of (\ref{eBEgj}), (\ref{eAEgj}) and 
(\ref{ediffeq}), we conjecture for the self-adjoint extension of the 
Hamiltonian (\ref{eradialschr}) at negative coupling $-g$
(with $g > 0$) the quantization condition
\begin{eqnarray}
\label{equantAnharmOsc}
& & {1\over \Gamma \left[ \ud( 1+j-B_{\nu}(E, -g, j)) \right] \,
\Gamma \left[ \ud(1-j-B_{\nu}(E, -g, j)) \right] } \nonumber\\
& & \qquad \qquad \qquad \qquad \qquad
+ \left(-{2\over g}\right)^{B_{\nu}(E, -g, j)} 
\frac{\exp\left(- A_{\nu}(E, -g)\right)}{2 \pi} = 0 \,.
\end{eqnarray}
In leading order in $g$, this equation may be derived via 
instanton calculus [see equation (\ref{eOnuAnharmonicSelfAdjoint})]:
\begin{equation}
{1\over \Gamma \bigl (\ud( 1+j-E)  \bigr) \Gamma \bigl (\ud(1-j-E) \bigr) }
+ \left(-{2\over g}\right)^{ E} {\e^{-1/3g}\over 2\pi} \approx 0 \,.
\end{equation}

We now turn our attention to an important special case.
Setting $j = -1$ in (\ref{ediffeq}) 
and shifting $E \to 2 E$, we obtain the
Fokker--Planck Hamiltonian
\begin{subequations}
\label{hfp}
\begin{equation}
H_{\rm FP} = -{ g\over2} \, \frac{\d^2}{\d p^2} +
{1\over 2g} \, \left(p^2 - \frac{1}{4}\right)^2 + p\,,
\end{equation}
with the associated eigenvalue problem
\begin{equation}
H_{\rm FP} \, \varphi(p) = E \, \varphi(p) \,, 
\end{equation}
\end{subequations}
which is known to have a peculiar property: the perturbative expansion 
of the ground-state energy vanishes to all orders in the coupling.
Of course, the symmetry $j \to -j$, which is present in (\ref{esymj}),
implies that the the spectrum of (\ref{hfp}) is invariant 
under a change of the sign of the linear symmetry-breaking term.

({\em Remark.}) With the shift $p = q - 1/2$ in (\ref{hfp}),
we may rewrite the Fokker--Planck Hamiltonian as
\begin{equation}
\label{hfpii}
H_{\rm FP} = -{g \over 2} \, \frac{\d^2}{\d q^2} +
{1\over g} \, \left[ \frac12 \, q^2 \, \left(1 - q\right)^2 + g \, q \right]
- \frac12\,.
\end{equation}
We may thus identify the Fokker--Planck Hamiltonian as originating 
from the double-well Hamiltonian (\ref{epotdw}), with an additional
linear symmetry-breaking term of relative order $g$, and an additional
global shift of $-1/2$ which compensates the ground-state energy of the 
harmonic oscillator.

In view of the correspondence implied by (\ref{ediff}) and the necessary
shift $E \to 2 E$ in going from (\ref{ediffeq}) to (\ref{hfp}), 
we conjecture the following secular equation for the
Fokker--Planck Hamiltonian as a special case of the 
anharmonic oscillator (\ref{equantAnharmOsc}),
\begin{equation}
\label{equantfp}
{1\over \Gamma \left( - B_{\rm FP}(E, g)\right)\, 
\Gamma \left(1 - B_{\rm FP}(E, g) \right)}
+ \left(-{2\over g}\right)^{2 B_{\rm FP}(E, g)} \,
\frac{\exp\left(- A_{\rm FP}(E, g)\right)}{2 \pi} = 0 \,.
\end{equation}
The perturbative quantization condition is 
\begin{equation}
\label{equantpertFP}
B_{\rm FP}(E, g) = N
\end{equation}
where $N$ is a nonnegative integer.
The functions $B_{\rm FP}(E, g)$ and $A_{\rm FP}(E, g)$ are given by
\begin{subequations}
\begin{eqnarray}
\label{eBEgFPj}
B_\nu(E, -g, j= \pm 1) &=& 2 B_{\rm FP}\left(\frac{E}{2}, g\right) \,,
\\
\label{eAEgFPj}
A_\nu(E, -g, j= \pm 1) &=& A_{\rm FP}\left(\frac{E}{2}, g\right) \,,
\end{eqnarray}
\end{subequations}
where either sign of $j$ leads to the same result.
The first terms read
\begin{subequations}
\label{eABEgFP}
\begin{eqnarray} 
\label{eBEgFP}
B_{\rm FP}\left(E, g\right) &=& E + 
3 \, E^2 \, g +
\left( 35\, E^3 + \frac52\,E \right)\, g^2 +
{\mathcal O}(g^3)\,,
\\
\label{eAEgFP}
A_{\rm FP}\left(E, g\right) &=& \frac{1}{3 g} +
\left( 17 \, E^2 + \frac56 \right)\, g + 
\left( 227\, E^3 + \frac{55}{2}\,E  \right)\, g^2 +
{\mathcal O}(g^3)\,.
\end{eqnarray}
\end{subequations}
Indeed, the quantization condition (\ref{equantfp}) 
may be derived, at leading order in $g$, using 
instanton calculus [see equation (\ref{eninstFP}) with $C = 1$]:
\begin{equation}
\Delta(E) = {1\over \Gamma (- E)\,\Gamma (1 -E)}
+ \left(-{2\over g}\right)^{2E} \, {\e^{-a/g}\over 2\pi}  \,.
\end{equation}

%
%
\subsection{${\mathcal O}(\nu)$--Anharmonic Oscillator: Resonances}
\label{ssAOresonances}

In the current chapter, we investigate the quantization 
condition for the resonances of the 
${\mathcal O}(\nu)$-anharmonic oscillator at negative coupling.
These may be found by analytic continuation of the 
eigenvalues to negative coupling. The notation is the same 
as in chapter~\ref{ssAOFP}.

First, we briefly investigate the special case $j=0$ at $g > 0$.
The correspondence (\ref{ecorrespondence}) implies the 
perturbative 
condition $B_{\nu}(E,g,j=0) = 2 \, B_{\rm dw}(E/2,-g) = 2 N + 1$
for eigenvalues of the ${\mathcal O}(\nu)$-anharmonic oscillator at 
positive coupling. 

The generalization to arbitrary $j$ at $g > 0$
implies a Borel-summable, alternating series 
for $E$ as a function of $g$.  Indeed, the perturbative quantization 
condition (\ref{equantpertOnu}) becomes exact in this case,
\begin{equation}
\label{eAOperturbative}
B_{\nu}(E,g,j) = 1 + j + 2 N\,, \qquad N \geq 0\,.
\end{equation}

We now turn to the resonances for arbitrary $j$ at $g < 0$.
We recall that 
the $A_\nu$ and $B_\nu$-functions (\ref{eBEgj}) and (\ref{eAEgj}) 
are determined by the formal properties of the 
potential alone. However, the actual quantization
condition may also depend on the boundary conditions.
The secular equation for resonances at $g < 0$
is conjectured to be 
\begin{eqnarray} 
& & {\rm i}\,\e^{-A_{\nu}(E,g)}\,
\left(-{2\over g }\right)^{B_{\nu}(E,g)} \,
\exp\left[{\rm i}\,\pi \, \frac{j+1+B_{\nu}(E,g)}{2}\right] \,
\nonumber\\
& & 
\qquad 
\times {\Gamma \left[\frac{1}{2}\,(j+1-B_{\nu}(E,g))\right] \over 
\Gamma\left[\frac{1}{2}\,(j+1+B_{\nu}(E,g)) \right]} = 1\,. 
\label{eOdeinst} 
\end{eqnarray}
In leading order in $g$, the corresponding relation 
may be derived on the basis of instanton calculus
[see equations (\ref{eOdeinsti}) and (\ref{einstImEj})].

%
%
\section{Summary of the Quantization Conditions}
\label{sSummary}

We summarize the conjectured quantization conditions.
The general convention for the potential is indicated in 
equation~(\ref{econvention}). The $A$ and $B$-functions 
are determined by equation~(\ref{econj}), once the potential is fixed.
All quantization conditions discussed here have been derived, in 
leading order in $g$, by instanton calculus.

({\em Double-well potential.})
For the double-well potential (\ref{epotdw}),
\begin{equation}
H = - {g\over2} \left({ \d  \over \d  q} \right)^2 + 
{1\over g} \left[ {1\over2} \, q^2 (1-q )^2\right]\,,
\end{equation}
we have (\ref{equantization}),
\begin{equation}
{1 \over \sqrt{2\pi}} \, \Gamma\left( {1 \over 2} - B_{\rm dw}(E,g) \right) \,
\left(- {2 \over g} \right)^{B_{\rm dw}(E,g)} \, 
\exp\left[-{A_{\rm dw}(E,g) \over 2}\right] = 
\varepsilon {\rm i}\,.
\end{equation}
The functions $A_{\rm dw}$ and $B_{\rm dw}$ are defined in
(\ref{defABdw}), and the perturbative quantization
condition, from which the perturbative expansion of 
an eigenvalue may be obtained, is given in (\ref{equantpertdw}).
 
({\em General symmetric potential.})
We now consider a general symmetric potential with degenerate minima of the 
form~(\ref{eomega}),
\begin{equation}
V(q) = \ud \, q^2 + {\mathcal O}(q^3)\,,\quad 
V(q) = \ud \, (q-q_0)^2 + {\mathcal O}\left((q-q_0)^3\right)\,,
\end{equation}
where we have used (\ref{eobda}). The quantization condition is 
(\ref{equantsym}),
\begin{equation}
{1 \over \sqrt{2\pi}} \, 
\Gamma\left( {1 \over 2} - B_{\rm sym}(E,g) \right) \,
\left(- {2 C \over g} \right)^{B_{\rm sym}(E,g)} \, 
\exp\left[-{A_{\rm sym}(E,g) \over 2}\right] = 
\varepsilon {\rm i}\,.
\end{equation}
By contrast, the purely
perturbative quantization condition is (\ref{equantpertsym}).
The constant $C$ is given in (\ref{edefC}),
\begin{equation}
C = q_0^2 \,
\exp\left[\int_0^{q_0} \,
\left({1\over \sqrt{2 V(q)}} - {1\over q} - {1\over q_0-q}\right)\right]\,.
\end{equation}
 
({\em Asymmetric wells, degenerate minima.})
For general asymmetric wells with degenerate minima
[see equation~(\ref{eomega1omega2})], we have
\begin{equation}
V(q) = \ud \, q^2 + {\mathcal O}(q^3)\,,\quad 
V(q) = \ud \, \omega^2 \, (q-q_0)^2 + {\mathcal O}\left((q-q_0)^3\right)\,.
\end{equation}
The generalized Bohr--Sommerfeld quantization formula takes the form
(\ref{egeneral}),
\begin{eqnarray}
& & \frac{1}{\Gamma(\ud-B_1(E, g)) \, \Gamma(\ud-B_2(E, g))} 
\nonumber\\
& & \qquad + {1 \over 2\pi}\,
\left(-{2 \, C_\omega \over g} \right)^{B_1(E,g)}\,
\left(-{2 \, C_\omega \over \omega \, g} \right)^{B_2(E,g)} \,
\e^{-A(g,E)}=0\,, 
\end{eqnarray}
The perturbative quantization condition is (\ref{equantpertasym}).
Here, we have [see equation (\ref{edefCgenii})]
\begin{equation} 
C_\omega = q^2_{0} \, \omega^{2/(1+\omega)} \,
\exp \left\{{ 2\omega \over 1+\omega} \,
\left[ \int^{q_{0}}_{0} \d q \,
\left({ 1 \over \sqrt{ 2V (q )}}-{1 \over q}-
{1 \over\omega (q_{0}-q )} \right)\right] \right\}\,.
\end{equation}

({\em Periodic cosine potential.})
For the periodic cosine potential 
$V(q) = {1 \over16} \, (1- \cos 4q)$,
the conjecture then takes the form 
\begin{equation}
\left({2 \over g}\right)^{-B_{\rm pc}(E,g)}{\e^{A_{\rm pc}(E,g)/2}\over 
\Gamma[\ud - B_{\rm pc}(E,g)]} +
\left({-2 \over g}\right)^{B_{\rm pc}(E,g)}\,
{\e^{-A_{\rm pc}(E,g)/2}\over \Gamma[\ud + B_{\rm pc}(E,g)]} =
{2 \cos\varphi \over \sqrt{2\pi}}.
\end{equation} 
The functions $A_{\rm pc}$ and $B_{\rm pc}$ are defined in
(\ref{ecos}), and the perturbative quantization condition
can be found in (\ref{equantpertpc}).

({\em Self-adjoint extension of the 
anharmonic oscillator with ${\mathcal O}(\nu)$-symmetry,
at negative coupling.})
For the ${\mathcal O}(\nu)$-anharmonic oscillator with the 
radial Hamiltonian (\ref{eradialschr}),
\begin{equation}
H_l \equiv H_l(g) = - {1\over 2}\left(\d\over\d r\right)^2
- {1\over 2}\,{\nu-1\over r}{\d\over\d r} 
+ {1\over 2}\,{l\,(l+\nu-2)\over r^2}
+ {1\over 2}\,r^2 + g\,r^4.
\end{equation}
For $g < 0$, the radial Hamiltonian being endowed 
with a self-adjoint extension, the quantization 
condition is (\ref{equantAnharmOsc})
\begin{eqnarray}
& & {1\over \Gamma \left[ \ud( 1+j-B_{\nu}(E,g,j)) \right] \,
\Gamma \left[ \ud(1-j-B_{\nu}(E,g,j)) \right] }
\nonumber\\
& & \qquad + \left(-{2\over g}\right)^{B_{\nu}(E, g, j)} 
\frac{\exp\left(A_{\nu}(E, g)\right)}{2 \pi} = 0 \,.
\end{eqnarray}
The functions $A_\nu$ and $B_\nu$ are defined in
(\ref{eABnu}).

({\em Resonances of the 
anharmonic oscillator with ${\mathcal O}(\nu)$-symmetry,
at negative coupling.})
For $g < 0$, the resonances of the radial Hamiltonian
are conjectured to follow the quantization condition
(\ref{eOdeinst}),
\begin{eqnarray} 
& & {\rm i}\,\e^{-A_{\nu}(E,g)}\,
\left(-{2\over g }\right)^{B_{\nu}(E,g)} \,
\exp\left[{\rm i}\,\pi \, \frac{j+1+B_{\nu}(E,g)}{2}\right] 
\nonumber\\
& & \qquad \times {\Gamma \left[\frac{1}{2}\,(j+1-B_{\nu}(E,g))\right] \over 
\Gamma\left[\frac{1}{2}\,(j+1+B_{\nu}(E,g)) \right]} = 1\,, 
\end{eqnarray}

({\em Fokker--Planck Hamiltonian.})
An important special case of (\ref{equantAnharmOsc})
is the Fokker--Planck Hamiltonian (\ref{hfp}) 
\begin{equation}
H_{\rm FP} = -{ g\over2} \, \frac{\d^2}{\d p^2} +
{1\over 2g} \, \left(p^2 - \frac{1}{4}\right)^2 + p\,,
\end{equation}
The corresponding conjecture for the quantization condition is given by
the equation~(\ref{equantfp}), 
\begin{equation}
{1\over \Gamma \left( - \frac12 B_{\rm FP}(E, g)\right)\, 
\Gamma \left(1 - \frac12 B_{\rm FP}(E, g) \right)}
+ \left(-{2\over g}\right)^{B_{\rm FP}(E, g)} \,
\frac{\exp\left(-A_{\rm FP}(E, g)\right)}{2 \pi} = 0 \,.
\end{equation}
The functions $A_{\rm FP}$ and $B_{\rm FP}$ are defined in
(\ref{eABEgFP}).

%
%
\chapter{Perturbative and WKB Expansions from Schr\"odinger Equations}
\label{BSWKB}

%
%
\section{Orientation}

In the simplest examples these conjectures, based on the semi-classical
evaluation of path integrals and instanton calculus, have obtained independent 
heuristic confirmations by considerations based 
on the Schr\"odinger equations, as we now explain.

We write the Schr\"odinger equation as 
\begin{equation}
H \, \psi(q) \equiv
-{ g\over 2} \, \psi''(q)+ {1\over g}\,V(q)\,\psi(q) = E\,\psi(q). 
\label{einstSchr}
\end{equation}
We consider only potentials $V$ that  are entire functions. 
This allows  extending
the Schr\"odinger equation and thus its solutions to the $q$ complex plane. 

Moreover, we assume that $V(q)$ has an absolute minimum, 
for convenience located at $q=0$, where $V(q)= \ud \, q^2 + {\mathcal O}(q^3)$. 

%
%
\section{General Considerations}
\label{sGeneralConsiderations}

%
%
\subsection{The Transition to Quantum Mechanics}
\label{ssTransition}

We recall in this chapter how the 
transition from classical to quantum mechanics can be
understood in terms of matter waves whose wave fronts
are given approximately by the 2-dimensional manifolds 
of constant classical action. 
The classical action $A_0$ for a particle 
whose Lagrangian is not explicitly time-dependent,
reads
\begin{equation}
A_0(t, \bbox{r}) = \int^t_0 \d t' \,
L(\bbox{r}(t'), \bbox{\dot r}(t'))\,.
\end{equation}
One immediately verifies the relations
\begin{equation}
\frac{\d A_0}{\d t} = L \,, \quad
\frac{\d A_0}{\d t} = \frac{\partial A_0}{\partial t} +
\bbox{\dot r} \, \bbox{p} \,, 
\quad \bbox{\nabla} A_0 = \bbox{p}\,,
\end{equation}
where $\bbox{p}$ is the momentum of the classical particle.
This can be rewritten as
\begin{equation}
\frac{\partial A_0}{\partial t} =
- (\bbox{p}\,\bbox{\dot r}-L) = -H\,,
\end{equation}
where $H$ is the Hamiltonian. Using $H = \bbox{p}^2/(2 m) + V$,
where $V$ is the potential and $m$ is the mass of the 
particle, the Hamilton--Jacobi equation is 
recovered,
\begin{equation}
\label{eHamJac}
-\frac{\partial A_0}{\partial t} =
\frac{(\bbox{\nabla} A_0)^2}{2 m} + V(r)\,.
\end{equation}
The momentum $\bbox{p} = \bbox{\nabla} A_0$ is thus 
perpendicular to the manifolds of constant classical 
action $A_0 = {\rm constant}$. That is to say, the 
classically allowed trajectories are perpendicular
to the surfaces of constant action. On the other hand, 
the wave fronts of a traveling wave are characterized by a
constant phase, and the wave moves perpendicular to the 
surfaces of constant phase. Therefore, choosing the 
ansatz
\begin{equation}
\label{eAnsatz0}
\psi_0(t, \bbox{r}) = \exp({\rm i}\, \xi_0(t, \bbox{r}) )
\end{equation}
for the wave function, we are led to the conclusion 
that $A_0$ should be constant on the surfaces 
of constant $\xi$. Because $\xi$ has to have dimension one,
but $A_0$ has the dimension of an action, the identification
\begin{equation}
\xi_0(t, \bbox{r}) = \frac{A_0(t, \bbox{r})}{\hbar}
\end{equation}
is suggested. If the dynamics of the matter wave were purely 
governed by the laws of classical mechanics, generalized to matter 
waves, then the equation (\ref{eHamJac}) would 
adequately describe the time evolution of the wave 
packet described by the matter wave function
$\psi_0$, and the transition from classical to quantum mechanics
would be completely equivalent to the transition from ray 
to wave optics. In order to verify how well (\ref{eHamJac})
is applicable to the description of a quantal wave packet,
we choose the ansatz
\begin{equation}
\label{eAnsatz}
\psi(t, \bbox{r}) = \exp\left(\frac{\rm i}{\hbar} \, A(t, \bbox{r})\right)
\end{equation}
for the quantum mechanical wave function and insert this into
the Schr\"{o}dinger equation. The result is
\begin{equation}
\label{eQuant}
-\frac{\partial A}{\partial t} =
\frac{(\bbox{\nabla} A)^2}{2 m} + V(r) -
\frac{{\rm i} \, \hbar}{2 m} \, \bbox{\nabla}^2 \, A\,.
\end{equation}
This is {\em not} identical to (\ref{eHamJac}).
The third term on the right-hand side characterizes a 
diffusive process in the context of the Brownian motion
of the quantal particle~\cite{Ro1986} and is responsible
for the spreading of quantum mechanical wave packets.
It may be interpreted as a quantum mechanical correction
to the classical equations of motion of matter waves.
The correction is of order $\hbar$.
An expansion in powers of $\hbar$ is a semiclassical
expansion, or WKB expansion~\cite{Da1968}.

For a stationary state, the ansatz
\begin{equation}
A(t, \bbox{r}) = a(\bbox{r}) - E\,t
\end{equation}
leads to
\begin{equation}
\label{eQuantJac}
E = \frac{(\bbox{\nabla} a)^2}{2 m} + V(r) -
\frac{{\rm i} \, \hbar}{2 m} \, \bbox{\nabla}^2 \, a\,.
\end{equation}
We now specialize this equation to the one-dimensional
Hamiltonian given by equation (\ref{ehamdw}) where, 
upon multiplication by a overall factor $g$, 
the coupling $g$ takes the formal role of $\hbar$ with $m=1$.
We therefore have to replace $E \to g\,E$, 
as well as $\hbar \to g$, and 
we also set $a(q) = {\rm i} \int S(q') \, \d q'$.
Then,
\begin{equation}
g E = -\frac12\, S^2(q) + V(q) + \frac{g}{2}\,S'(q)\,.
\label{eRiccati}
\end{equation}

%
%
\subsection{Riccati Equation}
\label{ssRiccati}

According to the previous chapter,
a convenient way to generate 
semiclassical expansions is to use the Riccati
equation derived from the Schr\"odinger equation 
(\ref{einstSchr}) by setting 
\begin{equation}
\label{eLogDeriv}
S(q)=-g \, \psi'/\psi\,.
\end{equation}
The function $S$ then satisfies (\ref{eRiccati})
\begin{equation}
g\,S'(q)-S^2(q)+2V(q)-2gE=0\,.
\label{eRiccat}
\end{equation}
Following the discussion of the appendix \ref{appSchrRic}, 
we also introduce two independent solutions  
$\psi_{1,2}$ of equation (\ref{einstSchr}) such that
\begin{equation}
\psi_2(q)\mathop{\to}_{q\to+\infty }0\,, \quad 
\psi_1(q)\mathop{\to}_{q\to-\infty }0\,,
\end{equation}
and the decomposition (\ref{eresolSpm}) with
\begin{equation}
S_\pm(q)={g\over2}\left({\psi '_1\over \psi _1}
\mp{\psi '_2 \over \psi _2}\right). 
\label{eSpmdef}
\end{equation} 
[The derivations of a few simple identities needed in
this chapter are recalled in appendix
\ref{appSchrRic}, but with different normalizations.]

In the leading WKB expansion,
the two wave functions $\psi _1$ and $\psi_2$,
for large $q$, i.e. in regions where $V(q) > g E$, may be expressed as 
$\exp(-1/g \int \d q \, \sqrt{2 \, V(q) - 2 g E} )$
and $\exp(1/g \int \d q \, \sqrt{2 \, V(q) - 2 g E} )$,
respectively (the positive square root being implied). 

Then, equation (\ref{eRiccat}) translates into 
[see (\ref{eRiccatipm})]
\begin{subequations}
\label{eRiccatpm}
\begin{eqnarray}
& & g \, S'_{-}-S^2_{+}-S^2_{-} + 2\,V(q) - 2\,gE=0\,,
\label{eRiccatpma}\\
& & g \, S'_{+}-2S_+S_-=0\,.
\label{eRiccatpmb}
\end{eqnarray}
\end{subequations}
We note that in equations (\ref{eRiccatpm})
a change $(g,E) \mapsto (-g, -E)$ can be
formally compensated by the change $(S_+, S_-) \mapsto (S_+,-S_-)$.
That is to say, $S_+$ is even under 
$(g,E) \mapsto (-g, -E)$,
whereas $S_-$ is odd.

It follows that in the sense of a series expansion
\begin{equation}  
\label{eSym}
S_{\pm}(q,-g,-E) = \pm \, S_{\pm}(q,g,E)\,.
\end{equation}
The relations (\ref{eRiccatpm}) allow to write the wave function 
in terms of $S_+$ only,
\begin{equation}
\label{eSponly}
\psi(q) = (S_+)^{-1/2} \, \exp\left[ -\frac{1}{g} \,
\int^q \d q' \, S_+(q') \right]\,.
\end{equation}

%
%
\subsection{Spectral Equation and Fredholm Determinant}
\label{ssinstFredh}

We now consider the {\em resolvent operator}
\begin{equation}
R(E)=[ H-E]^{-1}\,. 
\label{einstResolv}
\end{equation}
The poles in $E$ of its trace
\begin{equation}
G(E)=\tr R(E)=\tr [ H-E]^{-1} \,, 
\end{equation}
give the complete spectrum of the Hamiltonian $H$. 
Note that this expression may require some regularization 
to deal with  the divergence of the sum over eigenvalues.  
The resulting ambiguity corresponds to adding a polynomial in 
$E$ and does not affect poles and residues, i.e.~it leaves the 
spectrum invariant.

The trace of the resolvent $G(E)$ is equal to the negative logarithmic 
derivative of the Fredholm determinant  
${\mathcal D}(E)=\det(H-E)$,
\begin{equation}
G(E)=-{\partial \ln{\mathcal D}(E)\over \partial E}\,,
\label{einstFred}
\end{equation}
in terms of which the spectral equation reads 
\begin{equation}
\label{eSpectralEquation}
{\mathcal D}(E)= \det (H - E) = 0\,.
\end{equation}
This can be seen as follows,
\begin{eqnarray}
\frac{\partial}{\partial E} \ln{\mathcal D}(E) &=& 
\frac{\partial}{\partial E} \ln \det (H - E)
\nonumber\\
&=& 
\frac{\partial}{\partial E} \tr \ln (H - E)
\nonumber\\
&=& 
- \tr \frac{1}{H - E}
\: = \: - G(E)\,.
\end{eqnarray}

%
%
\subsection{Spectral Equation and Logarithmic Derivative of the Wave Function}
\label{ssSpectral}

In the normalization 
\ref{einstSchr}, the Schr\"odinger equation can be written as,
\begin{equation}
\bigl(-\d_q^2 + 2\,V(q)/g^2 - 2E/g\bigr)\,\psi(q)=0\,, 
\end{equation}
which, in the notation of appendix \ref{appSchrRic},  
implies $L=2\,(H-E)/g$, $u(q)=2\,V(q)/g^2$, $z=2\,E/g$  and thus a 
different normalization for the resolvent (\ref{einstResolv}).
The corresponding diagonal elements $r(q)$ of the resolvent 
$[H-E]^{-1}$ satisfy
\begin{equation}
\frac12\, g^2 \, r \, r'' - \frac{1}{4} \, g^2 \, r'{}^2 =
2 \, \bigl(V(q) - Eg\bigr)\, r^2 - 1\,.
\label{einstresolv}
\end{equation}
The equation defines $r(q)$ up to a sign that is  fixed by
the behaviour for $E\to-\infty $.
The trace of $R(E)$ is then given by 
\begin{equation}
G(E)=\tr {1\over H-E}=\int\d q\, r(q), 
\end{equation}
an expression that requires some regularization to deal 
with the large $q$ divergences. 

Returning to the formalism of appendix~\ref{appSchrRic},  
one finds $r(q)\propto 1/S_+(q)$.
A comparison between the different normalizations leads to
\begin{equation}
r(q)=1/S_+(q).
\end{equation}
We then conclude that
\begin{equation} 
G(E) =\int \d q\,{1\over S_+(q)}\,, \quad 
\ln {\mathcal D}(E)  ={1\over g}\int\d q\, S_+(q). 
\end{equation}
The integral may diverge for $|q|\to\infty $ and 
has then to be regularized.

One way of writing  the {\em spectral equation}, or quantization 
condition, is then [see equation (\ref{eresolspectr})]
\begin{equation}
\frac{1}{2{\rm i}\pi g} \,
\lim_{\varepsilon\to0_+}\int\d q
\left[S_+(q, g, E_N-i\varepsilon)- 
S_+(q,g,E_N + i\varepsilon)\right] = N + \ud\,,
\end{equation}
for $N\ge 0$. 
In the case of analytic potentials, the contour of integration 
can be locally deformed in the $q$ complex plane, and the limit 
$\varepsilon\to $ can then be taken. 
The spectral equation in terms of the logarithmic 
derivative of the wave function (\ref{eLogDeriv}) becomes
\begin{equation}
\label{equantCii}
B(E, g) \equiv -\frac{1}{2 {\rm i}\pi g}\,
\oint_{C}\d z\, S_+(z, g, E) = N+\ud\,, 
\end{equation}
where $N$ has also the interpretation of the number of 
real zeros of the eigenfunction, and $C$ is a contour that encloses them. 

%
%
\subsection{Perturbative Expansion}
\label{ssPerturbative}

The purpose of the current chapter 
is to investigate the condition (\ref{equantCii}) in detail.
Before embarking on this endeavour, 
we remark that the rather elegant formulation discussed here
is restricted, however,
to one dimension and analytic potentials, which 
represents integrable systems from the 
point of view of classical mechanics. Our treatment bypasses the
difficulties generally associated with turning points.
This is the form that is useful here in the context of
perturbative and WKB expansions. Recently, it has
been pointed out by Rosenfelder~\cite{Ropriv} that nonanalytic 
energy shifts of the form $\exp(-a/g)$ also occur in potentials
which cannot be represented as polynomials.

In general, canonical considerations which lead to the 
Bohr--Sommerfeld quantization conditions (see, e.g.,
chapter III of~\cite{Da1968}), are replaced here by a
simpler rigorous approach, based on the Riccati equation,
that uses only the uniqueness of the wave function at $q=0$.
We thus assume that the potential has a unique minimum 
at $q=0$. In order to 
see the connection with the uniformity of the wave function,
we note that the condition (\ref{equantCii}) may be rewritten as 
\begin{equation}
\frac{1}{2 {\rm i} \pi} \, \oint \d z \, \frac{\psi'(z)}{\psi(z)} = N\,.
\end{equation}
At $q \to 0$, this implies that the coefficient of $q^{-1}$ in
the Laurent expansion of $S_+(q)$ is of the form
$-g N/q$ where $N$ is an integer.
{\em Idem est}, we have $\ln \psi(q) \sim N \, \ln q$
for $q \to 0$, a relation which is uniform under a sign change of $q$ only if 
$N$ is an integer.

The expansion for $g\to0$ at $E$ fixed of the equations 
(\ref{eRiccatpm}), or (\ref{einstresolv}), 
leads to the  expansion in powers of $g$ of the function
$B(E,g)$ that enters the left-hand side of (\ref{equantCii}).
This function enters in expressions like
(\ref{egenisum}) and other quantization conditions.
We first indicate how the function $B(E,g)$ can be 
calculated {\em (i)} as a series expansion in powers of $g$ at $E$ fixed, 
which leads to the perturbative expansion, and 
{\em (ii)} at $Eg$ fixed, which leads to the
WKB expansion. We also discuss the relation between the two expansions.

In view of (\ref{eRiccatpm}), (\ref{eSym}) and (\ref{eSponly}),
we conclude that
\begin{equation}
B(E,g)=-B(-E,-g)\,.
\end{equation}
The latter reflection symmetry is purely formal ---  
it is valid in the sense of power series --- 
since it corresponds to change the quantum number 
$N$ in $-1-N$, which is unphysical. 

The {\em perturbative expansion} is obtained 
by expanding the solution of Riccati's equation 
(\ref{eRiccat}) in powers of $g$ (at $E$ fixed):
\begin{equation}
S(q)=\sum_{k\ge 0} g^k s_k(q)\,. 
\end{equation}
Setting
\begin{equation}
\label{edefU}
U(q)=\sqrt{2\,V(q)},\quad U(q)=q + {\mathcal O}(q^2)\,,
\end{equation}
we first find
\begin{equation}
s_0(q)=U(q),\quad s_1(q)={1\over 2 U(q)}\bigl(U'(q)-2E\bigr). 
\end{equation}
The square root in (\ref{edefU}) is to be understood such 
that the resulting function is uniform at the two 
minima of $V$, and positive in the region between the minima.
For example, in the case of the double-well potential
$V(q) = q^2\,(1-q)^2/2$, we have $U(q) = q\,(1-q)$.

Higher orders are obtained from the recursion relation
\begin{equation}
s_k(q)={1\over 2 U(q)}\left(s'_{k-1}(q)
-\sum_{l=1}^{k-1}s_{k-l}(q) s_l(q)\right)\,. 
\label{eRiccpert} 
\end{equation}
For example,
\begin{equation}
s_2(q)={U''\over 4U^2}-{3U'{}^2\over 8U^3}
+{EU'\over 2U^3}-{E^2\over 2U^3}\,.
\end{equation}
Thus
\begin{equation}
S_+=U(q)-g{E\over U(q)}+g^2\left({U''\over 4U^2}
-{3U'{}^2\over 8U^3} -{E^2\over 2U^3}\right) + {\mathcal O}(g^3). 
\end{equation}
The contour integral (\ref{equantCii}) reduces to the residue 
at $q=0$ (where the potential has its unique absolute minimum). 
Thus, it depends only on the expansion of $U(q)$ at $q=0$. 
To find the perturbative spectrum, it is sufficient to expand the recursion 
relation (\ref{eRiccpert}) in powers of $q$. 
For two (or more) degenerate minima, this procedure gives rise to 
perturbative functions $B_1, B_2, \dots$ in the sense of 
chapter~\ref{BSqf} [see also equation (\ref{econj}) below].
 
For instance, at order $g$, one finds  $B(E,g)=E$. The next 
orders, for a general polynomial potential,
are given in equation (\ref{einstBgengii}).
The condition (\ref{equantCii}) then implies that the eigenfunctions 
are uniform at  $q=0$, as they should be.  

There is an interesting connection 
to {\em Borel summability}. Note that even in situations where 
the perturbative expansion of eigenvalues is Borel summable, 
it is not clear whether the function $B(E,g)$ is Borel summable 
in $g$ at $E$ fixed. Indeed, the wave function $\psi(q)$ is 
unambiguously defined only when the quantization condition is satisfied 
with $N$ a non-negative integer. When $E$ is not an eigenvalue, 
the solution of the Schr\"odinger
equation is an undefined linear combination of two particular solutions. 

%
%
\subsection{WKB Expansion}
\label{ssWKBspec}

As explained in chapter~\ref{ssTransition},
the WKB expansion~\cite{Vo1983long}
is an expansion for $g \to 0$ at $Eg$ fixed, 
in contrast to the perturbative expansion where $E$ is
fixed. In~\cite[chapter~6]{ItZu1980}, it has been stressed 
that the loop expansion in quantum field theory 
is actually an expansion in powers of $\hbar \sim g$,
i.e.~a generalized WKB expansion.
We thus expand equation (\ref{eRiccat}) in the complex $q$-plane in 
powers of $g$, at $Eg$ fixed, starting from 
\begin{equation}
S(q)=S_0(q)\,,\qquad S_0^2(q)=2\,V(q) - 2\,gE \,.
\label{eWKBlead}
\end{equation}
For $E>0$, the function $S_0$ has two branch points $q_-<q_+ $ on 
the real axis. We put the cut between $q_-$ and $q_+$ and choose the 
determination of $S_0$ to be positive for $q>q_+$. 
This ensures the decrease of  wave functions on the real axis 
for $|q|\to\infty $, at least in the WKB approximation.

Then [see equation (\ref{eSWKB})],
\begin{subequations}
\label{eWKBCalc}
\begin{equation}
S(q)=\sum_{k\ge 0} g^k S_k(q). 
\end{equation}
One first obtains
\begin{equation}
S_1={S_0' \over 2S_0}\,. 
\end{equation}
For $k>1$, one finds the recursion relation 
\begin{equation}
\label{eSrecursion}
S_k(q)={1\over 2 S_0(q)}\,
\left(S'_{k-1}(q)-\sum_{l=1}^{k-1}S_{k-l}(q)S_l(q)\right).   
\end{equation}
\end{subequations}
At order $g^2$, for example, 
\begin{equation} 
S_2={S_0'' \over 4 S_0^2}-{3 S_0'{}^2 \over 8 S_0^3} .  
\end{equation}
In the semi-classical limit, the contour $C$ in the expression 
(\ref{equantCii}) encloses the cut of $S_0(q)$
which joins the two solutions of $S_0(q)=0$
(classical turning points). 

Note that equation (\ref{equantCii}) can also be written as
\begin{equation}
\exp\left[-{1\over  g}\oint_{C }\d z\, S_+(z) \right]+1=0\,.
\label{equantCiii} 
\end{equation}
The contribution of $S_1$ as well as that of all $S_j$ with odd $j$
to the symmetric component $S_+$ vanishes. 
This is evident as we inspect the symmetry condition
(\ref{eSym}), the properties of the leading approximation (\ref{eWKBlead})
and take into account the additional odd-power-of-$g$ prefactor 
that prevails in all $S_j$ with odd $j$. Therefore, we may safely 
ignore all odd-order WKB approximants in the sequel; this 
simplifies the analysis carried out in appendix~\ref{appWKB}.

We now consider the {\em WKB expansion} of the 
{\em resolvent}.
In the WKB limit $E \, g = {\mathcal O}(1)$, the solution of equation 
(\ref{einstresolv}) at leading order, in the notation 
(\ref{eRiccat}), is
\begin{equation}
r(q)={1\over S_0(q)}={1\over\sqrt{2\,V(q)-2\,Eg}}\ 
\Rightarrow\ G(E)=\int{\d q \over \sqrt{2\,V(q)-2\,Eg}}. 
\end{equation}
Formally,
\begin{equation} 
\ln{\mathcal D}(E) = {1\over g}\int\d q \, S_0(q) = 
{1\over g}\int\d q \, \sqrt{2\,V(q)-2\,Eg}\,, 
\end{equation}
but the divergence of the integral for $|q|\to\infty $ has to be eliminated by
subtracting a $E$-independent infinite constant.\par
At order $g^2$, one finds [see equation (\ref{eWKBii})]
\begin{equation}
S_+(q)=S_0(q)+ g^2\left({S_0'' \over 4 S_0^2}-{3 S_0'{}^2 \over 8
S_0^3}\right) + {\mathcal O}(g^4),
\end{equation}
and thus, after an integration by parts,
\begin{equation}
\ln{\mathcal D}(E)= {1\over g}\int\d q \, S_0(q)+{g\over8}
\int\d q \,S_0'{}^2(q) \, S_0^{-3}(q) + {\mathcal O}(g^3). 
\end{equation}

In the present formulation, it is easy to {\em relate 
the WKB and the perturbative expansions}. 
The terms in the WKB expansion of the quantization condition  
(\ref{equantCii})  are  contour integrals that,  clearly, are 
regular functions of $E$ at $E=0$. To obtain the perturbative expansion 
from the WKB expansion, one thus expands the functions $S_{2k}(q)$
in powers of $E$ and calculates the residues.

For example, at leading WKB order [see also appendix~\ref{sSecond} 
and equation~(\ref{eS0powerii})], 
\begin{equation}
\label{eLeading}
S_0 = U \, \sum_{n=0} \left(2gE\,U^{-2}\right)^n
\frac{\Gamma(n-\ud)}{\Gamma(n+1)\Gamma(-\ud)}.  
\end{equation}
One has then to evaluate the residue at the origin of
\begin{equation}
-{1\over 2{\rm i}\pi}\oint_{C}\d q\, U^{1-2n}\,.
\end{equation}
More generally, replacing  $S_+$  by its WKB expansion and  
expanding each term  in a power series of $Eg$, one obtains 
the perturbative expansion of the function $B(E,g)$. 
The WKB expansion  corresponds to successive summations, to all 
orders in $g$, of the
terms of highest degree in $E$ of the perturbative expansion. 
A few terms are calculated in appendix~\ref{appinstcal}.

%
%
\section{Resolvent and Degenerate Minima in Specific Cases}
\label{ssresoldeg}

%
%
\subsection{Potentials with Degenerate Minima}

\begin{figure}[htb!]
\begin{center}
\begin{minipage}{12.0cm}
\begin{center}
\epsfxsize=91.7mm
\epsfysize=60.7mm
\centerline{\epsfbox{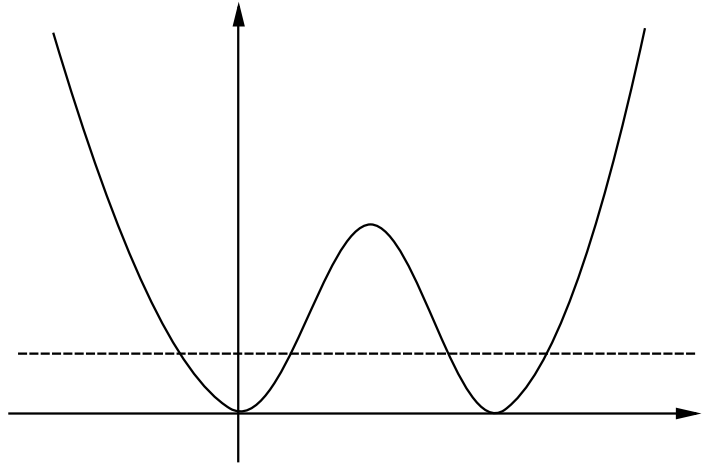}}
\caption{\label{figcirmi}
The four turning points.} 
\end{center}
\end{minipage}
\end{center}
\end{figure}

We now consider potentials of double-well type, with two 
(not necessarily symmetric) degenerate minima.
The symmetric case is displayed in figure~\ref{figcirmi},
whereas the discussion here is actually applicable 
also to the case of minima with unequal curvatures
[see equation~(\ref{eomega1omega2})].  
For $E$ small enough, the function $S_0(q)$ has four branch 
points $q_1,\ldots,q_4$ on the real axis.

One expects, on intuitive grounds, to find two perturbative functions
$B_1(E,g)$ and $B_2(E,g)$ obtained by integrating $S_+$ around the 
two cuts $[q_1,q_2]$ and $[q_3,q_4]$ of figure~\ref{figcirmi}. 

However, if one thinks in global terms,  an obvious difficulty arises: 
a function $S_0(q)$ with the corresponding two cuts is no longer 
an appropriate solution because it increases both for $q\pm \infty $, 
and thus the wave function cannot decrease both for $q\pm \infty $. 
Moreover, the contour integral of $S_0$ cannot give the correct answer. 
For example, if the potential is symmetric, the integral vanishes.

For symmetric potentials, the correct perturbative answer is obtained 
from the difference between the two integrals  when  the additional zero 
at the symmetry point for odd eigenfunctions is taken into account. 
One then finds  
\begin{equation}
2B(E,g)=1+ 2[N/2], 
\end{equation}
where $[N/2]$ is the integer part of $N/2$, 
and this is the correct perturbative
answer.

Moreover, quite generally the WKB expansion of the spectral condition 
has a singularity when $2E$ reaches the relative maximum of the
potential between the
degenerate minima, where two complex singularities of the $S_0(q)$
pinch the real axis and the number of real singularities 
passes from two to four.

Therefore, let us examine what happens if one starts from $E$ large enough, 
where $S_0$ has only two branch points on the real axis and no special 
difficulty arises,
and proceeds by analytic continuation of the spectral condition.
At leading order, before continuation, the function $B(E,g)$ can be written as
\begin{equation}
B(E,g)={\mathcal B}(E,g) \equiv{1\over \pi g}\int_{q_-(E)}^{q_+(E)}
\d q\,\sqrt{2\,Eg -2\,V(q)} + {\mathcal O}(g)\,,
\end{equation}
where $q_\pm $ are the two real zeros of $Eg-V$. To avoid the singularity 
when $E$ reaches the local maximum of the potential, we take $g$, and 
thus $Eg$ slightly complex. Eventually, the initial contour is deformed 
into a sum $C'$ of three contours around the cuts:
\begin{equation}
{\mathcal B}(E,g)=B_1(E,g)+B_2(E,g)+B_{12}(E,g) 
\end{equation}
with
\begin{equation}
B_{12}(E,g)={1\over \pi g}\int_{q_2}^{q_3}
\d q\,\sqrt{2\,Eg -2\,V(q)} + {\mathcal O}(g) \,,  
\end{equation}
where the cuts now are $[q_1,q_2]$ and $[q_3,q_4]$. 
The function  $B_{12}$ is imaginary and the determination of 
the square root in the  integral depends on the continuation. 
The integral can also be written as
\begin{equation}
B_{12}(E,g)={1\over 2\pi g}\oint_{C'}\d q\,\sqrt{2\,Eg -2\,V(q)} +
{\mathcal O}(g)\,, 
\end{equation}
where $C'$ surround the function cut along $[q_2,q_3]$.
Note that  $B_{12}(E,g)$ as a function of $E$ has a branch point at
$E=0$. In a rotation about an angle $2\pi$ around $E=0$, $B_{12}$ becomes 
$B_{12} \pm (B_1+B_2)$.

The quantization condition (\ref{equantCii}) or alternatively
(\ref{equantCiii}),
\begin{equation}
\label{equant}
-\frac{1}{2 {\rm i}\pi g}\, \oint_{C'}\d z\, S_+(z, g, E) = N+\ud\,, 
\qquad \qquad
\exp\left[-{1\over  g}\oint_{C'}\d z\, S_+(z) \right] + 1=0\,,
\end{equation}
then becomes
\begin{equation}
\exp\left[2 {\rm i}\pi\,
\left(B_1(E,g)+B_2(E,g)+B_{12}(E,g)\right)\right] +1=0 \,.
\end{equation}
The contribution  $B_{12}$ contains the effect of barrier penetration and 
we choose the sign in front of it, which depends on the analytic continuation, 
to obtain decreasing corrections to perturbation theory. Finally, the 
contribution $B_1+B_2$ can be absorbed into 
in a redefinition of $B_{12}$ by going to the proper sheet 
in the $E$ Riemann surface.

Comparing with the instanton result, one infers the decomposition 
[in the notation of equation (\ref{egeneral})]
\begin{eqnarray}
& & {1\over g}\oint_{C'}\d z\, S_+(z) =
A(E,g) + \ln(2\pi) \nonumber\\
& & \qquad - \sum_{i=1}^2
\left\{ \ln \Gamma\left(\frac12-B_i(E,g)\right) +
B_i(E,g)\ln\left(-\frac{g}{2C_i}\right) \right\}\,,
\label{econj}
\end{eqnarray} 
where the constants $C_i$ are adjusted in such a way 
that $A(E,g)$ has no term of
order $g^0$ and the coefficient of $g^0$ is proportional to the expansion
of $B_1+B_2$. This equation, adapted to specific cases,
almost directly leads to the conjectures discussed
in chapter~\ref{BSqf} and summarized in chapter~\ref{sSummary}
when combined with the 
quantization condition (\ref{equant}).

The expansion for $Eg$ small of its WKB expansion 
yields the function $A(E,g)$.
To identify with the WKB expansion, the function $\Gamma(\ud-B)$ has 
to be replaced by its asymptotic expansion for $B$ large:
\begin{equation}
-\ln\Gamma(\ud-B )\mathop{\sim}_{E\to\infty } 
B \ln(-B)-B\mathop{\sim}_{g\to 0}B\ln(-E)+\cdots 
\label{einstGamBE}
\end{equation}
Therefore, at leading order,
\begin{equation} 
{1\over g}\,
{\rm Re}\, \oint_{C'}\d z\, S_+(z) =
A(E,g) + \ln(2\pi) +
\sum_{i=1}^2 B_i(E,g) \left[\ln\left(\frac{gE}{2C_i}\right)-1\right]\,.
\end{equation}
Here, we have assumed that the determinations of all functions 
of $E$ are such that
$A(E,g)$ is real in perturbation theory, something that 
is possible since the ambiguities
are proportional to $2{\rm i}\pi(B_1+B_2)$: in the 
WKB expansion, the discontinuity comes from the asymptotic expansion 
(\ref{einstGamBE}) of the sum of the $\Gamma$ functions, 
which also has a discontinuity $-2{\rm i}\pi( B_1+B_2) $, showing the 
consistency of the
whole scheme.  Moreover, the contribution proportional to 
$\ln(-g)$ leads to the combination $\ln(gB_i)$, which is 
globally even in $E,g$ as it should be [because it results from an 
integral over $S_+$ with the symmetry property (\ref{eSym})].

A calculation of $A(E,g)$ at a
finite order in $g$ requires the WKB expansion and the asymptotic expansion
of the $\Gamma$-function only to a finite order. For example, the expansions
up to order $g^2$ (\ref{eAdble}) and 
(\ref{ecosa}) of the function $A(E,g)$ for the
double-well and cosine potentials, which have initially been determined in
part by numerical calculations, are reproduced by the expansion of the two
first WKB orders (the calculations for the double-well potential are given in
appendix~\ref{appinstcal}).

%
%

%
%
\subsection{General Symmetric Potentials} 

We first consider general symmetric potentials: 
$V(-q)=V(q)$.
As is evident from the discussion in chapter~\ref{sDouble} above
and chapter~\ref{sPartition} below,
it is necessary to introduce a further quantum number in this
particular case, which is the parity.
States are classified according to the principal 
quantum number $N$ and the parity $\varepsilon$.
We again combine the formalism of chapter 
\ref{ssWKBspec} and appendix \ref{appSchrRic}.
Using equation (\ref{eresolGasig}), and taking care of the 
different normalization, one infers
\begin{equation} 
G_a(E)\equiv \tr P \left[H-E\right]^{-1}=
2\int_0^\infty  {\d q\over S_+(q)}
\exp\left[-{2\over g}\int_0^q\d y\,S_+(y)\right], 
\label{einstGadef}
\end{equation}
where $P$ is the reflection operator that acts here according to
$P\,\psi(q) = \psi(-q)$ [cf.~ equation~(\ref{eparity}) 
where the reflection symmetry is about $q=1/2$ in contrast to the 
case of a center of symmetry at $q=0$ discussed here].

An alternative integrated form is 
[equation (\ref{eresolPFredh})]
\begin{equation}
\label{ecorr2}
{\mathcal D}_a(E) =
\exp\left[-\tr P\ln(H-E)\right] \approx
-\frac{1}{S_+(0)} \, \frac{\partial}{\partial E}S_+(0) =
-\frac{g}{[S_+(0)]^2}\,.
\end{equation}
At leading order in the WKB limit, one finds $-(2E)^{-1}$,
the harmonic oscillator result for $E\to-\infty $ 
[equation (\ref{eresolGahar})].

In the case of the {\em symmetric double-well potentials,} 
this situation again is more subtle, because
one has to consider two different WKB approximations near the two wells. 
number in this case 
A strategy consists in calculating~\cite{ZJ1981jmpremark}
\begin{equation}
G_a\bigl(\ud(E_{+,N}+E_{-,N})\bigr)\sim -{4\over E_{-,N}-E_{+,N}} .
\label{einstGaN}
\end{equation}
We thus consider the Schr\"odinger
equation in the normalization 
(\ref{einstSchr}) with $E=N+\ud$ and call $q_0>0$ the position of 
one of the minima of the potential: $V(q )\sim \ud(q - q_0)^2$.\par 
For $  q \ge 0  $, at leading order in the semi-classical limit, 
\begin{equation}
\label{eSplus}
S_+(q)=\sqrt{U^2(q)-2Eg}=U(q)-{Eg \over U(q)} +
{\mathcal O}(g^2)\,,
\end{equation}
where $U(q)$ is defined as [see equation (\ref{edefU})]
\begin{equation}
U(q)=\sqrt{2\,V(q)},\quad U(q)=q + {\mathcal O}(q^2)\,.
\end{equation}
We choose the branch cut of the 
square-root function such that 
$U(q)$ becomes a function regular at $q=q_0$ and positive for $q>q_0$.
The approximation (\ref{eSplus}) is valid for $|q-q_0|$ large.

The solution $\psi_N(q)$ can be written  as
\begin{eqnarray}
\psi_N(q) &\sim& q_0^{-N-1/2}(q_0-q)^N\,
\sqrt{q-q_0\over U(q)}
\exp\left[-{1\over g} \int_0^q \d q'\, U(q')\right] 
\nonumber \\
& & \times
\exp\left[ (N+\ud)\int_0^q\d q'\left({1\over U(q')}
-{1\over q'-q_0}\right) \right]\,.
\end{eqnarray}
We introduce the definitions
\begin{eqnarray}
\label{estrangeA}
a &=& -4 \, \int_0^{q_0}\d q\, U(q) \,,\\
C &=& q_0^2 \, 
\exp\left[-2\int_0^{q_0} 
\d q\left({1\over U(q)} + {1\over q+q_0} - {1\over q-q_0}\right)\right]\,.
\end{eqnarray}
When comparing to (\ref{einstAdegen}), it is necessary to remember that
we are dealing with a symmetric potential $V(-q) = V(q)$ 
in the current section, and we integrate from the center of symmetry
at $q = 0$ to one of the minima of the potential.
By contrast, in (\ref{edefA}) and in (\ref{einstAdegen}), 
the integration extends from one of the minima to the other.
Also, the convention for the square root is the opposite 
$\tilde U = - U$ for $q$ inside the region between the 
two minima. 

Based on (\ref{edefra}), we can write $r_a(q)$ as
\begin{equation}
r_a(q)=C^{-N-1/2} \, \e^{a/2g} \, \eta^2 _N(q)
\end{equation}
with
\begin{equation} 
\eta_N(q) \sim  (q_0-q)^N \, \e^{-(q-q_0)^2/2g}\,,  
\end{equation}
an expression valid for $1\ll |q-q_0|\ll \sqrt{g}$.\par
The leading contribution to the integral comes from the neighbourhood of
$q=q_0$, where the harmonic oscillator approximation is relevant.
The corresponding eigenfunction  with norm $\sqrt{\pi}$ is
\begin{equation}
\varphi_N(q)={1\over 2^{N/2}\sqrt{N!}}\,
(q-\d_q)^N \, \e^{-q^2/2}
\mathop{\sim}_{q\to \infty}
{ 2^{N/2}\over\sqrt{N!}}q^N\e^{-q^2/2}\,.
\end{equation}
Therefore, $\eta_N(q)$ has to be approximated by
\begin{equation}
\eta_N(q)\sim\sqrt{N!} (g/2)^{N/2} 
\varphi_N\bigl((q-q_0)/\sqrt{g}\bigl).
\end{equation}
We conclude
\begin{eqnarray}
G_a \bigl(\ud(E_{N+}+E_{N-})\bigr)
&\sim&
2 \, C^{-N-1/2} \, N! \, (2/g)^N \, \e^{a/2g}
\int\d q\, \varphi_N^2 \, \bigl((q-q_0)/\sqrt{g}\bigl)
\nonumber\\
&\sim& 2\sqrt{\pi g}\,C^{-N-1/2} \, N! \, (g/2)^N \, \e^{a/2g}\,.
\end{eqnarray}
Using equation (\ref{einstGaN}), one obtains the 
one-instanton contribution to the energy difference 
\begin{equation}
\delta E_N(g)\sim  
{2 \over\sqrt{2\pi}}{1\over N!} \,
\left(2C\over g\right)^{N+1/2} \,
\e^{-a/2g},
\end{equation}
a result  consistent with the expansion of the zeros of expression 
(\ref{einstsyge}).

Note that to go beyond leading order by this method is simple for $N$ 
fixed, but not for generic $N$ because the WKB expansion yields the
perturbative eigenfunction under the form of an expansion only valid for 
large arguments. 

To compare with instanton calculations, we calculate
the {\em contribution of the twisted resolvent at leading order},
\begin{equation}
\tr P\ln(  H-E)=
\sum_N\ln\left( E_N-\delta E_N/2-E \over E_N+\delta E_N/2-E\right)
\sim
-\sum_N{\delta E_N \over E_N-E }\,, 
\end{equation}
where $E$ has been assumed to be not too close to an eigenvalue.
At leading we can replace $E_N$ by $N+\ud$. Thus,
\begin{eqnarray}
& & \tr P\ln(H-E) 
\nonumber\\
&\sim&
-  {2 \over\sqrt{2\pi}}\,
\e^{-a/2g} \, \sum_N {1\over N!}
\left(2C\over g\right)^{N+1/2}
\int_0^\infty \d \beta\, \e^{\beta( E-N-1/2)}
\nonumber\\
&\sim& - 2 \left(  C\over \pi g\right)^{ 1/2}
\e^{-a/2g}
\int_0^\infty \d \beta\, 
\exp\left[\beta (E-\ud)+2\,C\e^{-\beta}/g\right].
\end{eqnarray}
We note that if we evaluate the integral in the limit 
$g\to0_-$, we obtain 
\begin{equation}
\tr P\ln(H-E) \sim
-2 \, \frac{\e^{-a/2g}}{\sqrt{2\pi}} \,
\exp\left[{\rm i}\,\pi(E-\ud)\right]\,
\left(\frac{2C}{g}\right)^E\,
\Gamma (\ud-E),
\end{equation}
an expression that appears in equation (\ref{epoles}) and has the correct 
poles and residues. Of course, the real part,
\begin{equation}
\tr P \ln(H-E)\sim - 
2 {\e^{-a/2g}\over\sqrt{2\pi}} \, \cos\bigl(\pi(E-1/2)\bigr) \,
\left({2C\over g}\right)^  E\Gamma (\ud-E),
\end{equation}
has the same poles with the same residues, but has zeros for $E<0$.

%
%
\subsection{${\mathcal O}(\nu)$--Symmetric Hamiltonians in the Radial Coordinate}

A general ${\mathcal O}(\nu)$-symmetric Hamiltonian, after diagonalization of the 
angular momentum, decomposes into a set of Hamiltonians $H_l$, 
where $l$ is the angular momentum, of the form
\begin{equation}
H_l= {g\over 2}\left[-\left(\d\over\d r\right)^2- 
{\nu-1\over r}{\d\over\d r} + 
{l(l+\nu-2)\over r^2}\right]+{1\over g}V(r). 
\label{eOnuhamg} 
\end{equation}
We assume that $V(r)$ is an analytic function of $r^2$, in 
such a way that the Schr\"odinger equation can be continued to the complete 
real axis.   Eigenfunctions are then odd or even, and they 
behave like $r^l$ for $r\to 0$.
The eigenfunction $\psi_{l,N}$ corresponding to angular momentum 
$l$ and quantum number $N$ thus has $l+2N$ zeros on the real axis. 
The quantization condition (\ref{equantCii}) is replaced  by
\begin{equation}
{1\over 2{\rm i}\pi}\oint_C\d z\, {\psi_{l,N}'(z)\over \psi_{l,N}(z)}=l+2N\,,
\label{equantCpsil}
\end{equation}
where the contour $C$ now encloses all zeros of the 
eigenfunction $\psi(r)$ on the real axis (this includes $r\le 0$).

The transformation
\begin{equation}
{\psi'(r)\over\psi(r)} = -{1\over g}S(r)-{\nu-1 \over 2r},
\end{equation}
leads to the Riccati equation 
\begin{equation}
gS'(r)-S^2(r) + 2V(r)-2Eg+g^2 \,{j^2-1/4\over r^2}=0 
\end{equation}
with $j=\nu/2+l-1$. \par
Introducing the two functions $S_\pm$, we can write the 
quantization condition as
\begin{equation}
-{1\over 2 {\rm i}\pi \, g}\oint_{C}\d z\, S_+(z)=j+2N+1\,.
\label{equantCl}
\end{equation}

In the {\em perturbative expansion of
the ${\mathcal O}(\nu)$-symmetric Hamiltonian},
the effect of angular momentum appears 
only at order $g$. 
In the parameterization (\ref{einstUpU}) [see also 
equations (\ref{eBEgjg4}),
(\ref{einstBgengii}) and (\ref{einstBgengij})],
this implies the following relation between
$B_\nu(E, g, j)$ and $B_{\rm dw}(E, g)$:
\begin{equation}
\label{einstj}
B_\nu(E,g,j) = 2\,B_{\rm dw}\left(\frac{E}{2},-g\right) +
\frac{1}{12}\, \alpha_2 \,
\left(j^2 - 1\right)\,g +
{\mathcal O}(g^2)\,.
\end{equation}

The {\em WKB expansion of the ${\mathcal O}(\nu)$-symmetric Hamiltonian} 
is now an expansion at $E\,g$ and $g^2\,(j^2-1/4)$ fixed. The leading order is
\begin{equation}
S(r) = \sqrt{U^2(r) - 2E \, g + \frac{g^2}{r^2} \,
\left(j^2-\frac{1}{4}\right)} \,.
\end{equation}
Again, the 
perturbation series is recovered by simply expanding in powers of $g$. 

%
%
\chapter{Instantons in the Double--Well Problem}
\label{ssninstdw}

%
%
\section{Orientation}

The conjectures presented in chapter \ref{BSqf} have 
been initially motivated by semi-classical evaluations of  
path integrals, more precisely by summation of leading order multi-instanton 
contributions.  

Therefore, we now explain, first in the example of the double-well 
potential, how multi-instanton contributions can be summed explicitly, 
yielding results consistent with the conjectured form (\ref{egenisum}) of the 
secular equations. The reason for still emphasizing instanton calculus in 
situations now rather well understood by other methods, is that the 
arguments can be generalized to other examples where our understanding is 
more limited.
%

%
%
\section{General Considerations}

%
%
\subsection{Partition Function and Resolvent}
\label{sPartition}

The path integral formalism allows calculating
directly the quantum partition function, which  
for Hamiltonians with discrete spectrum has the expansion
\begin{equation}
{\mathcal Z}(\beta)\equiv \tr\e^{-\beta H} =
\sum^\infty_{N=0}\e^{-\beta E_N}\,.
\end{equation}
Eigenvalues are labeled $E_N$ ($N \in \mathbb{N}_0$),
with $E_0$ being the ground-state energy and 
$E_N \leq E_{N+1}$.
The trace $G(E)$ of the resolvent of $H$ is related to the 
partition function by Laplace transformation:
\begin{equation} 
G(E) =\tr{1\over H-E} =
\int_0^\infty \d\beta\, \e^{\beta E}{\mathcal Z}(\beta)\,, 
\label{eninstGZ}
\end{equation}
where we have to 
assume initially that $E<E_0$ (with $E_0$ being the ground state energy),  
to ensure the convergence of the integral for $\beta \to\infty $.

The poles of $G(E)$ then yield the spectrum of the Hamiltonian $H$. 
From $G(E)$ one can also derive the Fredholm determinant 
${\mathcal D}(E)=\det(H-E)$, which vanishes on the spectrum 
(chapter \ref{ssinstFredh}). 

Note that the integral (\ref{eninstGZ}) does not necessarily converge at 
$\beta =0$ because the eigenvalues may not increase fast enough, as 
the example of the harmonic oscillator shows:
\begin{equation}
G_{\rm osc.} (E)=\int_0^\infty \d\beta\, {\e^{\beta E}
\over 2\sinh(\beta /2)}, 
\end{equation}
which diverges. However, $G(E)-G(0)$ is defined, and fixing the 
irrelevant constant term we can set [see equation (\ref{goscii})]
\begin{equation}
\label{gosc}
G_{\rm osc.} (E) = - \psi\left(\ud - E\right) = 
\frac{\partial}{\partial E} \ln \Gamma\left(\ud - E\right)\,.
\end{equation}

Note that for the double-well potential, one can separate eigenvalues
corresponding to symmetric and antisymmetric eigenfunctions by considering
the two functions
\begin{equation}
{\mathcal Z}_{\pm}(\beta )=\tr\left[\ud \, (1\pm P) \, \e^{-\beta H}\right]=
\sum_{N=0}^\infty \e^{-\beta E_{\pm ,N}}\,,
\label{einstZpm}
\end{equation}
where $P$ is the parity operator (\ref{edwparity}). 
The eigenvalues are then poles of the Laplace transforms 
($\varepsilon =\pm$):
\begin{equation}
G_\varepsilon (E) = \int_0^\infty \d\beta\, 
\e^{\beta E}{\mathcal Z}_{\varepsilon}(\beta )\,. 
\label{eninstGZpm}
\end{equation}

%
%
\subsection{Path Integrals and Spectra of Hamiltonians}
\label{sPathIntegrals}

In the path integral formulation of quantum mechanics, 
the partition function is given by
\begin{equation}
{\mathcal Z}(\beta)\propto \int_{q (-\beta /2)=q (\beta /2 )} 
\left[\d  q (t) \right]
\exp\left[-{1\over g} \, {\mathcal S} \bigl(q (t) \bigr)\right], 
\label{epathint} 
\end{equation}
where the symbol $\int[\d q(t)]$ means summation over all paths
which satisfy the boundary conditions, that is closed paths, and
$ {\mathcal S} (q ) $ is the Euclidean action:
\begin{equation} 
{\mathcal S}(q)= \int^{\beta /2}_{-\beta /2} \left[
\ud \, {\dot q} ^2 (t)+V\bigl(q(t)\bigr) \right] \d  t\,. 
\end{equation}
In~\cite{Fo2000primer}, it has been stressed that the 
formulation in terms of the Euclidean action follows naturally 
from an analytic continuation of the usual 
definition of the path integral to imaginary time.

In the case of the symmetric potential 
(\ref{ehamdw})~which has degenerate minima,
it is actually necessary to also consider the quantity
\begin{equation}
{\mathcal Z}_{\rm a}(\beta)\equiv \tr\left(P\e^{-\beta H}\right)\propto
\int_{q(-\beta /2)+q (\beta /2 )=1} \left[\d  q (t)
\right] \exp\left[-{1\over g}{\mathcal S} \bigl(q (t) \bigr)\right],
\label{epathii}
\end{equation} 
where $P$ is the parity operator 
(\ref{edwparity}). Then, eigenvalues corresponding to symmetric and
antisymmetric eigenfunctions can be derived from the combinations
\begin{equation}
{\mathcal Z}_{\pm}(\beta )=
\tr\left[\ud \, (1\pm P)\,\e^{-\beta H}\right]= 
\ud\bigl({\mathcal Z}(\beta )\pm{\mathcal Z}_{\rm a}(\beta)\bigr). 
\end{equation}
This construction is not necessary (even impossible) in the 
case of asymmetric wells discussed below in chapter~\ref{ssninstg}.
Because there is no parity in the asymmetric case, the 
quantity ${\mathcal Z}_{\rm a}(\beta)$ does not find a natural 
representation in terms of eigenfunctions in contrast to 
the equations (\ref{ezaeigen}) and (\ref{eparity}) below, and 
it is only the quantity ${\mathcal Z}(\beta)$ as defined in (\ref{epathint})
which contributes to the partition function. Of course,
${\mathcal Z}(\beta)$ is composed exclusively of paths in which the 
starting and the endpoints are identical, and one-instanton 
effects therefore do not contribute to ${\mathcal Z}(\beta)$.

We adopt here a convention, in which
the proportionality sign in the 
notation (\ref{epathint}) implies that the final integration over 
$Q = q(\beta/2)$ is understood. 

Because the only irreducible representations of the parity 
operator are one-dimensional,
the nondegenerate eigenfunctions have definite parity,
\begin{equation}
\label{eparity}
P \phi_{\epsilon,N}(1-q) = \epsilon \, \phi_{\epsilon,N}(q) \,.
\end{equation}
Here, $\varepsilon = \pm 1$ is the eigenvalue of the parity operator
defined according to (\ref{edwparity}).
It is instructive to consider
the eigenfunction decomposition of the spectral partition 
function ${\mathcal Z}(\beta)$,
\begin{equation}
{\mathcal Z}(\beta) = 
\int {\rm d}q \,
\sum_{\varepsilon,N} 
\phi_{\varepsilon,N}(q) \, \phi^*_{\varepsilon,N}(q) \,
\exp(- \beta \, E_{\varepsilon,N})\,.
\end{equation}
Here, we assume a Hamiltonian
whose spectrum is discrete
(no continuous spectrum).
The quantity ${\mathcal Z}_a(\beta)$ then finds a representation as
\begin{equation}
\label{ezaeigen}
{\mathcal Z}_a(\beta) = \sum_{\varepsilon,N}
\int {\rm d}q \, \phi_{\varepsilon,N}(q) \, \phi^*_{\varepsilon,N}(1-q) \,
\exp(- \beta \, E_{\varepsilon,N})\,.
\end{equation}
Now, ${\mathcal Z}_+(\beta)$ is determined by the 
even eigenfunctions alone,
\begin{eqnarray}
{\mathcal Z}_+(\beta) &=&
\sum_{N=0}^\infty
\int {\rm d}q \, \phi_{+,N}(q) \, \phi^*_{+,N}(q) 
\exp(- \beta \, E_{+,N})
\nonumber\\
&=& \frac{{\mathcal Z}(\beta) + {\mathcal Z}_a(\beta)}{2} \,.
\end{eqnarray}
whereas ${\mathcal Z}_-(\beta)$ is determined by the 
odd eigenfunctions,
\begin{eqnarray}
{\mathcal Z}_-(\beta) &=&
\sum_{N=0}^\infty
\int {\rm d}q \, \phi_{-,N}(q) \, \phi^*_{-,N}(q) 
\exp(- \beta \, E_{-,N})
\nonumber\\
&=& \frac{{\mathcal Z}(\beta) - {\mathcal Z}_a(\beta)}{2} \,.
\end{eqnarray}

%
%
\subsection{Perturbation Theory}

Perturbative
expansions, that is expansions in powers of $g$ for $g\to0$, 
can be obtained by applying the steepest descent
method to the path integral. Saddle points are solutions 
$q_{\rm c}(t)$ to the Euclidean equations of motion (Euclidean equations 
differ from the normal equations of classical mechanics by 
the sign in front of the potential). When
the potential has a unique minimum, located for example at $q=0$, 
the leading
saddle point is  $q_{\rm c}(t)=0$. A systematic expansion around the saddle
point then leads to a purely perturbative expansion of the 
eigenvalues of the Hamiltonian of the form 
\begin{equation}
E(g)=\sum_{l=0}^{\infty} E_l \, g^l \,.
\end{equation}
In the case of a potential with degenerate minima, one must sum over
several saddle points: to each saddle point corresponds an eigenvalue
and thus several eigenvalues are degenerate at leading
order. Because the potential (\ref{ehamdw})~is symmetric, 
the lowest eigenvalue, and
more generally all eigenvalues which remain finite  when $g$ goes to zero,
are twice degenerate to all orders in perturbation theory
[see equation (\ref{esim})]:
\begin{equation} 
E_{\pm,N}(g) \approx 
E^{(0)}_N(g)\equiv\sum_{l=0}^{\infty} E^{(0)}_{N,l} \, g^l\,. 
\end{equation}

%
%
\subsection{Instanton Configurations}

\begin{figure}[htb!]
\begin{center}
\begin{minipage}{12.0cm}
\begin{center}
\epsfxsize=84.mm
\epsfysize=29.mm
\centerline{\epsfbox{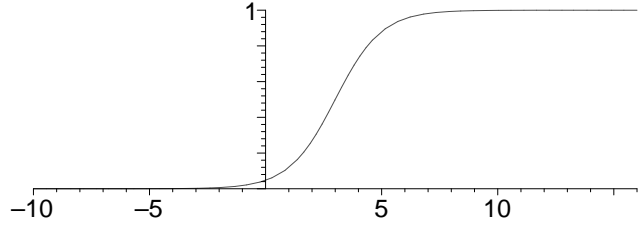}}
\caption{\label{figinsti}
The instanton configuration.} 
\end{center}
\end{minipage}
\end{center}
\end{figure}

Eigenvalues which remain finite when $g$ goes to zero can be extracted 
from the large $\beta $ expansion.
In the infinite $\beta$ limit, leading contributions to the 
path integral come from paths which are solutions of the Euclidean equations 
of motion that have a finite action. 
In the case of the path integral representation of 
${\mathcal Z}_{\rm a}(\beta)$, constant solutions of the equation of 
motion  do not
satisfy the boundary conditions. Finite action solutions 
necessarily correspond to paths which connect the 
two minima of the potential  (see figure~\ref{figinsti}). 

In the example of the double-well potential 
(\ref{ehamdw}),  such solutions are
\begin{equation}
q_{\rm c}(t)=\left(1+\e^{\pm (t-t_0)}\right)^{-1}, 
\qquad {\mathcal S}(q_{\rm c})=1/6\,. 
\end{equation} 
Since the two solutions depend on an integration constant $t_0$,
one finds two one-parameter families of degenerate saddle points.

Non-constant (in time) solutions with finite action are called 
{\em instanton}\/ solutions. 
Since the main contribution to the action comes from the region 
around $t=t_0$, one calls $t_0$ the {\em position}\/ of the instanton.

%
%

The corresponding contribution to the path integral is proportional, 
at leading order in $g$ and for $\beta\to\infty $, to $ \e^{-1/(6g)}$ 
and thus is non-perturbative. It is also proportional to $\beta$ 
because one has to sum over all degenerate saddle points  and the 
integration constant $t_0$ varies in $[0,\beta]$ for $\beta$ large 
but finite. The complete calculation involves taking the time $t_0$ 
as a collective coordinate, and integrating over the remaining 
fluctuations in the Gaussian limit. One finds that the two
lowest eigenvalues are  given by ($\varepsilon =\pm$)

\begin{eqnarray}
E_{ \varepsilon,0}(g) &=& \mathop{\lim}_{ \beta\to \infty}
-{1\over \beta }
\ln {\mathcal Z}_{\varepsilon}(\beta)\mathop{=}_{g\to0,\beta\to \infty}
E_0^{(0)}(g)-\varepsilon  E_0^{(1)}(g), 
\nonumber\\
E_0^{(1)}(g) &=&
{1\over \sqrt{ \pi g}}\e^{-1/6g} \bigl(1 + {\mathcal O}(g) \bigr).
\end{eqnarray} 

%
%
\section{Multi--Instantons}
\label{ssninst}

Taking into account  
$E^{(0)}(g)$ and $E^{(1)}(g)$, one obtains  for  
the functions (\ref{einstZpm}) an expansion of the form
\begin{equation} 
{\mathcal Z}_ \varepsilon (\beta) \mathop{\sim}_{\beta \to\infty } 
\e^{-\beta(  E_0^{(0)} -\varepsilon  E_0^{(1)}) }\sim  \e^{-\beta E_0^{(0)}}
\sum^{\infty}_{n=0}{ \left( 
\varepsilon\beta \right)^{n} \over n!}\,
\left(E^{(1)}_0\right)^{n}.
\label{epartit} 
\end{equation}
Thus, the existence of a one-instanton contribution to eigenvalues 
implies the existence of {$ n$-instanton} contributions to the  
functions (\ref{einstZpm}), proportional to $ \beta^{n}$. 

For $\beta$ finite, the path integrals indeed have  
other saddle points which correspond to oscillations in 
the well of the potential $-V(q)$. In the
infinite $\beta$ limit, the solutions with $n$ oscillations  have an
action $n \times 1/6$ and thus give contributions to the path 
integral of the expected form. 

However, there is a subtlety: naively one would expect 
these configurations to give a contribution of order $\beta$  
(for $\beta$ large) because a given classical trajectory depends 
only on one time integration constant. This has to be contrasted 
with the expansion
(\ref{epartit}) where the $n^{\rm th}$ term is of order $\beta^n$.

Indeed, one discovers that the Gaussian integration near the saddle 
point involves the determinant of an operator that has eigenvalues which 
vanish exponentially in the large-$\beta$ limit. The divergence has the 
following origin: in the large-$\beta$ limit, the classical solution 
decomposes into a succession of largely separated instantons and fluctuations 
which tend to change the distances between instantons induce an 
infinitesimal variation of the action. 
It follows that, to properly study the limit, one has to introduce 
additional {\em collective coordinates}\/ that parameterize all 
configurations close to solutions of the Euclidean equation of motion, 
even though they have a sightly different action. 
It is then easy to understand where in the expansion 
(\ref{epartit}) the factor $ \beta^{n} $ 
originates from:
Although a given classical trajectory can only generate a factor
$ \beta$, these new configurations depend on $ n $ independent 
collective coordinates over which one has to integrate.  

To summarize: we know that {$ n$-instanton} contributions do exist.
However, these contributions do not correspond, in general,  
to solutions of the classical equation of motion. They correspond to 
configurations of largely
separated instantons connected in a way which we shall discuss, which become
solutions of the equation of motion only asymptotically, in the limit of
infinite separation. These configurations depend on $ n $ times more
collective coordinates than the one-instanton configuration.

%
%
\section{Specific Calculations}
\label{sSpecific}

%
%
\subsection{Instanton Interaction}
\label{ssBSminst}

We now briefly explain, still with the example of the double-well 
potential, how multi-instanton contributions to the path integral 
can be evaluated at leading order for $g\to0$. 

In the infinite $\beta$ limit, the instanton solutions can be 
written as
\begin{subequations}
\label{efundef}
\begin{eqnarray}
q_{\pm}(t) &=& f\bigl(\mp(t-t_{0})\bigr)\,,
\\
f(t) &=& 1/\left(1+ \e^{t}\right)=1-f(-t),
\end{eqnarray}
\end{subequations}
where the integration constant $t_0$ characterizes the instanton position.

We first construct the {\em two-instanton
configuration}. Actually it is convenient to now call instanton a solution 
which goes from 0 to 1 and anti-instanton a solution which goes from 1 to 0. 
Then, the relevant configurations are
instanton-anti-instanton pairs. These configurations depend on 
one additional time 
parameter, the separation between instantons, decompose in the 
limit of infinite separation into two instantons and for
large separation must minimize the variation 
of the action \cite{BrPaZJ1977,Bo1980}. For this
purpose, we could introduce a constraint in the path integral 
fixing the separation between instantons (see chapter \ref{ssninstco}), 
and solve the equation of motion with a Lagrange multiplier for the 
constraint. Instead, we use a method which, at least at
leading order, is simpler and more intuitive. 

%
%
\begin{figure}[htb!]
\begin{center}
\begin{minipage}{12.0cm}
\begin{center}
\epsfxsize=120.mm
\epsfysize=30.mm
\centerline{\epsfbox{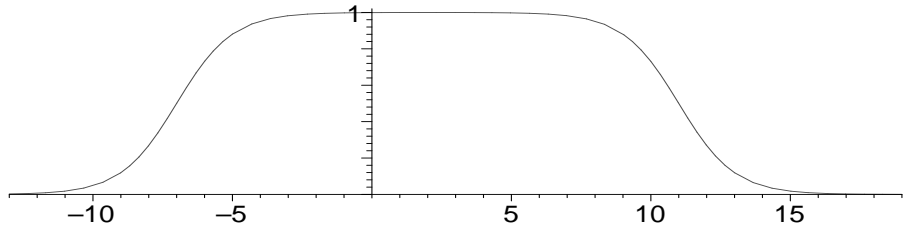}}
\caption{\label{figinstii}
The two-instanton configuration.}
\end{center}
\end{minipage}
\end{center}
\end{figure}

We consider a configuration $q_{\rm c}(t)$ that is the sum of instantons
separated 
by a distance $\theta$, up to an additive constant adjusted in such a way as
to satisfy the boundary conditions (figure \ref{figinstii}):
\begin{equation}
q_{\rm c} (t)= f(t-\theta/2) + f(-t-\theta/2) - 1 =
f(t-\theta/2) - f(t+\theta/2) \,,
\label{etwoinst}
\end{equation}
where $f(t)$ is the function 
(\ref{efundef}) (and of course all configurations
deduced  by time translation). This path has the following
properties: it is continuous and differentiable
and when $\theta$ is large it differs, 
near each instanton, from the instanton solution only by 
exponentially small terms of order $\e^{-\theta}$.
Although the calculation of the corresponding action is straightforward, we
give here some details to show that the ansatz 
(\ref{etwoinst}) applies to more general symmetric potentials.

It is convenient to introduce some additional notation: 
\begin{eqnarray} 
u(t) & = & f(t - \theta/2)\,,
\nonumber\\
v(t) & = & u(t + \theta)\,,
\end{eqnarray}
and thus $q_c = u - v$.
The action corresponding to the path 
(\ref{etwoinst}) can be written as
\begin{eqnarray}
{\mathcal S}(q_c) & = & 
\int \d t \, \left[\ud \, {\dot q}_c^2 + V(q_c) \right] 
\nonumber\\
& = & 
2\times {1 \over 6} + 
\int \d t \, \left[- {\dot u}\,{\dot v}
+ V(u - v) - V(u) - V(v) \right]\,.
\end{eqnarray}
The parity of $q_c$ allows us to restrict the integration  to the region
$t>0$, where $v$ is at least of order $\e^{-\theta /2}$. 
After an integration by parts of the term
$ \dot v \dot u $, one finds
\begin{equation} 
{\mathcal S}(q_c) =
{1 \over 3} + 2 \left\{ v(0) \, {\dot u}(0)
+ \int_0^{+ \infty} \d t\,
\left[ v \, {\ddot u} + V(u - v) - V(u) - V(v) \right] \right\} .
\end{equation}
One then expands the integrand in powers of $v$.  
Since the leading correction to $\mathcal S$ is of order $\e^{-\theta}$, 
one needs the expansion only up to order $v^2$. 
The term linear in $v$ vanishes as a consequence of the 
$u$-equation of motion. One obtains
\begin{equation}
{\mathcal S}(q_c) - {1 \over 3} \sim 
2 \, v(0) \, {\dot u}(0) +
2 \left\{ \int_0^{+ \infty}  
\d t\, \left[ \ud \, v^2 \, V''(u) - \ud \, V''(0) \, v^2 \right] 
\right\}.
\end{equation}
The function $v$ decreases exponentially away from the origin so the main
contributions to the integral come from the neighbourhood of $t=0$, where
$u = 1 + {\mathcal O}(\e^{-\theta /2}$ and thus
$V''(u) \sim V''(1) = V''(0)$. Therefore, at leading order the two 
terms in the integral  cancel. At leading order, 
\begin{equation}
v(0) \, {\dot u(0)} \sim - \e^{-\theta} 
\end{equation}
and thus
\begin{equation} 
{\mathcal S}(q_c) = \frac{ 1}{ 3} - 2\e^{-\theta} +
{\mathcal O}\left( \e^{-2\theta} \right)\,. 
\label{eSclass}
\end{equation}
It will become clearer later why  the classical action is  
needed only up to order $ \e^{-\theta}$. In analogy with the partition 
function of a classical gas (instantons being identified with particles), 
one calls 
the quantity $ -2\e^{-\theta}$ the
 interaction potential between instantons.

Actually, it is simple to extend the result to  $ \beta $ 
large but finite. Symmetry
between $ \theta $ and $ \beta -\theta $ then implies
\begin{equation} 
{\mathcal S}(q_c) = \frac{1}{3} -
2 \e^{-\theta}-2 \e^{-(\beta -\theta)} 
+\ \hbox{negligible contributions.} 
\label{eScqt}
\end{equation} 
This expression can be verified by calculating its extremum  
as a function of $\theta $. One obtains 
\begin{equation}
\theta_{c} = \beta /2  ,\ \Rightarrow\ 
{\mathcal S}(q_c) = \frac{1}{3} - 4 \e^{-\beta /2} +
{\mathcal O}\left( \e^{-\beta} \right)\,. 
\end{equation} 
For the same Hamiltonian, one can calculate for 
$ \beta $ finite the action corresponding to a solution of the 
equation of motion with one oscillation. One finds
%
\begin{equation}  
{\mathcal S}(q_c) =\frac{1}{6} - 2 \, \e^{-\beta} +
{\mathcal O} \left( \e^{-2\beta} \right) \,.
\label{eScldwb}
\end{equation}
Both results are consistent. Indeed, to compare them one has to 
replace $\beta$ by $\beta/2$ in equation 
(\ref{eScldwb}) and multiply the action by a
factor 2, since the action corresponds to a trajectory described twice in the
total time $ \beta $.

We now examine the {\em variation of the action} due to the 
multi-instanton configuration.
Specifically, we show that if we modify 
infinitesimally (for $\theta$ large) the
configuration to further decrease the variation of the action, the change
$r(t)$ of the path is of order $\e^{-\theta}$ and the variation of 
the action of order $\e^{-2\theta}$ at least. Setting
\begin{equation} 
q (t) = q_{c} (t) + r(t) 
\end{equation}
and expanding the action up to second order in $r(t)$, one finds 
\begin{eqnarray} 
{\mathcal S}(q_c + r) &=& {\mathcal S}(q_{c}) + \int
\left[\dot q_{c} (t) \, {\dot r}(t) + 
V'\bigl(q_{c} (t) \bigr) \, r(t) \right] \, \d t
\nonumber\\
& & +{1 \over 2} \int \d  t \left[ \dot r^2(t) + V'' (q_{c} ) \, r^2(t)
\right] + {\mathcal O} \left(\left[ r(t) \right]^3 \right)\,.
\label{eactsord}
\end{eqnarray}  
In the term linear in $r (t)$, one integrates by parts $ \dot r(t)$, 
in order to
use the property that $ q_{c} (t) $ approximately  satisfies the equation of
motion. In the term quadratic in $ r (t)$, one replaces $
V'' $ by 1, since $r (t) $ is expected  to
be large only far from the instantons. One then verifies that 
the term linear in
$r$ is of order $\e^{-\theta}$ while  the quadratic term is of order 1.
A shift of $r$ to eliminate the linear term would then give 
a negligible contribution, of
order $\e^{-2\theta}$. 

We now consider 
an {\em $n$-instanton configuration}, i.e.~a succession of $ n $
instantons (more precisely, alternatively instantons and anti-instantons) 
separated by times $ \theta_{i}$ with
\begin{equation}
{\textstyle \sum^{n}_{i=1}}\theta_{i}=\beta . 
\end{equation}
At leading order, we need only consider ``interactions'' between 
nearest neighbour instantons. Other interactions are negligible 
because they are of higher order in $ \e^{-\theta}$. 
This is an essential simplifying feature of quantum
mechanics compared to quantum field theory. 
The classical action 
$ {\mathcal S}_{\rm c} (\theta_{i} ) $ can then be 
directly inferred from 
expression (\ref{eScqt}):  
\begin{equation}
{\mathcal S}_{\rm c} (\theta_{i} )={ n \over 6}-2
\sum^{n}_{i=1} \e^{-\theta_{i}} +
{\mathcal O} \left( \e^{-(\theta_{i}+\theta_{j})} \right)\,. 
\end{equation} 
Note that for $ n $ even, the {$ n$-instanton} configurations
contribute  to $ \tr\e^{-\beta H} $, while for $ n $ odd they
contribute to $ \tr\left(P \e^{-\beta H}\right) $ ($P$ is the parity
operator). But all contributes to the combination (\ref{einstZpm}).

({\em Remark.}) 
Since we keep in the action all terms of order $\e^{-\beta}$, we
expect to find  the contributions not only to the two lowest energies but
also to all energies which remain finite when $g$ goes to zero.

%
%
\subsection{The $n$--Instanton Contribution}
\label{ssInstantonContribution}

We have calculated the $ n $-instanton action. We now evaluate, at leading
order, the contribution to the path integral of the neighbourhood of the  
$n$-instanton configuration \cite{ZJ1981npblong,ZJ1996ch43}. 
We expand the action up to second order in the deviation 
from the classical path. Although the path is not a
solution of the equation  of motion, it has been chosen it in 
such a way that  the linear terms in the expansion can be  neglected. The
Gaussian integration involves then the determinant of the second
derivative of the action at the classical path
\begin{equation} 
M (t', t) = 
\left[- \left({ \d  \over \d t}\right)^2
+ V''\bigl(q_{\rm c} (t)\bigr) \right] \,
\delta (t-t' ).
\label{eMdw} 
\end{equation} 
The operator $M$ has the form of a Hamiltonian with a potential that
consists of $ n $ wells asymptotically identical to the well arising in the
one-instanton problem, and which 
are largely separated. At leading order the corresponding spectrum is,
therefore, the spectrum arising in the one-instanton problem {$ n$-times}
degenerate. Corrections are exponentially small in the separation. 
Simultaneously, by introducing $n$ collective time variables, 
we have suppressed $n$ times the zero eigenvalue and generated 
the Jacobian of the one-instanton case to the power $n$. 
Therefore, the {$n$-instanton} contribution to the combination
(\ref{einstZpm})
\begin{equation} 
{\mathcal Z}_{\varepsilon}(\beta )=
\ud \, \tr\left[\left(1+\varepsilon P \right)
\e^{-\beta H}\right],  
\end{equation}  
($ \varepsilon =\pm1 $),
can be written as
\begin{equation}
{\mathcal Z}_{\varepsilon}^{(n)}(\beta ) = 
\e^{-\beta /2} \, {\beta\over n} \, \left(\varepsilon { \e^{-1/6g}
\over \sqrt{\pi g}} \right)^{n}\int_{\theta_{i}\geq 0}\delta \left( \sum
\theta_{i}-\beta \right) \prod_{i}\d\theta_{i} \exp\left[{2 \over g}
\sum^{n}_{i=1} \e^{-\theta_{i}}\right]. 
\label{eninstant} 
\end{equation}
All factors have already been explained, except the factor $ \beta $, 
which comes from the integration over a global time translation, and 
the factor $1/n$, which arises because the configuration is invariant 
under a cyclic permutation of the $ \theta_{i}$. Finally, the 
normalization factor $\e^{-\beta /2} $ corresponds to the partition 
function of the harmonic oscillator. Odd-$n$ instanton effects contribute
positively to ${\mathcal Z}_+^{(n)}(\beta)$, and negatively
to ${\mathcal Z}_-^{(n)}(\beta)$.

If the instanton interactions are neglected (but instanton configuration
with an arbitrarily large number of tunnelings are allowed), 
the integration  over the $\theta_i$'s is straightforward and the sum  
of the leading order {$ n $-instanton} contributions 
\begin{equation} 
{\mathcal Z}_\varepsilon  (\beta, g) \approx
\Sigma_\varepsilon  (\beta, g) = \e^{-\beta/2}+\sum^{\infty}_{n=1}
{\mathcal Z}^{(n)}_{\varepsilon}(\beta,g)
\label{esumSig} 
\end{equation}  
can be calculated:
\begin{equation} 
{\mathcal Z}_\varepsilon  (\beta, g) \approx
\Sigma_\varepsilon  (\beta ,g )= \e^{-\beta /2} \left[1+{\beta  \over n}\sum
^{\infty}_{n=1}\left(\varepsilon { \e^{-1/6g}
\over \sqrt{\pi g}} \right)^{n}{\beta^{n-1} \over \left(n-1
\right)!} \right] = \e^{-\beta  E_{\varepsilon,0}(g)  }
\end{equation}
with
\begin{equation} 
E_{\varepsilon,0}(g)=
{1 \over 2} +
{\mathcal O}\left(g\right) -
{\varepsilon \over \sqrt{ \pi g}} \e^{-1/6g} 
\left(1 + {\mathcal O}\left(g\right) \right).
\end{equation}
We recognize the perturbative and one-instanton contribution, at leading
order, to $E_{\varepsilon,0}(g)$, the ground state and the first excited state
energies.

In view of problematic issues related to 
the instanton interaction, the summation of these effects and 
necessary analytic continuations which have to be performed 
in order to calculate eigenvalues, we add here
a small discussion. To go beyond the one-instanton approximation, 
it is necessary to take into account the interaction between instantons
(see chapters~\ref{ssBSminst} and~\ref{ssInstantonContribution}). 
Unfortunately, if one examines expression (\ref{eninstant}), 
one discovers that the interaction between
instantons is {\em attractive}. Therefore, for $ g $ small, the 
dominant contributions to
the integral come from configurations in which the instantons are
close. For such configurations, the concept of instanton 
is no longer meaningful,
since the configurations cannot be distinguished from fluctuations
around the constant or the one-instanton solution. 

We should have expected such a difficulty. Indeed, the large-order
behaviour analysis has shown that the perturbative expansion in the case
of potentials with degenerate minima is not Borel summable. An ambiguity is
expected at the two-instanton order. But if the perturbative expansion is
ambiguous at the two-instanton order, contributions of the same order or even
smaller are ill-defined. To proceed any 
further, we must first give a meaning to the sum of the 
perturbative expansion. 

In the example of the double-well potential, it is possible to show that the 
perturbation series is Borel summable for $ g $ negative, by 
relating it to the perturbative expansion 
of the ${\mathcal O}(2)$ anharmonic oscillator.
Therefore, we {\em define}\/  the sum of the perturbation series as the
analytic continuation of this Borel sum from $g$ negative to 
$g =\left| g \right| \pm i0$. This corresponds in the Borel 
transformation to an integration above or below the real positive axis.
We then note  that, for $ g $ {\em negative}, the interaction between
instantons is {\em repulsive} and, simultaneously, the expression 
(\ref{eninstant}) becomes meaningful.
Therefore, we first calculate, for $ g $ small and negative, both the sum of
the perturbation series and the instanton contributions, and 
perform an analytic continuation to $ g $ positive of all quantities
consistently. In the same way, the perturbative expansion around each
multi-instanton configuration is also non-Borel summable 
and the sum is defined by the same procedure. 

At the same time, it should be remembered that the various
analytic continuations and cancellations of imaginary parts,
while illustrating the internal consistency of the approach,
are not really necessary in order to obtain the energy
eigenvalue via generalized resummation of the resurgent
expansion: it would suffice to say that the energy eigenvalue
is obtained by assigning to every divergent series in $g$,
i.e.~to every perturbative expansion about the $n$-instanton
contributions, the real part of the Borel sum associated with
that (factorially divergent) series. The real part of the Borel sum
is also obtained by evaluating the Laplace--Borel integral
using the principal-value prescription.
Although intuitively clear, we would like to remark that this
principal-value prescription
---of course--- corresponds to the integration contour $C_0$
as outlined in~\cite{Je2000prd}.

While, on the one hand,
the necessity for an analytic continuation may seem a little obscure
at first sight (see for example remarks on p.~241 of\cite{Sh1994}),
this procedure is, on the other hand, a mathematically well defined
concept. In the context of the double-well problem
it gives a defined meaning to each of the factorially divergent,
nonalternating power series occurring in equation (\ref{ecomexp}).
In the end, given the fact that even mathematical
proof~\cite{DeDiPh1990,DeDi1991} may be questioned by inquisitive 
minds, the soundness of the concept must be judged by its predictive
power. {\em Exempli gratia}, one may ask the question
whether the energy levels predicted by the
analytically continued instanton expansion are able to accurately
reproduce numerically determined levels. This is indeed the case,
as shown in chapter~\ref{Explicit} below.

%
%
\subsection{The Sum of the Leading--Order Instanton Contributions}

Let us now introduce two useful parameters [cf.~equations 
(\ref{defxi}) and (\ref{defchi})]
\begin{equation}
\label{einstmulampar}
\mu \equiv -{2 \over g} = \exp\left( \chi(g) \right)\,,\quad 
\lambda \equiv {1 \over \sqrt{2 \pi}} \, \e^{-1/6g} , 
\end{equation}
in such a way that the ``fugacity'' $\xi(g)$ of the instanton gas, 
which is half  the one-instanton contribution at leading order,
can be written as [see (\ref{defxi})]
\begin{equation}
\xi(g) = {1 \over \sqrt{\pi g}} \, \e^{-1/6g} = 
\lambda \, \sqrt{-\mu} = -{\rm i}\,\lambda \, \sqrt{\mu}\,, 
\end{equation}
where the quantity $\xi(g)$ also occurs in a form 
multiplied by the parity $\varepsilon$,
and a sign change in the determination of $\sqrt{-\mu}$  
is therefore equivalent to a change $\varepsilon\mapsto -\varepsilon$
[see also equation (\ref{deltaepsilon}) below].

As discussed before, we now assume that initially $g$ is negative.
The Laplace transform 
\begin{equation}
\label{estart}
G^{(n)}_\varepsilon (E) =
\int_0^\infty \d\beta \,\e^{\beta E} {\mathcal Z}_{\varepsilon}^{(n)}(\beta )
\end{equation}
of the $n$-instanton contribution (\ref{eninstant}):
\begin{equation} 
{\mathcal Z}_{\varepsilon}^{(n)}(\beta )\sim
\frac{\beta \, \e^{-\beta/2}\,
\left(-{\rm i} \, \varepsilon \lambda \sqrt{\mu}\right)^{n}}{n}\,
\int_{\theta_{i}\geq 0}\delta \left( \sum \theta_{i}-\beta \right)
\prod^{n}_{i=1} \d \theta_{i} \exp\left[{2 \over g} \sum^{n}_{i=1}
\e^{-\theta_{i}}\right],
\label{eninstanb}
\end{equation} 
yields the leading contribution $G^{(n)}_\varepsilon (E)$ to 
the trace $G_\varepsilon (E)$ of the resolvent  
[equation (\ref{eninstGZpm})].
In order to factorize the integral over the $\theta_i$,
we introduce a complex
contour integral representation for the $\delta$-function,
\begin{equation}
\delta\left( \sum_{i=1}^n \theta_i - \beta \right) = 
\frac{1}{2 \pi {\rm i}}\, 
\int\limits_{-{\rm i} \infty}^{{\rm i} \, \infty}
{\rm d}s \exp\left[ -s \left( \beta - \sum_{i=1}^n \theta_i \right)\right]\,.
\end{equation}
In terms of the function
\begin{equation}
{\mathcal I}(s,\mu ) =
\sqrt{\mu} \int^{+\infty}_{0} 
\exp\left(s\,\theta-\mu \, \e^{-\theta}\right) 
\, \d \theta\, , 
\label{einstIsmu}
\end{equation}
${\mathcal Z}_{\varepsilon}^{(n)}(\beta )$ can be rewritten as
\begin{equation} 
{\mathcal Z}_{\varepsilon}^{(n)}(\beta )\sim
\frac{\beta \, \e^{-\beta/2}\,
\left(-{\rm i}\, \varepsilon \lambda \right)^{n}}
  {2 \pi {\rm i} n}\,
\int\limits_{-{\rm i} \infty}^{{\rm i} \infty}
{\rm d}s \,
\e^{-\beta \, s}\,
\left[{\mathcal I}(s,\mu ) \right]^n\,.
\end{equation}
In view of (\ref{estart}), we have
\begin{eqnarray}
G^{(n)}_\varepsilon (E) &=&
\int_0^\infty \d\beta \,\e^{\beta E} {\mathcal Z}_{\varepsilon}^{(n)}(\beta )
\nonumber\\
&=& \int_0^\infty \d\beta \,
\frac{\beta \, \e^{\beta (E - 1/2)}\,
\left(-{\rm i}\, \varepsilon \lambda \right)^{n}}
  {2 \pi {\rm i} n}\,
\int\limits_{-{\rm i} \infty}^{{\rm i} \infty}
{\rm d}s \,
\e^{-\beta \, s}\,
\left[{\mathcal I}(s,\mu ) \right]^n
\nonumber\\
&=& 
\frac{\partial}{\partial E} \,
\int\limits_{-{\rm i} \infty}^{{\rm i} \infty}
{\rm d}s \,
\frac{\left(-{\rm i}\, \varepsilon \lambda \right)^{n}}
  {2 \pi {\rm i} n}\,
\left[{\mathcal I}(s,\mu ) \right]^n \,
\int_0^\infty \d\beta \,
\e^{\beta \left(E -s - 1/2\right)}
\nonumber\\
&=& 
\frac{\partial}{\partial E} \,
\int\limits_{-{\rm i} \infty}^{{\rm i} \infty}
{\rm d}s \,
\frac{\left(-{\rm i}\, \varepsilon \lambda \right)^{n}}
  {2 \pi {\rm i} n}\,
\left[{\mathcal I}(s,\mu ) \right]^n \,
\frac{1}{s + 1/2 - E}
\nonumber\\
&=& 
\frac{\partial}{\partial E} \,
\frac{\left(-{\rm i}\, \varepsilon \lambda \right)^{n}}{n}\,
\left[{\mathcal I}(E - \ud,\mu ) \right]^n \,.
\end{eqnarray}
Alternatively, one may observe that in (\ref{estart}),
the integral over $\beta $ is immediate and the integrals over the
$\theta_i$ then factorize. One obtains  
\begin{equation}
\label{eOneObtains}
G^{(n)}_\varepsilon (E) \sim 
{(-{\rm i}\, \varepsilon \lambda)^n \over n}\,
\frac{\partial}{\partial E} \, 
\left[{\mathcal I}(E-\ud,\mu)\right]^n\,.
\end{equation}

The function $ {\mathcal I}(s, \mu) $ is the only function 
needed for these leading-order calculations in one dimension. 
A first way to calculate it, is to expand
\begin{equation} 
\label{ederivation1}
{\mathcal I}(s,\mu )  =  
\int^{+\infty}_{0} \sum_{N=0}^\infty 
{(-1)^N \over N!} \mu^{N+1/2}\,\e^{(s-N)\theta} \d \theta =
\sum_{N=0}^\infty \frac{(-1)^N}{N!} \, \frac{\mu^{N+1/2}}{N-s}\,. 
\end{equation}
The expansion defines a real meromorphic function
of $s$ with poles at non-negative integers.  
The poles and residues can be trusted  because they depend only on 
the large $\theta $ behaviour.
However, after analytic continuation
to $g>0$, the function  grows as $\e^{-\mu}=\e^{2/g}$, 
a behaviour due to the integration near $\theta =0$ that 
cannot be trusted and cannot be correct. 

Let us now evaluate the integral (\ref{einstIsmu})  
in the limit $\mu \to +\infty $, and thus $g\to 0_{-}$. 
We change variables, setting $ \mu \, \e^{-\theta}=t  $, 
and the integral becomes 
\begin{equation} 
\label{ederivation2}
{\mathcal I} (s,\mu )=
\mu^{s+1/2} \int^{\mu}_{0}  \d t \, t^{-1-s} \, \e^{-t}=
\mu^{s+1/2}\int^{+\infty}_{0} \d t \, t^{-1-s} \, \e^{-t}  +
{\mathcal O}\left(\e^{-\mu}/\sqrt{\mu}\right).
\end{equation}
We thus obtain ($s = E - 1/2$)
\begin{equation}
{\mathcal I}(s,\mu) \approx \mu^{s+1/2}\,\Gamma(-s) =
\left(-\frac{2}{g}\right)^E \,\Gamma\left(\frac12 - E\right)\,,
\label{einstIsmub}
\end{equation}
a meromorphic function with the same poles and residues as the initial
expression.  For $\mu \to+\infty $, the difference is exponentially small
and both expressions are equivalent, but after analytic continuation to $g>0$,
the asymptotic form (\ref{einstIsmub}) has now an acceptable behaviour.
Therefore, our ansatz is that the estimate 
(\ref{einstIsmub}) gives the correct
leading behaviour of the true function.
 
In view of (\ref{eOneObtains}),
the generating function $ {\mathcal G} _\varepsilon (E ,g) $ of the 
leading-order
multi-instanton contributions (\ref{esumSig}) then is given by
[we use $\sum_n (-x)^n/n = - \ln(1+x)$]
\begin{equation}
{\mathcal G}_\varepsilon (E, g) 
= \sum_{n=1}^\infty G^{(n)}_\varepsilon (E)  
= -{\partial \over \partial E}\ln  \Delta_\varepsilon (E)  
\end{equation}
with
\begin{equation} 
\label{deltaepsilon}
\Delta_\varepsilon (E) \; = \;
1+ \varepsilon \, {\rm i}\,\lambda \,\, {\mathcal I}\left(E-\frac12,\mu\right) 
\; \approx \;
1+ \varepsilon \, {\rm i}\,\frac{e^{-1/6g}}{\sqrt{2 \pi}}\,
\left( -\frac{2}{g} \right)^E \, \Gamma\left(\frac12 - E\right)\,,  
\end{equation}
where we now explicitly see that
a change in the determination of $\sqrt{-\mu}$  
is equivalent to a change $\varepsilon\mapsto -\varepsilon$.
Because ${\mathcal G}_\varepsilon (E, g)$ approximates the trace of the 
resolvent,
\begin{equation}
{\mathcal G}_\varepsilon (E, g) \approx
{\rm Tr} \left( \frac{1}{H_\epsilon - E} \right)\,, \qquad
\end{equation} 
we have
\begin{eqnarray}
{\mathcal G}_\varepsilon (E, g) &\approx& 
- \frac{\partial}{\partial E} \, {\rm Tr}[ \ln (H_\epsilon - E) ]
\nonumber\\
&=& - \frac{\partial}{\partial E} \, \ln[ \det (H_\epsilon - E) ]
\nonumber\\
&\approx& - \frac{\partial}{\partial E} \, \ln \Delta_\varepsilon (E)\,. \qquad
\end{eqnarray}
Here, $H_\epsilon$ denotes the Hamiltonian restricted on the 
Hilbert space spanned by the eigenvectors of $H$ with parity $\epsilon$.
Therefore, the functions 
$\Delta_\varepsilon (E) \approx \det (H_\epsilon - E)$ are directly the sums 
of the leading order multi-instanton contributions to the products 
$ {\mathcal D}_\varepsilon (E)$ of even or odd eigenvalues, 
\begin{equation}
{\mathcal D}_\varepsilon (E) \propto
\prod_N \left(1- \frac{E}{E_{\varepsilon, N}}\right)\,, 
\end{equation}
but, unlike  ${\mathcal D}_\varepsilon (E)$, they are meromorphic functions 
in $E$ because the zero-instanton contribution has not yet been 
included at all; an expansion of (\ref{deltaepsilon}) in powers
of $\lambda$ starts with a term of order ${\mathcal O}(\lambda)$. 
Correcting for this effect,
we add to ${\mathcal G}_\varepsilon (E, g)$ the trace of the resolvent 
of the harmonic oscillator, and we thus divide $\Delta_\varepsilon (E)$
by $\Gamma(\ud - E)$. This amounts to the replacement 
\begin{eqnarray}
{\mathcal G}_\varepsilon (E, g) &\to& 
{\mathcal G}_\varepsilon (E, g) + G_{\rm osc.} (E) 
\nonumber\\
&\approx& 
- \frac{\partial}{\partial E} \, \ln \Delta_\varepsilon (E) 
+ \frac{\partial}{\partial E} \ln \Gamma\left(\frac12 - E\right)
\nonumber\\
&\approx& 
- \frac{\partial}{\partial E} \, \ln \frac{\Delta_\varepsilon (E)}
{\Gamma\left(\frac12 - E\right)}\,.
\end{eqnarray}

This cancels the poles, and $\Delta_\varepsilon (E)$ 
becomes an entire function:
\begin{equation} 
\Delta_\varepsilon (E)={ 1 \over\Gamma (\ud -E)}+ 
\varepsilon {\rm i} \,
\left(-{2\over g}\right)^E  {\e^{-1/6g}\over \sqrt{2\pi}} .
\label{epoles}
\end{equation}
The first term is simply the Fredholm determinant (\ref{einsthoscdet})
corresponding to the harmonic oscillator. We note here that the 
sum of all instanton contributions, in the leading approximation,
i.e.~neglecting subleading terms in the instanton interaction,
simply yields some sort of one-instanton (but $E$-dependent) correction 
to the spectral equation.

Since $ \lambda $ is small,   zeros  of the equation 
$\Delta_\varepsilon (E)=0$
are close to eigenvalues of the harmonic oscillator:
\begin{equation} 
E_{\varepsilon ,N} = N + \frac12 + {\mathcal O}(\lambda)\,,
\qquad N\geq 0\,. 
\end{equation}
The zeros of the function (\ref{epoles}) can 
then be expanded in a power series in $ \lambda $:
\begin{equation} 
E_{\varepsilon ,N}(g)= \sum_n E^{(n)}_{N} (g) \, (-\varepsilon \lambda)^n. 
\end{equation} 
One obtains from a unique equation the multi-instanton contributions
to all energy eigenvalues $E_{\varepsilon ,N}(g) $ of the double-well
potential at leading order in $g$, that is to say
all the coefficients $e_{N,nk0}$ in the notation
of the equations (\ref{ecomexp}) and (\ref{eInstn}). 
Concrete results for the double-well
potential are presented in chapter~\ref{ssLeading}.

The appearance of a factor $ \ln g $ in equation 
(\ref{einstEnn})~can now be simply
understood by noting that the interaction terms are only relevant for 
$g^{-1}\,\e^{-\theta}$ of order 1, that is $ \theta $ of order $ - \ln g $.

Assuming that an equation of the form 
(\ref{epoles}) holds beyond leading order, we conclude that the argument
of the $\Gamma$-function must be such that the equation reproduces at
zero instanton order the perturbative expansion: 
$E$ has to be replaced by the function $B(E,g)$. 
If we further assume that for $B$ large there is some
connection with the WKB expansion, we infer that $(-2/g)^E$ must also
be replaced by $(-2/g)^B$ to reconstruct a term of the form $B\ln(Eg)$.
This is the origin of the conjecture (\ref{equantization}).

This concludes the first part of the treatise of multi-instanton
effects in quantum mechanics. In the second part~\cite{ZJJe2004II},
we intend to generalize the instanton calculations to 
a wider family of potentials.

\appendix

\renewcommand{\chaptermark}[1]%
{\markboth{\MakeUppercase{APPENDIX~\thechapter: #1\\[2ex]}}{}} 

%
%
\chapter{Schr\"odinger and Riccati Equations: Some Useful Results}
\label{appSchrRic}

%
%
\section{Inverse Schr\"odinger Operator: Matrix Elements}

In this appendix,
we recall here the derivation of a few classical results that we have used
throughout this article.
We first consider the hermitian positive differential operator
\begin{equation}
L=-\d_x^2+u(x),
\end{equation}
where $u(x)$ has the form of a potential, which remains strictly positive
for $|x|\to\infty $.

We recall the derivation of the differential
equation satisfied by the diagonal matrix elements of its 
inverse $R=L^{-1}$.\par
Matrix elements $R(x,y)$ of $R$ satisfy the Schr\"odinger equation 
\begin{equation}
\bigl(-\d_x^2+u(x)\bigr) R(x,y) = \delta(x-y)\,. 
\label{eresol}
\end{equation}
We recall that $R(x,y)$ can be expressed in terms of two independent
solutions of the homogeneous equation
\begin{equation}
\bigl(-\d_x^2+u(x)\bigr) \varphi(x)=0\ .
\label{eSchrhom}
\end{equation}
We denote by $\varphi_1$, $\varphi_2$ two solutions, 
partially normalized by
\begin{equation}
\varphi'_{1}\,\varphi_2 - \varphi_1\,\varphi'_2 = 1\ ,
\label{einstwrons} 
\end{equation}
and, moreover, satisfying the boundary conditions
\begin{equation}
\varphi_1(x)\to 0\ {\rm for}\ x\to -\infty,\quad
\varphi_2(x)\to 0\ {\rm for}\ x\to +\infty \,.
\end{equation}
Then, it is easily verified that $R(x,y)$ is given by
\begin{equation}
R(x,y)=
\varphi_1(y)\,\varphi_2(x)\,\theta(x-y)+
\varphi_1(x)\,\varphi_2(y)\,\theta(y-x)\ ,
\label{einstRxy}
\end{equation}
where $\theta (x)$ is the usual Heaviside step function.
The diagonal matrix elements
\begin{equation}
r(x) \equiv R(x,x) = \varphi_1(x)\,\varphi_2(x) 
\label{eresolvd}
\end{equation}
satisfy
\begin{subequations}
\begin{eqnarray}
r'(x) &=& \varphi'_1(x) \, \varphi_2(x) + \varphi_1(x) \, \varphi'_2(x)\,,
\nonumber\\
r''(x) &=& 2\,\bigl(\varphi'_1(x)\,\varphi'_2(x) + u(x)\,r(x)\bigr)\,,
\end{eqnarray}
\end{subequations}
where we have used the equation (\ref{eSchrhom}).

One then verifies that, as the consequence of equation 
(\ref{einstwrons}), $r(x)$ satisfies the  non-linear differential equation  
\begin{equation}
2r(x) \, r''(x) - r'{}^2(x) - 4 \, u(x) \, r^2(x)+1=0\,. 
\label{eGxxquad}
\end{equation}
A quantity of special interest is the trace $G$ of the $R$:
\begin{equation}
G = \int\d x\, r(x).
\end{equation}

%
%
\section{Riccati's Equation} 

We now set
\begin{equation} 
\sigma (x) = -{\varphi'(x) \over \varphi(x)},
\end{equation}
in equation (\ref{eSchrhom}). We find
\begin{equation}
\sigma'(x) - \sigma^2(x) + u(x) = 0\,.
\end{equation}
To the two solutions $\varphi_{1,2}$ correspond two functions $\sigma _{1,2}$.
We introduce
\begin{equation}
\sigma_+ = \ud\, (\sigma _2-\sigma _1),\quad 
\sigma _-=\ud (\sigma _2+\sigma _1). 
\label{eresolSpm}
\end{equation}
The Wronskian condition (\ref{einstwrons}) implies
\begin{equation}
\label{eresolsigr}
\sigma _+(x) = {1\over 2\,\varphi_1(x)\,\varphi_2(x)} =
{1\over 2r(x)}\,. 
\end{equation}
The equation satisfied by $\sigma _{1,2}$ implies
\begin{subequations}
\label{eRiccatipm}
\begin{eqnarray} 
& & \sigma'_-  -  \sigma _+^2 - \sigma _-^2 + u(x) = 0\,, 
\\
& & \sigma '_+ - 2 \, \sigma _+ \, \sigma _- =0 \,.
\end{eqnarray}
\end{subequations}
The second allows expressing $\sigma_-$ in terms of $\sigma_+$ and,
thus,
\begin{subequations}
\label{eresolphisig}
\begin{eqnarray}
\varphi_2 &=& {1\over\sqrt{2\sigma_+}}
\exp\left[-\int_{x_0}^x \d y\, \sigma _+(y)\right]\,,
\label{eresolphisiga}
\\
\varphi_1 &=& {1\over\sqrt{2\sigma_+}}\,
\exp\left[ \int_{x_0}^x \d y\, \sigma_+(y)\right]\,.
\label{eresolphisigb} 
\end{eqnarray}
\end{subequations}

%
%
\section{Resolvent and Spectrum}

We now substitute $u(x) \mapsto u(x)-z$, set $L=H-z$  and apply these results. 
The operator $R(z)=L^{-1}(z)$ becomes the resolvent of the Hamiltonian $H$.
Riccati's equation becomes
\begin{equation}
\sigma'(x) - \sigma^2(x) + u(x) - z = 0\,.
\end{equation}
Then,
\begin{equation}
{\partial \sigma '\over \partial z}-
2\sigma {\partial \sigma  \over \partial z}=1\,.
\end{equation}
One first useful consequence is the identity
\begin{equation}
{\partial \sigma _+(x)\over \partial z}
+\varphi_1(x)\varphi_2(x)=
{\partial \over \partial x}
\left[{1\over 2\sigma _+(x)}{\partial \sigma _-(x)\over \partial z}\right]. 
\end{equation}
Then,
\begin{equation}
G(z) = \tr L^{-1}(z) =
\int\d x\, r(x) =
-\int \d x\, {\partial \sigma _+(x)\over \partial z} \,.
\label{eresoltrGz}
\end{equation}
Integrating, one infers
\begin{equation}
\ln{\mathcal D}(z) \equiv \ln \det L(z) =
\int \d x\, \sigma _+(x) = {1\over2} 
\left. \ln\left({\varphi_1(x) \over \varphi_2(x)}
\right)\right|^{+\infty }_{-\infty }\,. 
\label{eresolFredz} 
\end{equation}
The function ${\mathcal D}(z)$ is the Fredholm 
determinant of the operator $H-z$:
Assuming a discrete spectrum, we have
\begin{equation} 
{\mathcal D}(z) \propto \prod_{N=0}(1-z/z_N). 
\end{equation}
Note that the infinite product may require some convergence factor, and
the expressions in equation 
(\ref{eresolFredz}), in general, require some large-$x$ regularization.

The spectrum of $H$ is given by the solutions of 
the equation ${\mathcal D}(z)=0$. 
An alternative and useful way of writing  this spectral equation is
\begin{equation} 
\lim_{\varepsilon\to0_+}
\ln{\mathcal D}(z_N-i\varepsilon) - \ln{\mathcal D}(z_N+i\varepsilon) = 
i\pi(2N+1)\,,
\end{equation}
and, therefore,
\begin{equation}
{1\over2i\pi}\lim_{\varepsilon\to0_+} 
\int\d x\left[ \sigma (x,z_N-i\varepsilon) - 
\sigma _+(x,z_N+i\varepsilon) \right] = N+\ud.
\label{eresolspectr}
\end{equation}
Finally , the large $x$ behaviour then determines the solutions 
completely and one finds
\begin{subequations}
\label{eresolsigiiz}
\begin{eqnarray}
{\partial \sigma_2(x) \over \partial z} &=&
- {1\over \varphi_2^2(x)} \,
\int_x^{+\infty } \d y\, \varphi_2^2(y) ,
\\
{\partial \sigma_1(x) \over \partial z}
& = & 
{1\over \varphi_1^2(x)}\int^x_{-\infty }
\d y\,\varphi_1^2(y)
\end{eqnarray}
\end{subequations}

({\em General symmetric potentials.}) 
We now assume $u(x)=u(-x)$. We then choose
\begin{equation}
\varphi_1(-x) = \varphi_2(x) \equiv \varphi(x). 
\end{equation}
Instanton calculus suggests that it is then also 
interesting to consider the quantities
\begin{subequations}
\begin{eqnarray}
\label{edefra}
r_a(x) &=& R(x,-x) = 
\varphi^2(x) \, \theta (x) + \varphi^2(-x) \theta (-x)\,,
\\
G_a(z) &=& \tr PL^{-1}(z) =
\int \d x\, r_a(x)=2\int_0^\infty\d x\, \varphi^2(x)\,,
\end{eqnarray}
\end{subequations}
where $P$ is the $x\mapsto -x$ reflection operator.

Using equations (\ref{eresolphisiga}) and 
(\ref{einstwrons}) taken at $x=0$ to normalize, 
we obtain a first expression
\begin{equation}
G_a(z) = \int_0^\infty {\d x \over \sigma _+(x)}
\exp\left[-2\int_0^x\d y\, \sigma _+(y)\right]. 
\label{eresolGasig} 
\end{equation}
We then combine with equation (\ref{eresolsigiiz}) to
obtain
\begin{equation}
G_a(z) = {\partial \over \partial z}
\ln \left(\varphi(0)\over\varphi'(0)\right)\,.  
\end{equation}
In particular, integrating over $z$, one finds an expression 
for the generalization of the Fredholm determinant:
\begin{equation}
\exp\left[-\tr P\ln(H-z)\right] = 
\prod_{N=0}\left(z-z_{2N}\over z-z_{2N+1}\right)
\propto{\varphi' (0,z)\over \varphi(0,z)}=
{\sigma'_+ (0) \over \sigma_+(0)}=
-{1\over [\sigma _+(0)]^2},
\label{eresolPFredh}
\end{equation}
an expression that can be easily understood since for even eigenfunctions
$\varphi'(0)$ vanishes, while $\varphi(0)$ vanishes for odd eigenfunctions.

%
%
\section{Example: Harmonic Oscillator}

As an illustration, 
let us apply this formalism to the harmonic oscillator 
with $u(x)=x^2$ and $z=2E$.

The trace of the resolvent for the harmonic oscillator  formally reads
\begin{equation}
G_{\rm osc.}(E) = \sum_{N=0}{1\over N+1/2-E}\,, 
\end{equation}
which diverges. However, the first derivative of 
$G(E)$ is defined. Integrating one can choose
\begin{equation}
\label{goscii}
G_{\rm osc.} (E) = - \psi(\ud - E), 
\end{equation}
where $\psi(z)$ is the logarithmic derivative of the $\Gamma $-function.
Finally, from $G_{\rm osc.} (E)$ one derives the Fredholm determinant  
\begin{equation}
{\mathcal D}_{\rm osc.}(E) =
1/\Gamma (\ud-E),
\label{einsthoscdet}
\end{equation}
which vanishes on the poles of the $\Gamma $-function.

The solution  of the Schr\"odinger equation that 
decreases for $x\to+\infty $ is proportional to
\begin{equation}
\varphi(x,E) = {1\over2i\pi} \,
\oint_C {\d p \over p^{E+1/2}} \, \e^{-x^2/2+px-p^2/4}\,,
\end{equation}
where the contour encloses the real negative axis. 
The asymptotic form for $x\to-\infty $ is given by steepest descent:
\begin{equation}
\varphi(x,E)
\mathop{\sim}_{x\to-\infty } {2\over\sqrt{\pi}} \,
(-2x)^{-E-1/2} \, \sin\pi(E+\ud)\e^{x^2/2}\,.
\end{equation}
For $x\to+\infty $, the integral is dominated by 
the neighbourhood of the origin:
\begin{equation}
\varphi(x,E)
\mathop{\sim}_{x\to+\infty }
x^{E-1/2}\,{1\over \Gamma (E+\ud)}\, \e^{-x^2/2}\,.
\end{equation}
Then,
\begin{equation}
{\mathcal D}(E)
\propto 
{\sin\pi(E+\ud) \over  \Gamma (E+\ud )}
\propto
{1\over \Gamma (\ud-E)}\,.
\end{equation}
Moreover,
\begin{subequations}
\begin{eqnarray}
\varphi(0,E) &=&
{1\over \pi} \, 2^{-E-1/2} \, \sin[\pi(E+\ud )] \, 
\Gamma \left(\frac{1}{4} - \ud E\right)\,,
\\
\varphi'(0,E) &=&
-{1\over \pi} \, 2^{1/2-E} \, \sin[\pi(E+\ud )] \, 
\Gamma \left(\frac{3}{4} - \ud E\right)\,,
\end{eqnarray}
\end{subequations}
and thus the expected result is obtained.

For example, in the case of the harmonic oscillator,
\begin{equation}
G_a(E) = \psi\left(\frac{1}{4}-\frac12\, E\right) - 
\psi\left(\frac{3}{4} - \frac12 \, E\right) =
- {1\over 2E} + {1\over 8 E^3} + {\mathcal O}(1/E^5),
\label{eresolGahar}
\end{equation}
an expansion that is reproduced by the WKB approximation. 

%
%
\chapter{Spectral Equation: A Few Terms of Perturbative and WKB Expansions}
\label{appinstcal}
 
%
%
\section{Perturbative Expansion}
\label{appPerturbative}

For illustration purpose, we first calculate a few 
terms of the perturbative expansion of the 
function $B(E,g)$, using the expansion (\ref{eRiccpert}).
We then verify the relation between perturbative and WKB expansions.

We expand the Riccati equation (\ref{eRiccat})
in powers of $g$,
\begin{equation}
S(q) = \sum_{k\ge 0} g^k s_k(q)\,.  
\end{equation}
Setting 
\begin{equation}
U(q) = \sqrt{2V(q)}\,, 
\end{equation}
we first find
\begin{equation}
s_0(q)=U(q),\quad 
s_1(q)={1\over 2 U(q)}\bigl( U'(q)-2E \bigr)\,. 
\end{equation}
Higher orders are given by solving the recursion relation
\begin{equation}
s_k(q) =
{1\over 2 U(q)}\left(s'_{k-1}(q)-\sum_{l=1}^{k-1}s_{k-l}(q) s_l(q)
\right).
\end{equation} 
At order $g^2$, one finds
\begin{equation}
s_2(q) = {1\over 8 U^3(q)}
\left(2U''U-3U'{}^2+4EU'-4E^2\right)\,.
\end{equation} 
We then simply need the residues at $q=0$.  

As we show later, it is often convenient to parameterize the potential
as the solution of the equation
\begin{equation} 
U'{}^2 = \rho(U) = 1 +  {\alpha_1}\,U + {\alpha_2}\,U^{2} + 
{\alpha_3}\,U^{3} + {\alpha_4}\,U^{4} + {\mathcal O}\left(U^5\right).
\label{einstUpU}
\end{equation}
Beyond perturbation theory, this implies simple well 
or symmetric double well potentials. For example, the quartic 
double-well potential corresponds
to $\rho =1-4u$ and the cosine potential to $\rho =1-4u^2$.

Up to the order of $g^6$,
the solution has the expansion
\begin{eqnarray}
V(q) & = &
\frac{1}{2} q^2 +  
\frac{1}{4} \, \alpha_1 \, q^3 +
\left( \frac{1}{6}\, \alpha_2+ \frac{1}{32}\alpha_1^2 \right) \, q^4 +
\left( \frac{1}{16} \, \alpha_1 \, \alpha_2 + 
\frac{1}{8} \, \alpha_3 \right) q^5 
\nonumber\\
& & + \left[ 
\frac{3}{80} \, \alpha_1 \, \alpha_3+
\frac{1}{45}\,\alpha_2^2+
\frac{1}{10}\,\alpha_4+
\frac{1}{4} \, \alpha_1 \, 
\left(
\frac{1}{48} \, \alpha_1 \, \alpha_2 +
\frac{1}{8}\, \alpha_3
\right) \right] \, q^6\,.
\label{eVqexp}
\end{eqnarray}
This suggests  an alternative method of calculation. 
Introducing the parameterization
\begin{equation}
S = \sigma_+(U) + g U' \sigma_-(U) = U + {\mathcal O}(g)\,,
\end{equation}
one obtains another form of Riccati's equation:
\begin{subequations}
\label{einstRiccasig}
\begin{eqnarray}
& & \sigma_+'(u)-2\sigma _+(u)\sigma _-(u) = 0\,, 
\label{einstRiccasiga}
\\
& & g^2 \left[\rho (u)\bigl( \sigma _-'(u)-\sigma _-^2(u)\bigl)+ 
\ud\rho'(u)\, \sigma _-(u)\right] -
\sigma_+^2(u) + u^2 - 2gE = 0\,.
\label{einstRiccasigb}
\end{eqnarray}
\end{subequations}
(One may first calculate $\sigma _+^2$.)
Then
\begin{equation}
B(E,g) =
-{1\over 2i\pi g} \, \oint\d u {\sigma _+(u) \over \sqrt{\rho (u)}},
\end{equation}
and the residue at $q=0$ becomes the residue at $u=0$.

In terms of the parameters (\ref{einstUpU}), one obtains
up to the order $g^2$,
\begin{eqnarray}
& & B(E,g) =
E + \left[\left( - \frac {1}{4} \alpha_2  + 
\frac {3}{16} \alpha_1 ^{2}\right) E^{2} - 
\frac {1}{16} \alpha_2 + 
\frac {1}{64} \alpha_1 ^{2}\right] \, g 
\nonumber\\
& &
+ \left[
\left( - \frac {1}{4} \, \alpha_4  + 
\frac {3}{8} \, \alpha_1 \, \alpha_3  + 
\frac {3}{16} \, \alpha_2^2  - 
\frac {15}{32} \, \alpha_1^{2} \, \alpha_2 + 
\frac {35}{256} \, \alpha_1^{4}\right) \, E^{3}  \right.
\nonumber\\
& & \left. + \left( - \frac {5}{16} \, \alpha_4 + 
\frac {5}{32} \, \alpha_1 \, \alpha_3 + 
\frac {5}{64} \, \alpha_2^{2} - 
\frac {15}{128} \, \alpha_1^{2} \, \alpha_2 + 
\frac {25}{1024} \, \alpha_1^{4}\right) E\right]\,
g^{2}\,.
\label{einstBgengii}
\end{eqnarray}
One immediately verifies  that it agrees with the results obtained for 
various special potentials. In the case of an even potential 
($\alpha_1 = \alpha_3 = 0$) and finite 
angular momentum, the expression at the same order becomes
[see also equation (\ref{einstj})]
\begin{eqnarray}
& & B(E,g) = E+ \alpha_2 \,
\left[ - \frac {1}{4}  E^{2} +
\frac{1}{12}\,\left(j^2-1\right) \right] g 
+   \left[\left( - \frac {1}{4}  \alpha_4   + 
\frac {3}{16} \alpha_2 ^ 2 \right) \, E^3 \right.
\nonumber\\
& & \quad \left.
+ \left(\left\{j^2-1\right\}\,
\left(-\frac{17}{240}\,\alpha_2^2 + \frac{3}{20}\,\alpha_4\right) - 
\frac {1}{5} \, \alpha_4 + 
\frac {1}{40} \, \alpha_2 ^{2} \right) \, E\right] \, g^{2}\,.
\label{einstBgengij}
\end{eqnarray}

%
%
\section{WKB Expansion: Second--Order Calculation}
\label{sSecond}

({\em General considerations.})
The WKB expansion of $B(E,g)$ is an expansion in powers of $g$ at $gE$ fixed.
Therefore, we expand $S(q)$ in powers of $g$ at $gE$ fixed 
(or alternatively equation (\ref{einstresolv})): 
\begin{equation}
S(q)=\sum^\infty_{k = 0} g^k S_k(q). 
\label{eSWKB}
\end{equation}
Then,
\begin{equation}
S_1={S_0' \over 2S_0} \,,
\end{equation}
and the recursion relation becomes
\begin{equation}
S_k(q)={1\over 2 S_0(q)}\,
\left(S'_{k-1}(q) -
\sum_{l=1}^{k-1} S_{k-l}(q)\,S_l(q) \right)\,. 
\end{equation}
At order $g^2$, one finds
\begin{equation} 
S_2 = {S_0'' \over 4 S_0^2} - {3 S_0'{}^2 \over 8 S_0^3}\,.
\label{eWKBii}
\end{equation}
For the order $g^2$ contribution to the functions
$A(E,g)$ and $B(E,g)$, we need 
only the integral of $S_2$. Integrating by parts, one finds
\begin{equation}
\label{eByParts}
\int\d q\,S_2(q)=\int \d q\,
{S_0'{}^2(q) \over 8 S_0^3(q)}\,.
\end{equation}  

({\em Perturbative and WKB expansions.}) One recovers the  
contributions to the perturbative expansion
by expanding $S_k(q)$ in powers of $Eg$ and calculating the residues
at this minimum of the potential.

We expand the first WKB term in powers of $gE$:
\begin{equation}
\label{eS0powerii}
S_0 = U \sum^\infty_{n=0} \left(2gEU^{-2}\right)^n\,
{\Gamma(n-\ud)\over\Gamma(n+1)\Gamma(-\ud)}.
\end{equation}
One has thus to evaluate [see also the equations (\ref{einstIsi})
and (\ref{eI0Res})]:
\begin{equation}
-{1\over 2\,{\rm i}\,\pi} \, \oint_{C}\d q\, U^{1-2n} =
-{1\over 2\,{\rm i}\,\pi}\oint \d u\, {u^{1-2n} \over\sqrt{\rho (u)}}\,, 
\label{eperti}
\end{equation}
where the parameterization (\ref{einstUpU}) has been introduced.

For the next WKB order, we calculate the integrand
of~(\ref{eByParts}) more explicitly
\begin{equation}
\label{eS2powerii}
{ S_0'{}^2(q) \over 8 S_0^3(q)} = 
{U'{}^2 \, U^2\over 8 S_0^5}\,.
\end{equation}
Expanding $S_0$, one finds
[see also the equations (\ref{ebarii})
and (\ref{eI2Res})]:
\begin{equation} 
\label{econtourii}
\int\d q\,S_2 =
{1\over 8} \sum^\infty_{n=-1} {(2gE)^n\over \Gamma(n+1)}\,
{\Gamma(n+5/2)\over\Gamma(5/2)} \,
\int\d q\,U'{}^2 \, U^{-2n-3}\,.
\end{equation}
One thus has to evaluate 
\begin{equation}
-{1\over 2\,{\rm i}\,\pi} 
\oint_{C}\d q\,U'{}^2U^{-2n-3} =
-{1\over 2\,{\rm i}\,\pi} \oint\d u\,  
\sqrt{\rho (u)}u^{-2n-3}.
\label{epertii}
\end{equation}

({\em A special class of potentials.})
We follow~\cite{ZJ1984jmp}.  
It is possible to calculate many terms of the
WKB expansion for the class of potentials such that, 
in the parameterization (\ref{einstUpU}), 
\begin{equation} 
\rho (u)=1-4u^m\,.
\label{eclassm}
\end{equation}
Here, $m=1$ corresponds to the double-well potential, 
$m=2$ to the cosine potential, for $m=3,4$ the potentials are 
meromorphic elliptic functions. Higher values no longer correspond to
meromorphic functions (except for $m=6$ where $U^2$ is still meromorphic). 
The case
$m\ge 3 $ odd corresponds to potentials which diverge on the real axis, 
$m$ even to periodic and finite potentials on the real axis.

One verifies that the term of order $g^l$ in the expansion is of the form
$P_l(u)/u^{2l-1}$, where $P_l$ is polynomial. The expansion of $B$ thus
involves only the integral  
\begin{equation}
L(\nu)=-{1\over 2\,{\rm i}\,\pi}\oint_{C}\d q\, U^{1-2\nu}.
\label{efuL}
\end{equation}
We change variables $q\mapsto u=U(q)$ in (\ref{efuL}) and find
\begin{equation}
L_m(\nu)=-{1\over 2\,{\rm i}\,\pi} \, 
\oint_{C}\d u\left(1-4u^m\right)^{-1/2}
u^{1-2\nu}\,.
\end{equation}
Expanding the root in powers of $u$, we obtain
\begin{equation}
L_m(\nu) =
-2^{4(\nu-1)/m}{\Gamma\left(\ud+2{(\nu-1)\over m}\right)\over
\Gamma\left(\ud\right)\,
\Gamma\left(1+2{(\nu-1)\over m}\right)}\quad{\rm for}\quad  
2(\nu-1)=0\pmod m \,.
\end{equation}
For example,
\begin{eqnarray}
& & -{1\over 2i\pi}\oint_{C}\d q\, S_0 = -\sum_{n\ge 0}\left(2gE\right)^n 
{\Gamma\left(n-\ud\right)\over
\Gamma\left(n+1\right)\Gamma\left(-\ud\right)}L(n), 
\nonumber\\
& & = -\mbox{\Large $\Sigma$}'_{k\ge 0} \,\, 
{(gE)^{1+km/2} \, 2^{1+k(2+m/2)} \,
\Gamma\left({km\over2}+\ud\right) \, 
\Gamma\left(k+\ud\right) 
\over\Gamma\left({km\over2}+2\right) \, 
\Gamma\left(-\ud\right) \, 
\Gamma\left(k+1\right) \, \Gamma\left(\ud\right)}\,,
\end{eqnarray}
where $\sum'$ means the sum over $k$, including 
only those terms for which $k\,m$ is even.

%
%
\chapter{Multi--Instantons}

%
%
\section{Determinant}
\label{apninstdet}

In this chapter,
we give some indication about a few additional technical ingredients that
are involved in the calculation of multi-instanton contributions.
We can write the operator $ M $ defined by equation 
(\ref{eMdw}) as
\begin{equation} 
M=- \left({ \d  \over \d  t} \right)^2+1+ 
\sum^{n}_{i=1} v(t-t_{i}), 
\end{equation}
in which $ v (t) $ is a potential localized around $ t=0$:
\begin{equation} 
v (t) = {\mathcal O}\left( \e^{- | t |} \right), 
\quad \left\vert t \right\vert \rightarrow \infty  
\end{equation}
and $ t_{i} $ are the positions of the instantons. 

We want to calculate
\begin{equation} 
\det M\, M^{-1}_{0}=
\det \left\lbrace 1+ \left[ - \left(\d 
\left/ \d  t\right. \right)^2+1 \right]^{-1} \sum^{n}_{i=1}v
(t-t_{i} ) \right\rbrace . 
\end{equation}
Using the identity $ \ln\det =\tr\ln $, we expand the r.h.s.\
in powers of $v(t)$:
\begin{eqnarray} 
& & \ln\det MM^{-1}_{0} = 
\sum^{\infty}_{k=1}{ (-1 )^{k+1} \over k}\,
\int \prod^{k}_{j=1} \d  u_{j} \bigg[ \Delta (u_1-u_2) \,
\nonumber\\
& & \quad \sum^{n}_{i_{1}=1} v(u_2-t_{i_{1}} )
\Delta (u_2-u_3)\cdots 
\Delta (u_{k}-u_{1} ) 
\sum^{n}_{i_{k}=1}
v (u_{1}-t_{i_{k}} ) \bigg]
\end{eqnarray}
with the definition
\begin{equation} 
\Delta (t)= \Bigl< 0 \Bigl | 
\left[ - \left( \d  / \d  t\right)^2 +
1 \right]^{-1} \Bigr| t \Bigr> \sim
{ 1 \over 2} \, \e^{- \left\vert t \right\vert}
\quad {\rm for} \quad 1 \ll t \ll \beta\,.
\end{equation}
It is clear from the behaviour of $ v (t) $ and of
$ \Delta (t) $, that when the instantons 
are largely separated, only the
terms in which one retains from each potential the same instanton contribution
survive. Therefore,
\begin{eqnarray} 
& & \ln\det MM^{-1}_{0} = \nonumber\\
& & \qquad n \sum^{\infty}_{k=1}{ (-1 )^{k+1} \over k} \,
\int \prod^{k}_{j=1} \d u_{j} 
\Delta (u_{1}-u_2 ) \, v(u_2) \cdots 
\Delta(u_{k}-u_{1} ) \, v(u_1) \,,
\nonumber\\
& & \qquad{\rm for}\quad
\left\vert t_{i}-t_{j} \right \vert \gg 1\,. 
\end{eqnarray}  
We recognize $ n $ times the logarithm of the one-instanton
determinant. 

%
%
\section{Instanton Interaction}
\label{ssninstint}

We assume, as in chapter~\ref{ssninstg}, that the
potential has two degenerate minima at 
the points $x=0$ and $x=x_{0}$ with
\begin{subequations}
\begin{eqnarray}
V (x ) & = & \ud \, x^2 + {\mathcal O}\left(x^3 \right) \,,
\\
V (x ) & = & \ud \, \omega^2 \, \left(x-x_{0} \right)^2 +
{\mathcal O}\left( (x - x_{0} )^3 \right)\,. 
\end{eqnarray}
\end{subequations}
Let us write the one-instanton solution $ q_{c} (t) $
which goes from 0 to $q_0=x_{0}/\sqrt{g}$ as
\begin{equation} 
q_{c} (t)=f (t)/\sqrt{ g}\,.
\end{equation}
We choose the function $ f (t) $ in such a way that it
satisfies
\begin{subequations}
\label{eascon}
\begin{eqnarray}
x_{0}- f (t) & \sim & 
\sqrt{C_\omega} \, 
\e^{-\omega t} / \omega \quad {\rm for}\ t  \rightarrow + \infty 
\\
f (t) & \sim & 
\sqrt{C_\omega} \, \e^{t} \quad{\rm for } \ t \rightarrow -\infty\, . 
\end{eqnarray} 
\end{subequations}
By solving the equation of motion, 
it is easy to calculate the constant $C_\omega$:
\begin{equation} 
\label{edefCgeniii}
C_\omega  = x^2_{0} \, \omega^{2/(1+\omega)} \exp \left\lbrace{ 2\omega
\over 1+\omega} \left[ \int^{x_{0}}_{0} \d  
x \left({ 1 \over \sqrt{ 2V (x )}}-{1 \over x}-{1 \over
\omega (x_{0}-x )} \right)\right] \right\rbrace .
\end{equation}
We recognize the constant (\ref{edefCgen}).

We now construct instanton--anti-instanton pair configurations $q(t)$
which correspond  to trajectories starting from, and returning to, $q=q_{0}$ 
or $q=0$. Since we want also to consider the case of two successive
instantons, we assume, but only in this last case, that $V(x)$ is an
even function and has therefore a third minimum at $x=-x_{0}$.\par
According to the discussion of chapter \ref{ssninstdw}, we can take as a
two-instanton configuration
\begin{equation}
q_{1}(t)={1 \over \sqrt{ g}} \left(f_{+}(t)+ \varepsilon  
f_{-}(t)\right) , \qquad \varepsilon =\pm 1  
\end{equation}
with
\begin{equation}
f_{+}(t) =f (t-\theta / 2 ),\qquad
f_{-}(t) =f(-t-\theta / 2 )\,.
\end{equation}
Here, $\theta$ is a measure of the instanton separation. 
The case $ \varepsilon = 1 $ corresponds to an instanton--anti-instanton pair starting
from $q=q_{0}$ at time $-\infty$, approaching $q=0$ at intermediate times
and returning to $q_{0}$. The case $ \varepsilon =-1 $ corresponds to a
sequence of two instantons going from $-q_{0}$ to $q_{0}$. Finally, for the
classical trajectory which  goes, instead, from the origin to $q_{0}$ and back,
we can take
\begin{equation} 
q_2(t)=\left[f(t+{\theta/2} )+ f ({\theta/2}-t)-x_{0} \right]/
\sqrt{ g}\,. 
\end{equation}
We now calculate the  classical action corresponding to 
$q_{1}(t)$. We separate the action into two parts, corresponding at
leading order to the two instanton contributions:
\begin{equation} 
{\mathcal S} ( q_{1} ) = {\mathcal S}_{+}( q_{1} )+ 
{\mathcal S}_{-}( q_{1} )   
\end{equation}
with
\begin{eqnarray}
{\mathcal S}_{+}( q_{1} ) & = &
\int_{0}^{+\infty}
\left[{1 \over 2} \, \dot{q}_{1}^2 + 
{1 \over g} \, V \, \bigl(\sqrt{g} \, q_{1}(t) \bigr) \right]
\d t, 
\nonumber\\
{\mathcal S}_{-}(q_{1}) 
& = & \int^{0}_{-\infty}
\left[{1 \over 2} \, \dot{q}_{1}^2 + 
{1 \over g} \, V \, \bigl( \sqrt{g} \, q_{1}(t)\bigr)\right]
\d t\,.
\end{eqnarray}
The value $t=0$ of the separation point is somewhat arbitrary and can be
replaced  by any value which remains finite when $\theta$ becomes infinite.
We then use the properties that for $\theta$ large 
$f_{+}(t) $ is small for $t<0 $, and $f_{-} (t) $ is small for 
$t>0 $, to expand both terms. 
For example, for ${\mathcal S}_{+}$ we find
\begin{eqnarray} 
{\mathcal S}_{+}( q_{1} ) & = & 
{1 \over g} \, 
\int_{0}^{+\infty} \d t\, 
\left\lbrace \left[\frac12 \, [\dot{f}_{+}^2(t)]
+ V \, \bigl( f_{+}(t) \bigr)\right]  \right.
\nonumber\\
& & \qquad \left. + \varepsilon \, 
\left[ \dot{f}_{-}(t) \, \dot{f}_{+}(t) +
V'\bigl(f_{+}(t) \bigr) \, f_{-}(t) \right]
\right.
\nonumber\\
& & \qquad \left. + \frac12\,\left[ 
\dot{f}_{-}^2(t) + V''\bigl( f_{+}(t) \bigr) \, 
f_{-}^2(t) \right] \right\rbrace\,.
\label{eSminq}
\end{eqnarray} 
Since $f_{-}(t)$ decreases exponentially, only values of $t$ small compared
to $\theta/2$ contribute to the last term of equation 
(\ref{eSminq}) which is
proportional to $V''$. For such values of $t$, we have
\begin{equation}
\ud \, V''\bigl( f_{+}(t)\bigr) \, f_{-}^2 
\sim 
V\bigl( f_{-}(t) \bigr)\,. 
\end{equation}
For the terms linear in $f_{-}(t)$, we integrate by parts the kinetic term
and use the equation of motion
\begin{equation} 
\ddot f (t)=V' \left[ f (t) \right] \,.
\end{equation}
Only the integrated term survives and yields
\begin{equation}
\int_{0}^{+\infty}\d t \,
\left[ \dot{f}_{-}(t) \, \dot{f}_{+}(t) +
V'\bigl(f_{+}(t)\bigr) \, f_{-}(t) \right] = 
-\dot f(-\theta/2) \, f(-\theta/2) \,.
\end{equation}
The contribution ${\mathcal S}_{-}$ can be evaluated by exactly the same method. We
note that the sum of the two contributions reconstructs twice the classical
action $a$. We then find
\begin{equation} 
{\mathcal S} ( q_{1} ) = {1 \over g}\,
\left[ 2 \, a - 2 \, \varepsilon \,
f(-\theta/2) \dot f(-\theta/2) + \cdots \right] 
\end{equation}
with
\begin{equation} 
a = \int^{x_{0}}_{0} \sqrt{ 2V (x )} \d x\,.  
\end{equation} 
Replacing, for $\theta$ large, $f$ by its asymptotic form 
(\ref{eascon}), we finally obtain the classical action
\begin{equation} 
{\mathcal S} ( q_{1} ) =
g^{-1} \left[ 2 a - 2 C \varepsilon \, \e^{-\theta} +
{\mathcal O}\left( \e^{-2\theta} \right) \right]  
\end{equation}
and thus the instanton interaction. 

Following the same steps, we can calculate the classical action corresponding
to $q_2(t)$. The result is
\begin{equation} 
{\mathcal S}( q_2 ) = {1 \over g}\,
\left\{ 2 a - 2 \left[ f(\theta/2) - x_{0} \right] \dot
f(\theta/2) + \cdots \right\} \,,
\end{equation}
which for $\theta$ large is equivalent to
\begin{equation} 
{\mathcal S} ( q_2 ) = {1 \over g}\,
\left[ 2 a - 2 (C/\omega) \e^{-\omega \theta} \right]\,.  
\end{equation}
Finally, if we consider the case of a finite time interval $ \beta
$ with periodic boundary conditions, we can combine both results to find the
action of a periodic trajectory  passing close to $q=0$ and $q=q_{0}$:
\begin{equation} 
{\mathcal S} (q ) =g^{-1} \left[ 2 a - 2 C \left( \e^{-\beta+\theta} + 
\e^{-\omega\theta}/\omega \right) \right] \,,
\end{equation}
in agreement with equations (\ref{eAgenpot})
and (\ref{eCunCde}).
%

%
%
\section{Multi--Instantons from Constraints}
\label{ssninstco}

Although multi-instanton configurations do not correspond to solutions
of the equation of motion, it is nevertheless possible to modify the
classical action by introducing constraints and integrating over all possible
constraints. The main problem
with such a method is to find a system of constraints which are both
theoretically reasonable, and convenient for practical
calculations.

One can, for instance, fix the positions of the
instantons by introducing in the path integral (in the example of the
double-well)
\begin{equation} 
1 = \int \prod^{n}_{i=1} \,
\left[\int \d t\, \dot q_{\varepsilon_i}^2 (t-t_i) \right] \,
\delta\left[\int \d t\, \dot q_{\varepsilon_i}(t-t_i)\,
\bigl(q(t) - q_{\varepsilon_i}(t-t_i) \bigr) \right] 
\d t_{i}\,, 
\end{equation}
where $t_i$ are the instanton positions and $\varepsilon_i$ a succession of
$\pm$ indicating instantons and anti-instantons. One then uses an integral
representation of the $ \delta $-functions, so that 
the path integral becomes
\begin{eqnarray}
& & \left(\| \dot q_{+}\|^2 \over 2i\pi \right)^n 
\int \prod^{n}_{i=1} 
\d t_{i}\,
\d\lambda_i \,
\int[ \d  q (t)] \,
\prod^{n}_{i=1}\,
\exp\left[-{\mathcal S}( q,\lambda_i )\right]
\quad {\rm with} 
\nonumber\\
& & \quad {\mathcal S} ( q,\lambda_i ) = {\mathcal S}(q) +  
\sum^{n}_{i=1}
\lambda_{i}\,
\int \d t\, \dot q_{\varepsilon_i} (t-t_i) \,
\bigl(q(t)-q_{\varepsilon_i}(t-t_i)\bigr)\,. 
\end{eqnarray}
%
%

%
%
\chapter{Dispersion Relations and (non-)Borel Summability}
\label{sDisp}

\section{Dispersion Relations}
\label{ssDispersion}

({\em The simplest example.})
The nonalternating ($g > 0$) factorially divergent
series
\begin{equation}
f(g) \sim \sum_{n=0}^\infty n!\, g^n
\end{equation}
is generated by expanding in powers of $g$ the following
integral,
\begin{equation}
f(g) = \int_0^{\infty \pm {\rm i}\,\epsilon} 
\d t \, \frac{1}{1 - g t} \, \exp(-t)\,.
\end{equation}
The imaginary part due to the (half-)pole at $t = 1/g$ is 
\begin{equation}
{\rm Im} \, f(g) = \pm \frac{\pi}{g} \, \exp(-1/g)\,.
\end{equation}
Because $f(g)$ has a cut along the positive real axis,
one may write a dispersion relation in an obvious way as 
\begin{equation}
0 = - 2 \pi {\rm i} \, \Res{z = z_0} \frac{f(z)}{z - z_0} +
\int^\infty_0 \d x \, \left[ \frac{1}{x - z_0} \,
(f(x + {\rm i}\,\epsilon) - f(x - {\rm i}\,\epsilon)) \right]\,.
\end{equation}
({\em Dispersion relation.}) The equality
\begin{equation}
\label{edispersion}
f(z_0) = \frac{1}{\pi}\,
\int^\infty_0 \d x \, \frac{1}{x - z_0} \, {\rm Im} \, f(x)
\end{equation}
follows immediately. Let us now start from this 
dispersion relation as the fundamental property of a function 
which has a cut along the positive real 
axis and fulfills sufficient conditions in order to ensure
the convergence of the integral.
Let us assume furthermore, that $f(z_0)$ can be 
expanded as a formal power series,
\begin{equation}
f(z_0) \sim \sum_{n=0}^\infty a_n \, z_0^n\,.
\end{equation}
Expanding the right-hand side of (\ref{edispersion}) into a
power series in $z_0$, 
\begin{equation}
\frac{1}{x - z_0} = \frac1x \,
\sum_{n=0}^\infty \left( \frac{z_0}{x} \right)^n\,,
\end{equation}
we immediately have
\begin{equation}
\label{eCalcan}
a_n = \frac{1}{\pi}\,
\int^\infty_0 \d x \, \frac{1}{x^{n+1}} \, {\rm Im} \, f(x)\,.
\end{equation}
In many cases, possible convergence problems near $x = 0$ are
eliminated due to a nonanalytic factor of the form $\exp(-1/x)$ 
that enters into ${\rm Im} \, f(x)$.

({\em A calculational example.}) We now replace
$x \to g$. For a function whose imaginary part is 
\begin{subequations}
\label{eCalcExample}
\begin{equation}
\label{eCalcExamplea}
{\rm Im} \, f(g) = \frac{1}{g} \, 
\exp\left(-\frac{1}{a\, g}\right)\,
(1 + b g + c g^2 + d g^3 + \dots)\,,
\end{equation}
the evaluation of the integral on the right-hand side of
(\ref{eCalcan}) leads to 
\begin{equation}
\label{eCalcExampleb}
a_n = \frac{a^{n+1}}{\pi} \, n! \,
\left( 1 + \frac{b}{n} +\frac{c}{n\,(n+1)} +
\frac{d}{n\,(n+1)\,(n+2)} + \dots \right)\,.
\end{equation}
\end{subequations}
This example shows that correction terms to the 
imaginary part of order $g$ to the imaginary 
part along the positive real axis (the cut) 
correspond to subleading corrections to the factorial 
growth of the ``perturbative coefficients'' $a_n$ of 
relative order $n^{-1}$. The correction terms can 
be summarized in a natural way in terms of an inverse
factorial series. This consideration is the basis
for the ``matching'' discussed in chapter~\ref{ssCorrAsymp}.

%
%
\begin{figure}[htb!]
\begin{center}
\begin{minipage}{12.0cm}
\begin{center}
\epsfxsize=91.3mm
\epsfysize=39.8mm
\centerline{\epsfbox{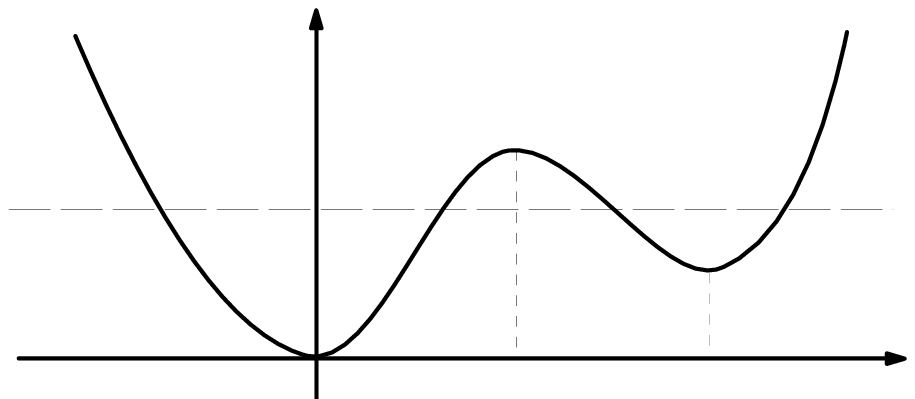}}
\caption{\label{figninstii}
The four roots of equation (\ref{eqxit}).}
\end{center}
\end{minipage}
\end{center}
\end{figure}

%
%
\section{A Simple Example of non--Borel Summability}
\label{sSimple}

Let us try to illustrate the problem of non-Borel summability with the
example of a simple integral, which shares some of the features of the problem
in quantum mechanics which we have studied in chapter~\ref{ssninstdw}. 
We consider the function
\begin{equation} 
I(g)={1 \over \sqrt{ 2\pi}} \int^{+\infty}_{-\infty}
\d  q\, \exp\left[-{1 \over g}V (q \sqrt{ g} )\right],
\label{eintIg} 
\end{equation} 
where $ V (x) $ is an entire function with an absolute minimum at $x=0$,
$V(0)=0$.  For $g$ small $I(g)$ can be calculated by steepest descent,
expanding $V$ around $q=0$:
\begin{equation} 
I(g)=\sum_{k\ge 0}I_{k}\,g^{k}. 
\end{equation}
It is easy to write a finite dimensional integral of the form 
(\ref{eintIg}) as a
generalized Borel or Laplace transform,
\begin{equation} 
I(g) = {1 \over \sqrt{ 2\pi}} \int \d  q\, \d  t\, 
\delta \, \left[ V (q \sqrt{ g} )-t \right] \, \e^{-t/g}.
\end{equation}
We integrate over $ q $:
\begin{equation} 
I(g) = { 1\over\sqrt{2\pi g}} \,
\int^{\infty}_{0} \d t\, \e^{-t/g} \,
\sum_{i}{1 \over \left\vert V'\left[ x_{i}(t) \right] 
\right\vert} \,,
\label{eintIgb} 
\end{equation}
in which $ \{x_{i} (t) \}$ are the solutions of the equation
\begin{equation} 
V \left[ x_{i} (t) \right] =t\,. 
\label{eqxit} 
\end{equation}
When the function $ V(x) $ is monotonic both
for $ x $ positive and negative, the equation 
(\ref{eqxit}) has two solutions
for all values of $ t $ and the equation (\ref{eintIgb}) 
is directly the Borel
representation of the function $ I(g) $, which has a
Borel summable power series expansion. 

%
%
\begin{figure}[htb!]
\begin{center}
\begin{minipage}{12.0cm}
\begin{center}
\epsfxsize=105.3mm
\epsfysize=50.5mm
\centerline{\epsfbox{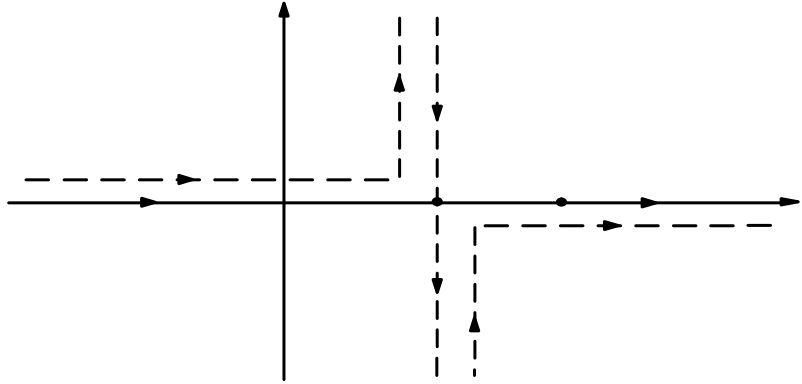}}
\caption{\label{figninstiii}
The different contours in the $x$-plane.}
\end{center}
\end{minipage}
\end{center}
\end{figure}

We now assume, instead, that $V(x)$ has a second local minimum which gives a
negligible contribution to $ I(g) $ for $g$ small. A simple example is
\begin{equation} 
V (x) = \ud x^2 - 
\frac{1}{3a} \, x^3 \, (1+a) +
\frac{1}{4a} \, x^{4}, \quad 
\ud<a<1\,, 
\label{exampot}
\end{equation}
which has a minimum at $x=1$.
Between its two minima the potential $ V(x) $ has a maximum, located
at $x=a$, whose contribution dominates the large order behaviour of the 
expansion in powers of $ g $: 
\begin{equation} 
I_{k} \mathop{\propto}_{k\to\infty} \Gamma(k) \,
\left[ V (a) \right]^{-k} \,, \qquad V(a) > 0 
\end{equation}
(in the example (\ref{exampot}) $V(a) = a^2 \, (1-a/2) / 6$) 
and the series is not Borel summable. 

The {\em naive}\/ Borel transform of $I(g)$ is obtained by 
retaining in equation
(\ref{eintIgb}) only the roots of equation 
(\ref{eqxit}) which exists for $ t $ small.
The singularities of the Borel transform then 
correspond to the zeros of $ V''(x) $. 

For the potential (\ref{exampot}) the expression 
(\ref{eintIgb}) has the form
\begin{eqnarray} 
I(g) &=& {1\over\sqrt{2\pi g}} \,
\int^{+\infty}_{0} \d  t\, \e^{-t/g} \,
\left[{1 \over \left| V' \bigl(x_{1}(t) \bigr) \right|} +
{\theta\bigl(V(a)-t\bigr) \over \left| V' \bigl(x_2(t) \bigr) \right|}
\right. 
\nonumber\\
& &
\left. \quad + {\theta\bigl(V(a)-t\bigr) \, \theta\bigl(t-V(1)\bigr) \over 
\left| V'\bigl(x_3(t)\bigr) \right|} + {\theta\bigl(t-V(1)\bigr)
\over \left| V' \bigl(x_{4}(t) \bigr) \right|} \right]  
\label{eIgexam} 
\end{eqnarray}
with the definitions (see figure \ref{figninstii}):
$ x_{1} (t) \leq 0 \leq x_2 (t) \leq a \leq 
  x_3 (t)\leq 1\leq x_{4} (t)$.

The idea of the analytic continuation is to integrate each
contribution up to $ t = +\infty $ following a contour which passes below or
above the cut along the positive real axis. 
This means that we consider $ x_2 (t) $ to solve the equation:
\begin{equation} 
V \left[ x_2 (t) \right] = t \pm {\rm i} \, \varepsilon \,. 
\end{equation}
The sign is arbitrary. Let us, for instance, choose the positive sign.
We then have to subtract this additional contribution. We  proceed in
the same way for $ x_3 (t) $ for $ t > V (a ) $.
Since $ x_2 (t) $ and $ x_3 (t) $ meet at $t = V(a)$, 
the analytic continuation will correspond to take for $x_3(t) $ the
other solution
\begin{equation} 
V \left[ x_3 (t) \right] =t \mp {\rm i}\, \varepsilon\, .
\end{equation}
We thus  have to subtract from the total expression the
contributions of two roots of the equation. But it is easy to verify that this
is just the contribution of the saddle point located at $ x=a $, which
corresponds to a maximum of the potential.

Therefore, we have  succeeded in writing expression 
(\ref{eIgexam}) as the sum
of three saddle point contributions 
(see figure \ref{figninstiii}). There
is some arbitrariness in this decomposition which here corresponds to the
choice $ \varepsilon =\pm 1 $. 

In the complex $ x $ plane, we have replaced the initial contour $ C $ on
the real positive axis, by a sum of three contours $ C_{1} $, $ C_2 $
and $ C_3 $ corresponding to the three saddle 
points located at $0, a, 1$.

%
%
\chapter{Degenerate Minima: Energy Splitting and Schr\"odinger Equation}
\label{ssSchrodsplit}

We consider the Hamiltonian
\begin{equation}
H = -\ud \d^2_x + V(x) 
\label{eSchrodsta}
\end{equation}
where the potential is analytic on the real axis, even: 
$V(x) = V(-x)$ and has
an absolute minimum located at  $x = \pm x_m$ 
(choosing $x_m>0$) where it vanishes. Eigenfunctions and eigenvalues 
can be calculated in a perturbative expansion in each well of the potential.
There are twice degenerate to all orders. We call $\psi_{N}(x)$ the
WKB eigenfunction corresponding to the expansion near the well around
$x=x_m$:
\begin{equation}
-\ud \psi''_{N}(x) 
+ V(x)\psi_{N}(x) = E_N \psi_{N}(x)\,.
\end{equation}
Then the second eigenfunction with the same perturbative energy is
$\psi_{N}(-x)$ and it is a WKB approximation near $-x_m$. We 
know that the true eigenfunctions are even or odd.
To calculate the energy difference 
$\delta E_N = E_{+,N} - E_{-,N}$ we construct 
trial wave functions $\phi_{N,\varepsilon}(x)$, which are even
or odd depending on the value $\varepsilon=\pm 1$, 
\begin{equation}
\phi_{N,\varepsilon}(x) =  
\left\{ \begin{array}{cc}
\psi_{N}(x) \, \left(
1 + \varepsilon \,
{\psi_{N}(-\alpha) \over \psi_{N}(\alpha)}
\right) & 
\mbox{for} \quad \alpha < x\,, \\
\psi_{N}(x) + \varepsilon \psi_{N}(-x), &
\mbox{for} \quad -\alpha < x < \alpha\,, \\
\psi_{N}(-x) \left(\varepsilon + 
{\psi_{N}(-\alpha) \over \psi_{N}(\alpha)}
\right) & 
\mbox{for} \quad x < -\alpha\,, 
\end{array}
\right.
\end{equation}
where $x_m > \alpha>0$. At the points $x = \pm \alpha$ the 
function $\psi_N(x)$ is
exponentially small in the semi-classical limit and 
$\psi_N(-\alpha) \ll \psi_N(\alpha)$.

The trial wave functions are continuous, satisfy locally the Schr\"odinger
equation, but are not differentiable at the points $x \pm \alpha$. We now
calculate 
\begin{equation}
E_{N,\varepsilon}=
{\left<\phi_{N,\varepsilon}\right|H\left|\phi_{N,\varepsilon}\right> 
\over
\left<\phi_{N,\varepsilon}|\phi_{N,\varepsilon}\right>}. 
\end{equation}
Integrating by parts, one finds
\begin{eqnarray}
\lefteqn{\int_{-\infty}^{+\infty} \d x \,
\left[\ud \bigl( \phi'_{N,\varepsilon}(x) \bigr)^2 +
V(x) \, \phi^2_{N,\varepsilon}(x) \right]}
\nonumber\\
& = & E_N \, \int_{-\infty}^{+\infty} \d x\,
\left\{ \phi^2_{N,\varepsilon}(x)
+ \phi_{N,\varepsilon}(\alpha)\,
\left[ \phi'_{N,\varepsilon}(\alpha_-)
- \phi'_{N,\varepsilon}(\alpha_+) \right] \right\}\,.
\end{eqnarray} 
It follows, that up to smaller exponential corrections,
\begin{equation}
\delta E_N=-{\psi_N(\alpha)
\psi'_N(-\alpha)+
\psi_N(-\alpha)\psi'_N(\alpha)
\over 
\int\d x\,\psi_N^2(x)}, 
\label{eSchrodDE}
\end{equation}
where only the neighbourhood of $x=x_m$ contributes to the integral in the
denominator. We note that since $  \psi_N(\pm x)$ are two eigenfunctions with
the same energy, the quantity  
\begin{eqnarray}
W(\alpha) &=& -\psi_N(\alpha) \, \psi'_N(-\alpha)
-\psi_N(-\alpha) \, \psi'_N(\alpha) 
\nonumber\\
&=& -\psi_N(\alpha)\, \psi_N(-\alpha) \,
\left({\psi'_N(-\alpha)\over \psi_N(-\alpha)} +
{\psi'_N(\alpha) \over \psi_N(\alpha)} \right) 
\end{eqnarray} 
is a Wronskian and thus independent of $\alpha$. 
The calculation then follows closely the lines of chapter 
(\ref{ssresoldeg}), where the inverse quantity is calculated.
Equation (\ref{eSchrodDE})~can easily be generalized to the 
Schr\"odinger equation in
higher dimensions.

We now apply this result to the Schr\"odinger
equation in the normalization (\ref{ehamdw}).
At leading order in the semi-classical limit, the
eigenfunction $\psi_N(q)$ can be written, for 
$0 \le q < q_0$, as
\begin{eqnarray}
\psi_N(q) &=& 
q^N \, \sqrt{q/U(q)} \, \exp\left[-{1\over g} 
\int_0^q\d q'\, U(q')\right] 
\nonumber\\
& & \quad \times
\exp\left[ (N+\ud) \, \int_0^q \d q' \,
\left( {1\over U(q')} - {1\over q'}\right) \right] \,.
\end{eqnarray}
At this order $\psi'_N(q)/ \psi_N(q) \sim - U(q)/g$.
The value of the Wronskian $W(q)$ thus is
\begin{eqnarray}
W(q) &=& -\psi_N(q) \, \psi_N(q_0-q) \,
\left( {\psi'_N(q_0-q) \over \psi_N(q_0-q)} +
{\psi'_N(q) \over \psi_N(q)}\right) 
\nonumber\\
&=& {2\over g} \, C^{N+1/2} \, \e^{-a/2g} \,,
\end{eqnarray}
with [see equations~(\ref{edefC}), (\ref{edefA}) and~(\ref{edefU})]
\begin{subequations}
\begin{eqnarray}
\label{edefAii}
a &=& 2\, \int_0^{q_0}\d q\, U(q)\,, \\
\label{edefCii}
C &=& q_0^2 \, \exp\left[\int_0^{q_0} \,
\left({1\over U(q)} - {1\over q} - {1\over q_0-q}\right)\right]\,.
\end{eqnarray}
\end{subequations}
For the denominator, we need the perturbative eigenfunction near $q=0$ with the
same normalization. We thus set $q=x\sqrt{g}$ and obtain 
\begin{equation}
\psi_N(x\sqrt{g})\mathop{\sim}_{g\to0}g^{N/2}x^N\e^{-x^2/2}.
\end{equation}
This result, however, yields the coefficient of $\e^{-x^2/2}$ in the WKB limit
$q=x\sqrt{g}$ fixed,  that is $x$ large. Actually, 
we know that at leading order
the wave function is simply the eigenfunction of the harmonic oscillator.
The eigenfunction with norm $\sqrt{\pi}$ is
\begin{equation}
\varphi_N(x) =
{1\over 2^{N/2}\sqrt{N!}}\,
(x-\d_x)^N \, \e^{-x^2/2} \,
\mathop{\sim}_{x\to \infty}
{ 2^{N/2}\over\sqrt{N!}}x^N\e^{-x^2/2} .
\end{equation}
We conclude
\begin{equation}
\psi_N(x\sqrt{g})\sim \sqrt{N!}\,(g/2)^{N/2}.
\end{equation}
Therefore,
\begin{equation}
\int\d q\,\psi_N^2(q)=\sqrt{\pi g}\, N! \, (g/2)^N\,.
\end{equation}
Finally, we notice that in the $q$ variable and with the form 
(\ref{eSchrodsta})~of
the Schr\"odinger operator the spectrum is $E/g$. 
The one-instanton contribution to the energy difference follows:
\begin{equation}
\delta  E_N(g) \sim  
{2 \over\sqrt{2\pi}}{1\over N!} \,
\left(2 C\over g\right)^{N+1/2} \, \e^{-a/2g},
\end{equation}
a result  consistent with the expansion of the 
zeros of expression (\ref{einstsyge}).

Note that to go beyond leading order by this method is simple for $N$ 
fixed, but not for generic $N$ because the WKB expansion yields the
perturbative eigenfunction under the form of an expansion only valid for large
arguments. 

\newpage

\clearpage\fancyhead[R]{}

\end{document}